%% file: brs.tex
\documentstyle[a4,12pt]{report}

\newcommand{\nit}{\noindent}
\newcommand{\nl}{\newline}
\newcommand{\np}{\newpage}
\newcommand{\dsp}{\displaystyle}
\newcommand{\vs}[1]{\vspace{#1 ex}}
\newcommand{\hs}[1]{\hspace{#1 em}}
\newcommand{\bfr}{\begin{flushright}}
\newcommand{\efr}{\end{flushright}}
\newcommand{\bc}{\begin{center}}
\newcommand{\ec}{\end{center}}
\newcommand{\ben}{\begin{enumerate}}
\newcommand{\een}{\end{enumerate}}

\newcommand{\be}{\begin{equation}}
\newcommand{\ee}{\end{equation}}
\newcommand{\ba}{\begin{array}}
\newcommand{\ea}{\end{array}}
\newcommand{\ct}{\cite}
\newcommand{\bit}{\bibitem}
\newcommand{\dd}[2]{\frac{\partial{#1}}{\partial{#2}}}
\newcommand{\ag}{\alpha}
\newcommand{\bg}{\beta}
\newcommand{\gam}{\gamma}
\newcommand{\del}{\delta}
\newcommand{\eps}{\epsilon}
\newcommand{\ve}{\varepsilon}

\newcommand{\thg}{\theta}
\newcommand{\kg}{\kappa}
\newcommand{\lb}{\lambda}
\newcommand{\sg}{\sigma}
\newcommand{\rg}{\rho}
\newcommand{\fg}{\phi}
\newcommand{\vf}{\varphi}
\newcommand{\og}{\omega}
\newcommand{\Gam}{\Gamma}
\newcommand{\Del}{\Delta}
\newcommand{\Fg}{\Phi}
\newcommand{\Sg}{\Sigma}
\newcommand{\Og}{\Omega}
\newcommand{\Lb}{\Lambda}

\newcommand{\bA}{{\bf A}}
\newcommand{\bd}{{\bf d}}
\newcommand{\bD}{{\bf D}}
\newcommand{\bF}{{\bf F}}
\newcommand{\bN}{{\bf N}}

\newcommand{\hp}{\hat{p}} 
\newcommand{\hq}{\hat{q}}
\newcommand{\hF}{\hat{F}} 
\newcommand{\hG}{\hat{G}}
\newcommand{\hH}{\hat{H}} 
\newcommand{\hU}{\hat{U}} 
\newcommand{\cA}{{\cal A}}
\newcommand{\cF}{{\cal F}}
\newcommand{\cH}{{\cal H}} 
\newcommand{\cO}{{\cal O}}
\newcommand{\cP}{{\cal P}}
\newcommand{\cD}{{\cal D}}
\newcommand{\cM}{{\cal M}}
\newcommand{\cN}{{\cal N}}

\newcommand{\lh}{\left(}
\newcommand{\rh}{\right)}
\newcommand{\ld}{\left.}
\newcommand{\rd}{\right.}

\newcommand{\cL}{{\cal L}}

\newcommand{\Slashed}{\hspace{-1.3ex}/\hspace{.2ex}}
\newcommand{\der}{\partial}
\newcommand{\Der}{D}
\newcommand{\sDer}{\Der\Slashed}

\begin{document}

\pagestyle{empty}
\begin{flushright}
NIKHEF 01-018
\end{flushright}

\begin{center}
{\Large{ \bf{Aspects of BRST Quantization$^\dagger$} }}\\
\vs{3}

{\large J.W.\ van Holten$^*$} \\
\vs{2}

{\large{NIKHEF and Vrije Universiteit}} \\
\vs{2}

{\large{ Amsterdam NL}} \\
\vs{2}
{december 20, 2001}
\vs{10}

{\small{ \bf{Abstract} }} \\
\end{center}

\nit
{\footnotesize{
BRST-methods provide elegant and powerful tools for the construction 
and analysis of constrained systems, including models of particles, 
strings and fields. These lectures provide an elementary introduction
to the ideas, illustrated with some important physical applications.}} \vfill
\nit

\vfill
\footnoterule

\noindent
{\footnotesize{ 
$^\dagger$ Lectures given at the Summerschool on {\em Geometry and 
Topology in Physics} \nl $~\hs{0.5}$ (Rot a/d/ Rot, Germany) sept 2001 \nl
$^*$ E-mail: v.holten@nikhef.nl }}

\np
~\hfill
\np

\nit
{\bf Conventions} \nl 

\nit
In these lecture notes we use the following conventions. Whenever two 
objects carrying a same index are multiplied (as in $a_i b_i$ or in 
$u_{\mu} v^{\mu}$) the index is a dummy index and is to be summed 
over its entire range, unless explicitly stated otherwise (summation 
convention). Symmetrization of objects enclosed is denoted by braces 
$\{...\}$, anti-symmetrization by square brackets $[...]$; the total 
weight of such (anti-)symmetrizations is always unity. 

In these notes we deal both with classical and quantum hamiltonian 
systems. To avoid confusion, we use braces $\{\, , \}$ to denote
classical Poisson brackets, brackets $[\, , ]$ to denote commutators
and suffixed brackets $[\, , ]_+$ to denote anti-commutators. 

The Minkowski metric $\eta_{\mu\nu}$ has signature $(-1,+1,...,+1)$, 
the first co-ordinate in a pseudo-cartesian co-ordinate system $x^0$ 
being time-like. Arrows above symbols $(\vec{x})$ denote purely 
spatial vectors (most often 3-dimensional).

Unless stated otherwise, we use natural units in which $c = \hbar = 1$. 
Therefore we usually do not write these dimensional constants explicitly. 
However, in a few places where their role as universal constants is not 
a priori obvious they are included in the equations. 

\np
~\hfill
\np
\pagestyle{plain}
\pagenumbering{arabic}

\include{brs1}

\include{brs2}

\include{brs2a}

\include{brs3}

\end{document}

%% file: brs1.tex
\chapter{Symmetries and constraints \label{ch1}}


The time-evolution of physical systems is described mathematically 
by differential equations of various degree of complexity, such as
Newton's equation in classical mechanics, Maxwell's equations 
for the electro-magnetic field, or Schr\"{o}dinger's equation 
in quantum theory. In most cases these equations have to be 
supplemented with additional constraints, like initial conditions 
and/or boundary conditions, which select only one ---or sometimes 
a restricted subset--- of the solutions as relevant to the 
physical system of interest.  

Quite often the prefered dynamical equations of a physical system 
are not formulated directly in terms of observable degrees of freedom, 
but in terms of more primitive quantities, such as potentials, from
which the physical observables are to be constructed in a second 
separate step of the analysis. As a result, the interpretation of
the solutions of the evolution equation is not always straightforward.
In some cases certain solutions have to be excluded, as they do not 
describe physically realizable situations; or it may happen that 
certain classes of apparently different solutions are physically 
indistinguishable and describe the same actual history of the system. 

The BRST-formalism \ct{brs,tyu} has been developed specifically to 
deal with such situations. The roots of this approach to constrained 
dynamical systems are found in attempts to quantize General Relativity 
\ct{feynman,dewitt} and Yang-Mills theories \ct{fp}. Out of these roots 
has grown an elegant and powerful framework for dealing with quite 
general classes of constrained systems using ideas borrowed from 
algebraic geometry.\footnote{Some reviews can be found in refs.\
\ct{kugo}-\ct{henn2}.}
 
In these lectures we are going to study some important examples of
constrained dynamical systems, and learn how to deal with them so 
as to be able to extract relevant information about their observable 
behaviour. In view of the applications to fundamental physics at 
microscopic scales, the emphasis is on quantum theory. Indeed, this 
is the domain where the full power and elegance of our methods become 
most apparent. Nevertheless, many of the ideas and results are 
applicable in classical dynamics as well, and wherever possible we 
treat classical and quantum theory in parallel.  

\section{Dynamical systems with constraints \label{s1.2}} 

Before delving into the general theory of constrained systems, 
it is instructive to consider some examples; they provide a 
background for both the general theory and the applications to 
follow later.  
\vs{1} 

\nit 
{\em 1.\ The relativistic particle.} \\
The motion of a relativistic point particle is specified completely
by its world line $x^{\mu}(\tau)$, where $x^{\mu}$ are the position 
co-ordinates of the particle in some fixed inertial frame, and $\tau$ 
is the proper time, labeling succesive points on the world line. All 
these concepts must and can be properly defined; in these lectures I 
trust you to be familiar with them, and my presentation only serves 
to recall the relevant notions and relations between them. 

In the absence of external forces, the motion of a particle w.r.t.\ 
an inertial frame satisfies the equation 
\be 
\frac{d^2 x^{\mu}}{d\tau^2} = 0. 
\label{1.2.1}
\ee 
It follows, that the four-velocity $u^{\mu} = dx^{\mu}/d\tau$ is 
constant, and the complete solution of the equations of motion is 
\be 
x^{\mu}(\tau) = x^{\mu}(0) + u^{\mu} \tau.
\label{1.2.2}
\ee 
A most important observation is, that the four-velocity $u^{\mu}$ 
is not completely arbitrary, but must satisfy the {\em physical} 
requirement 
\be 
u_{\mu} u^{\mu} = - c^2,
\label{1.2.3}
\ee 
where $c$ is a {\em universal} constant, equal to the velocity of
light, for all particles irrespective of their mass, spin, charge 
or other physical properties. Equivalently, eq.(\ref{1.2.3}) states
that the proper time is related to the space-time interval traveled
by 
\be 
c^2 d\tau^2 = - dx_{\mu} dx^{\mu} = c^2 dt^2 - d \vec{x}^{\,2},
\label{1.2.4}
\ee 
independent of the physical characteristics of the particle. 

The universal condition (\ref{1.2.3}) is required not only 
for free particles, but also in the presence of interactions. When
subject to a four-force $f^{\mu}$ the equation of motion (\ref{1.2.1}) 
for a relativistic particle becomes 
\be 
\frac{dp^{\mu}}{d\tau} = f^{\mu}, 
\label{1.2.5}
\ee 
where $p^{\mu} = m u^{\mu}$ is the four-momentum. Physical forces 
---e.g., the Lorentz force in the case of the interaction of a 
charged particle with an electromagnetic field--- satisfisy the 
condition 
\be 
p \cdot f = 0.
\label{1.2.6}
\ee 
This property together with the equation of motion (\ref{1.2.5}) 
are seen to imply that $p^2 = p_{\mu} p^{\mu}$ is a constant along
the world line. The constraint (\ref{1.2.3}) is then expressed by 
the statement that 
\be 
p^2 + m^2 c^2 = 0, 
\label{1.2.7}
\ee 
with $c$ the same universal constant. Eq.\ (\ref{1.2.7}) defines 
an invariant hypersurface in momentum space for any particle of 
given restmass $m$, which the particle can never leave in the 
course of its time-evolution. 

Returning for simplicity to the case of the free particle, we now
show how the equation of motion (\ref{1.2.1}) and the constraint 
(\ref{1.2.3}) can both be derived from a single action principle. 
In addition to the co-ordinates $x^{\mu}$, the action depends 
on an auxiliary variable $e$; it reads 
\be 
S[x^{\mu};e] = \frac{m}{2}\, \int_1^2\, \lh \frac{1}{e}\, 
 \frac{dx_{\mu}}{d\lb} \frac{dx^{\mu}}{d\lb} - e c^2 \rh d\lb. 
\label{1.2.8}
\ee 
Here $\lb$ is a real parameter taking values in the interval 
$[\lb_1,\lb_2]$, which is mapped by the functions $x^{\mu}(\lb)$ 
into a curve in Minkowski space with fixed end points 
$(x_1^{\mu},x_2^{\mu})$, and $e(\lb)$ is a nowhere vanishing 
real function of $\lb$ on the same interval. 

Before discussing the equations that determine the stationary 
points of the action, we first observe that by writing it in 
the equivalent form 
\be 
S[x^{\mu};e] = \frac{m}{2}\, \int_1^2\, \lh 
 \frac{dx_{\mu}}{e d\lb}\, \frac{dx^{\mu}}{e d\lb} - c^2 \rh
 e d\lb, 
\label{1.2.9}
\ee 
it becomes manifest that the action is invariant under a change 
of parametrization of the real interval $\lb \rightarrow 
\lb^{\prime}(\lb)$, if the variables $(x^{\mu},e)$ are transformed
simultaneously to $(x^{\prime\, \mu},e^{\prime})$ according to the 
rule 
\be 
x^{\prime\, \mu}(\lb^{\prime}) = x^{\mu}(\lb), \hs{2} 
e^{\prime}(\lb^{\prime})\, d\lb^{\prime} = e(\lb)\, d\lb. 
\label{1.2.10}
\ee 
Thus the co-ordinates $x^{\mu}(\lb)$ transform as scalar functions 
on the real line ${\bf R}^1$, whilst $e(\lb)$ transforms as the 
(single) component of a covariant vector (1-form) in one dimension. 
For this reason it is often called the {\em einbein}. For obvious 
reasons the invariance of the action (\ref{1.2.8}) under the 
transformations (\ref{1.2.10}) is called reparametrization invariance. 

The condition of stationarity of the action $S$ implies the functional 
differential equations 
\be 
\frac{\del S}{\del x^{\mu}} = 0, \hs{2} 
\frac{\del S}{\del e} = 0. 
\label{1.2.11}
\ee 
These equations are equivalent to the ordinary differential equations 
\be 
\frac{1}{e} \frac{d}{d\lb} \lh \frac{1}{e} \frac{dx^{\mu}}{d\lb} \rh 
 = 0, \hs{2} 
\lh \frac{1}{e} \frac{d x^{\mu}}{d\lb} \rh^2 = -c^2. 
\label{1.2.12}
\ee 
The equations coincide with the equation of motion (\ref{1.2.1}) 
and the constraint (\ref{1.2.3}) upon the identification 
\be
d\tau = e d\lb, 
\label{1.2.13}
\ee 
a manifestly reparametrization invariant definition of proper time. 
Recall, that after this identification the constraint (\ref{1.2.3}) 
automatically implies eq.(\ref{1.2.4}), hence this definition of 
proper time coincides with the standard geometrical one. 
\vs{1} 

\nit
{\bf Exercise 1.1} \nl
Use the constraint (\ref{1.2.12}) to eliminate $e$ from the action;
show that with the choice $e > 0$ ($\tau$ increases with 
increasing $\lb$) it reduces to the Einstein action 
\[
S_E = - mc \int_1^2\, \sqrt{- \frac{dx_{\mu}}{d\lb} 
 \frac{dx^{\mu}}{d\lb} }\, d\lb = - mc^2 \int_1^2 d\tau,
\]
with $d\tau$ given by eq.(\ref{1.2.4}). Deduce that the solutions 
of the equations of motion are time-like geodesics in Minkowski space.
Explain why the choice $e < 0$ can be interpreted as describing 
anti-particles of the same mass.
\vs{1} 

\nit
{\em 2.\ The electro-magnetic field.} \nl
In the absence of charges and currents the evolution of electric 
and magnetic fields $(\vec{E},\vec{B})$ is described by the equations 
\be 
\dd{\vec{E}}{t} = \vec{\nabla} \times \vec{B}, \hs{1}
\dd{\vec{B}}{t} = - \vec{\nabla} \times \vec{E}. 
\label{1.2.14}
\ee 
Each of the electric and magnetic fields has three components, but 
only two of them are independent: physical electro-magnetic fields 
in vacuo are transversely polarized, as expressed by the conditions 
\be 
\vec{\nabla} \cdot \vec{E} = 0, \hs{2} 
\vec{\nabla} \cdot \vec{B} = 0. 
\label{1.2.15}
\ee 
The set of four equations (\ref{1.2.14}) and (\ref{1.2.15}) represent 
the standard form of Maxwell's equations in empty space. 

Repeated use of eqs.(\ref{1.2.14}) yields 
\be 
\dd{^2 \vec{E}}{t^2} = - \vec{\nabla} \times (\vec{\nabla} \times 
 \vec{E}) = \Del \vec{E} - \vec{\nabla} \vec{\nabla} \cdot \vec{E}, 
\label{1.2.16}
\ee 
and an identical equation for $\vec{B}$. However, the transversality 
conditions (\ref{1.2.15}) simplify these equations to the linear
wave equations
\be 
\Box \vec{E} = 0, \hs{2} \Box \vec{B} = 0, 
\label{1.2.17}
\ee 
with $\Box = \Del - \der_t^{\, 2}$. It follows immediately that
free electromagnetic fields satisfy the superposition principle and
consist of transverse waves propagating at the speed of light ($c = 1$,
in natural units). 

Again both the time evolution of the fields and the transversality  
constraints can be derived from a single action principle, but it
is a little bit more subtle than in the case of the particle. For 
electrodynamics we only introduce auxiliary fields $\vec{A}$ and 
$\fg$ to impose the equation of motion and constraint for the 
electric field; those for the magnetic field then follow automatically.  
The action is 
\be 
\ba{l} 
S_{EM}[\vec{E},\vec{B};\vec{A},\fg] = \dsp{ \int_1^2 dt\, 
 L_{EM}(\vec{E},\vec{B};\vec{A},\fg), }\\
 \\
L_{EM} = \dsp{ \int d^3x \lh 
 - \frac{1}{2} \lh \vec{E}^{\,2} - \vec{B}^{\,2} \rh + \vec{A} \cdot 
 \lh \dd{\vec{E}}{t} - \vec{\nabla} \times \vec{B} \rh 
 - \fg\, \vec{\nabla} \cdot \vec{E} \rh. }
\ea
\label{1.2.18}
\ee 
Obviously, stationarity of the action implies 
\be 
\frac{\del S}{\del \vec{A}} = 
 \dd{\vec{E}}{t} - \vec{\nabla} \times \vec{B} = 0, \hs{2} 
\frac{\del S}{\del \fg} = - \vec{\nabla} \cdot \vec{E} = 0, 
\label{1.2.19}
\ee 
reproducing the equation of motion and constraint for the electric 
field. The other two stationarity conditions are 
\be 
\frac{\del S}{\del \vec{E}} = - \vec{E} - \dd{\vec{A}}{t} +
 \vec{\nabla} \fg = 0, \hs{2} 
\frac{\del S}{\del \vec{B}} = \vec{B} - \vec{\nabla} \times \vec{A}
 = 0, 
\label{1.2.20}
\ee 
or equivalently
\be 
\vec{E} = - \dd{\vec{A}}{t} + \vec{\nabla} \fg, \hs{2} 
\vec{B} = \vec{\nabla} \times \vec{A}. 
\label{1.2.21}
\ee 
The second equation (\ref{1.2.21}) directly implies the transversality of the 
magnetic field: $\vec{\nabla} \cdot \vec{B} = 0$. Taking its 
time derivative one obtains
\be 
\dd{\vec{B}}{t} = \vec{\nabla} \times \lh \dd{\vec{A}}{t} - 
 \vec{\nabla} \fg \rh = - \vec{\nabla} \times \vec{E}, 
\label{1.2.22}
\ee 
where in the middle expression we are free to add the gradient  
$\vec{\nabla} \fg$, as $\vec{\nabla} \times \vec{\nabla}\fg = 0$ 
identically. 

An important observation is, that the expressions (\ref{1.2.21}) 
for the electric and magnetic fields are invariant under a redefinition
of the potentials $\vec{A}$ and $\fg$ of the form 
\be 
\vec{A}^{\,\prime} = \vec{A} + \vec{\nabla} \Lb, \hs{2} 
\fg^{\prime} = \fg + \dd{\Lb}{t},
\label{1.2.23}
\ee 
where $\Lb(x)$ is an arbitrary scalar function. The transformations 
(\ref{1.2.23}) are the well-known gauge transformations of  
electrodynamics.

It is easy to verify, that the Lagrangean $L_{EM}$ changes only by 
a total time derivative under gauge transformations, modulo boundary 
terms which vanish if the fields vanish sufficiently fast at spatial 
infinity: 
\be 
L^{\prime}_{EM} = L_{EM} - \frac{d}{dt}\, \int d^3x\, \Lb 
 \vec{\nabla} \cdot \vec{E}. 
\label{1.2.24}
\ee 
As a result the action $S_{EM}$ itself is strictly invariant under 
gauge transformations, provided $\int d^3x \Lb \vec{\nabla} \cdot 
\vec{E}|_{t_1} = \int d^3x \Lb \vec{\nabla} \cdot \vec{E}|_{t_2}$; 
however, no physical principle requires such strict invariance
of the action. This point we will discuss later in more 
detail. 

We finish this discussion of electro-dynamics by recalling how to 
write the equations completely in relativistic notation. This is 
achieved by first collecting the electric and magnetic fields in 
the anti-symmetric field-strength tensor
\be 
F_{\mu\nu} = \lh \ba{cccc} 
      0 & -E_1 & -E_2 & -E_3 \\ 
      E_1 & 0 & B_3 & -B_2 \\
      E_2 & -B_3 & 0 & B_1 \\
      E_3 & B_2 & -B_1 & 0 \ea \rh, 
\label{1.2.25}
\ee 
and the potentials in a four-vector: 
\be 
A_{\mu} = (\fg, \vec{A}). 
\label{1.2.26}
\ee 
Eqs.(\ref{1.2.21}) then can be written in covariant form as 
\be 
F_{\mu\nu} = \der_{\mu} A_{\nu} - \der_{\nu} A_{\mu}, 
\label{1.2.27}
\ee 
with the electric field equations (\ref{1.2.19}) reading 
\be 
\der_{\mu} F^{\mu\nu} = 0. 
\label{1.2.28}
\ee 
The magnetic field equations now follow trivially from (\ref{1.2.27})
as 
\be 
\ve^{\mu\nu\kg\lb} \der_{\nu} F_{\kg\lb} = 0. 
\label{1.2.29}
\ee 
Finally, the gauge transformations can be written covariantly as 
\be 
A^{\prime}_{\mu} = A_{\mu} + \der_{\mu} \Lb.
\label{1.2.30}
\ee 
The invariance of the field strength tensor $F_{\mu\nu}$ under these 
transformations follows directly from the commutativity of the partial 
derivatives. 
\vs{1} 

\nit
{\bf Exercise 1.2} \nl
Show that eqs.(\ref{1.2.27})--(\ref{1.2.29}) follow from the action 
\[ 
S_{cov} = \int d^4x\, \lh \frac{1}{4}\, F^{\mu\nu} F_{\mu\nu} - 
 F^{\mu\nu}\, \der_{\mu} A_{\nu} \rh. 
\]
Verify, that this action is equivalent to $S_{EM}$ modulo a total 
divergence. Check that eliminating $F_{\mu\nu}$ as an independent 
variable gives the usual standard action 
\[
S[A_{\mu}] = - \frac{1}{4}\, \int d^4x\, F^{\mu\nu}(A) F_{\mu\nu}(A),
\]
with $F_{\mu\nu}(A)$ given by the right-hand side of eq.(\ref{1.2.27}). 

\section{Symmetries and Noether's theorems \label{s1.3}}

In the preceeding section we have presented two elementary examples 
of systems whose complete physical behaviour was described conveniently 
in terms of one or more evolution equations plus one or more constraints. 
These constraints are needed to select a subset of solutions of the 
evolution equation as the physically relevant solutions. In both examples 
we found, that the full set of equations could be derived from an 
action principle. Also, in both examples the additional (auxiliary) 
degrees of freedom, necessary to impose the constraints, allowed 
non-trivial local (space-time dependent) redefinitions of variables 
leaving the lagrangean invariant, at least up to a total time-derivative.  

The examples given can easily be extended to include more complicated
but important physical models: the relativistic string, Yang-Mills 
fields and general relativity are all in this class. However, instead 
of continuing to produce more examples, at this stage we turn to the 
general case to derive the relation between local symmetries and 
constraints, as an extension of Noether's well-known theorem relating
(rigid) symmetries and conservation laws. 

Before presenting the more general analysis, it must be pointed out 
that our approach distinguishes in an important way between time- 
and space-like dimensions; indeed, we have emphasized from the 
start the distinction between equations of motion (determining 
the behaviour of a system as a function of time) and constraints,
which impose additional requirements. e.g.\ restricting the spatial
behaviour of electro-magnetic fields. This distinction is very 
natural in the context of hamiltonian dynamics, but potentially 
at odds with a covariant lagrangean formalism. However, in the 
examples we have already observed that the not manifestly covariant 
treatment of electro-dynamics could be translated without too much 
effort into a covariant one, and that the dynamics of the 
relativistic particle, including its constraints, was manifestly 
covariant throughout. 
 
In quantum theory we encounter similar choices in the approach 
to dynamics, with the operator formalism based on equal-time 
commutation relations distinguishing space- and time-like behaviour 
of states and observables, whereas the covariant path-integral 
formalism allows treatment of space- and time-like dimensions on 
an equal footing; indeed, upon the analytic continuation of the 
path-integral to euclidean time the distinction vanishes alltogether. 
In spite of these differences, the two approaches are equivalent in 
their physical content. 

In the analysis presented here we continue to distinguish between
time and space, and between equations of motion and constraints. 
This is convenient as it allows us to freely employ hamiltonian
methods, in particular Poisson brackets in classical dynamics
and equal-time commutators in quantum mechanics. Nevertheless, 
as we hope to make clear, all applications to relativistic models 
allow a manifestly covariant formulation. 
\vs{1} 

\nit 
Consider a system described by generalized coordinates $q^i(t)$,
where $i$ labels the complete set of physical plus auxiliary degrees 
of freedom, which may be infinite in number. For the relativistic
particle in $n$-dimensional Minkowski space the $q^i(t)$ represent 
the $n$ coordinates $x^{\mu}(\lb)$ plus the auxiliary variable 
$e(\lb)$ (sometimes called the `einbein'), with $\lb$ playing the 
role of time; for the case of a field theory with $N$ fields 
$\vf^a(\vec{x};t)$, $a = 1,...,N$, the $q^i(t)$ represent the 
infinite set of field amplitudes $\vf^a_{\vec{x}}(t)$ at fixed 
location $\vec{x}$ as function of time $t$, i.e.\ the dependence on 
the spatial co-ordinates $\vec{x}$ is included in the labels $i$. 
In such a case summation over $i$ is understood to include 
integration over space. 

Assuming the classical dynamical equations to involve at most 
second-order time derivatives, the action for our system can now 
be represented quite generally by an integral
\be 
S[q^i] = \int_1^2\, L(q^i,\dot{q}^i)\, dt, 
\label{1.3.1}
\ee 
where in the case of a field theory $L$ itself is to be represented
as an integral of some density over space. An arbitrary variation
of the co-ordinates leads to a variation of the action of the form
\be
{\displaystyle \delta S = \int_{1}^{2} dt \: \delta q^{i} \left( 
 \dd{L}{q^{i}} - \frac{d}{dt} \dd{L}{\dot{q}^{i}} \right) + \left[ 
 \delta q^{i} \dd{L}{\dot{q}^{i}} \right]_{1}^{2} },
\label{1.3.2}
\ee
with the boundary terms due to an integration by parts. As usual we
define generalized canonical momenta as 
\be
p_{i} = \dd{L}{\dot{q}^{i}}. 
\label{1.3.3}
\ee
From eq.(\ref{1.3.2}) two well-known important consequences follow: 
\nl
- the action is stationary under variations vanishing at 
initial and final times: $\del q^i(t_1) = \del q^i(t_2) = 0$, if 
the Euler-Lagrange equations are satisfied: 
\be 
\frac{d p_{i}}{dt} = \frac{d}{dt}\, \dd{L}{\dot{q}^i} = \dd{L}{q^{i}}.
\label{1.3.4}
\ee
- for arbitrary variations around the classical paths $q^i_c(t)$ 
in configuration space: $q^i(t) = q^i_c(t) + \del q^i(t)$, with 
$q^i_c(t)$ and its associated momentum $p_{c\, i}(t)$ a solution 
of the Euler-Lagrange equations, the total variation of the action 
is 
\be 
\del S_c = \left[ \del q^i(t) p_{c\, i}(t) \right]_1^2.
\label{1.3.5}
\ee 
We now define an infinitesimal {\em symmetry} of the action as a set 
of continuous transformations $\del q^i(t)$ (smoothly connected 
to zero) such that the lagrangean $L$ transforms to first order into 
a total time derivative: 
\be 
\del L = \del q^i \dd{L}{q^i} + \del \dot{q}^i \dd{L}{\dot{q}^i}\, 
 = \frac{dB}{dt},
\label{1.3.6}
\ee 
where $B$ obviously depends in general on the co-ordinates and 
the velocities, but also on the variation $\del q^i$. It follows 
immediately from the definition that 
\be 
\del S = \left[ B \right]_1^2.
\label{1.3.7}
\ee 
Observe, that according to our definition a symmetry does {\em not} 
require the action to be invariant in a strict sense. 
Now comparing (\ref{1.3.5}) and (\ref{1.3.7}) we establish 
the result that, whenever there exists a set of symmetry 
transformations $\del q^i$, the physical motions of the system satisfy 
\be 
\left[ \del q^i p_{c\,i} - B_c \right]_1^2 = 0. 
\label{1.3.8}
\ee 
Since the initial and final times $(t_1,t_2)$ on the particular 
orbit are arbitrary, the result can be stated equivalenty in the 
form of a conservation law for the quantity inside the brackets. 

To formulate it more precisely, let the symmetry variations 
be parametrized by $k$ linearly independent parameters 
$\eps^{\ag}$, $\ag = 1,...,k$, possibly depending on time: 
\be
\delta q^{i} = R^{i}[\ag] = \eps^{\alpha} R^{(0)i}_{\alpha} + 
 \dot{\eps}^{\alpha} R^{(1)i}_{\alpha} + ...\, + 
 \stackrel{(n)}{\eps}{\hs{-.2}}^{\ag} R_{\ag}^{(n)\,i} + ...,
\label{1.3.9}
\ee
where $\stackrel{(n)}{\eps}{\hs{-.2}}^{\ag}$ denotes the $n$th 
time derivative of the parameter. Correspondingly, the lagrangean 
transforms into the derivative of a function $B[\eps]$, with 
\be
B[\eps] = \eps^{\alpha} B^{(0)}_{\alpha} + \dot{\eps}^{\alpha} 
B^{(1)}_{\alpha} + ...\, + \stackrel{(n)}\eps{\hs{-.2}}^{\ag} B_{\ag}^{(n)} 
 + ....
\label{1.3.10}
\ee
With the help of these expressions we define the `on shell' 
quantity\footnote{An `on shell' quantity is a quantity
defined on a classical trajectory.}
\be 
\ba{lll}
G[\eps] & = & p_{c\,i} R_c^i[\eps] - B_c[\eps] \\
 & & \\
 & = & \eps^{\ag} G_{\ag}^{(0)} + \dot{\eps}^{\ag} G_{\ag}^{(1)}
 + ...\, + \stackrel{(n)}{\eps}{\hs{-.2}}^{\ag}\, G_{\ag}^{(n)} + ...,
\ea
\label{1.3.11}
\ee 
with component by component $G_{\ag}^{(n)} = p_{c\,i}\, 
R_{c\,\ag}^{(n)\, i} - B_{c\,\ag}^{(n)}$. The conservation law 
(\ref{1.3.8}) can now be stated equivalently as 
\be 
\frac{dG[\eps]}{dt} = \eps^{\ag} \dot{G}_{\ag}^{(0)} + 
 \dot{\eps}^{\ag} \lh G_{\ag}^{(0)} + \dot{G}_{\ag}^{(1)} \rh 
 + ...\, + \stackrel{(n)}{\eps}{\hs{-.2}}^{\ag}\lh G_{\ag}^{(n-1)} 
 + \dot{G}_{\ag}^{(n)} \rh + ... = 0. 
\label{1.3.12} 
\ee 
We can now distinguish various situations, of which we consider only 
the two extreme cases here. First, if the symmetry exists only for 
$\eps =$ constant (a {\em rigid} symmetry), then all time derivatives 
of $\eps$ vanish and $G_{\ag}^{(n)} \equiv 0$ for $n \geq 1$, whilst 
for the lowest component
\be 
G_{\ag}^{(0)} = g_{\ag} = \mbox{constant}, \hs{2} 
G[\eps] = \eps^{\ag} g_{\ag},
\label{1.3.13}
\ee 
as defined on a particular classical trajectory (the value of $g_{\ag}$
may be different on different trajectories). Thus, rigid symmetries 
imply constants of motion; this is Noether's theorem. 

Second, if the symmetry exists for arbitrary time-dependent $\eps(t)$
(a {\em local} symmetry), then $\eps(t)$ and all its time derivatives
at the same instant are independent. As a result
\be 
\ba{l}
\dot{G}_{\ag}^{(0)} = 0,  \\ 
 \\
\dot{G}_{\ag}^{(1)} = - G_{\ag}^{(0)}, \\
 \\
...\\
 \\
\dot{G}_{\ag}^{(n)} = - G_{\ag}^{(n-1)}, \\
 \\
... 
\ea
\label{1.3.14}
\ee 
Now in general the transformations (\ref{1.3.9}) do not depend on
arbitrarily high-order derivatives of $\eps$, but only on a {\em 
finite} number of them: there is some finite $N$ such that 
$R^{(n)}_{\ag}  = 0$ for $n \geq N$. Typically, transformations 
depend at most on the first derivative of $\eps$, and $R^{(n)}_{\ag} 
= 0$ for $n \geq 2$. In general, for any finite $N$ all quantities 
$R^{(n)\,i}$, $B^{(n)}$, $G^{(n)}$ then vanish identically for 
$n \geq N$. But then $G_{\ag}^{(n)} = 0$ for $n = 0,...,N-1$ 
as well, as a result of eqs.(\ref{1.3.14}). Therefore $G[\eps] = 0$
at all times. This is a set of {\em constraints} relating the 
coordinates and velocities on a classical trajectory. Moreover, 
as $dG/dt = 0$, these constraints have the nice property that 
they are preserved during the time-evolution of the system. 

The upshot of this analysis is therefore, that local symmetries 
imply time-independent constraints. This result is sometimes refered 
to as Noether's second theorem.  
\vs{1} 

\nit
{\bf Exercise 1.3} \nl 
Show that if there is no upper limit on the order of derivatives
in the transformation rule (no finite $N$), one reobtains a 
conservation law 
\[ 
G[\eps] = g_{\ag}\, \eps^{\ag}(0) = \mbox{constant}.
\]
Hint: show that $G_{\ag}^{(n)} = ((-t)^n/n!)\, g_{\ag}$, with 
$g_{\ag}$ a constant, and use the Taylor expansion for $\eps(0) = 
\eps(t - t)$ around $\eps(t)$. 
\vs{1} 

\nit
{\em Group structure of symmetries.} \nl 
To round off our discussion of symmetries, conservation laws and 
constraints in the lagrangean formalism, we show that symmetry
transformations as defined by eq.(\ref{1.3.6}) possess an 
infinitesimal group structure, i.e.\ they have a closed commutator 
algebra (a Lie algebra or some generalization thereof). The proof 
is simple. First observe, that performing a second variation of 
$\del L$ gives
\be 
\ba{lll}
\del_2 \del_1 L & = & \dsp{ \del_2 q^j \del_1 q^i \dd{^2L}{q^j 
 \der q^i} + \del_2 \dot{q}^j \del_1 q^i \dd{^2L}{\dot{q}^j 
 \der q^i} + (\del_2 \del_1 q^i) \dd{L}{q^i} }\\
 & & \\
 & & \dsp{ \,+\, \del_2 \dot{q}^j \del_1 \dot{q}^i \dd{^2L}{\dot{q}^j 
 \der \dot{q}^i} + \del_2 q^j \del_1 \dot{q}^i \dd{^2L}{q^j 
 \der \dot{q}^i} + (\del_2 \del_1 \dot{q}^i) \dd{L}{\dot{q}^i} 
 = \frac{d (\del_2 B_1)}{dt}. }
\ea
\label{1.3.16}
\ee 
By antisymmetrization this immediately gives 
\be 
\left[ \del_2, \del_1 \right] L = \lh \left[ \del_1, \del_2 \right] q^i
 \rh \dd{L}{q^i} + \lh \left[ \del_2, \del_1 \right] \dot{q}^i \rh
 \dd{L}{\dot{q}^i} = \frac{d}{dt} \lh \del_2 B_1 - \del_1 B_2 \rh. 
\label{1.3.17}
\ee 
By assumption of the completeness of the set of symmetry transformations
it follows, that there must exist a symmetry transformation 
\be 
\del_3\, q^i = \left[ \del_2, \del_1 \right] q^i, \hs{2} 
\del_3\, \dot{q}^i = \left[ \del_2, \del_1 \right] \dot{q}^i, 
\label{1.3.18}
\ee 
with the property that the associated $B_3 = \del_2 B_1 - \del_1 B_2$. 
Implementing these conditions gives 
\be 
\left[ \del_2, \del_1 \right] q^i = 
 R^j_2 \dd{R^i_1}{q^j} + \dot{q}^k \dd{R^j_2}{q^k} \dd{R^i_1}{\dot{q}^j} 
 + \ddot{q}^k \dd{R^j_2}{\dot{q}^k} \dd{R^i_1}{\dot{q}^j} 
 - [1 \leftrightarrow 2]= R_3^i,
\label{1.3.19}
\ee 
where we use a condensed notation $R^i_{a} \equiv R^i[\eps_{a}]$, 
$a = 1,2,3$. In all standard cases, the symmetry transformations 
$\del q^i = R^i$ involve only the coordinates and velocities: 
$R^i = R^i(q,\dot{q})$. Then $R_3$ can not contain terms proportional 
to $\ddot{q}$, and the conditions (\ref{1.3.19}) reduce 
to two separate conditions 
\be 
\ba{l} 
\dsp{ R^j_2 \dd{R^i_1}{q^j} - R^j_1 \dd{R^i_2}{q^j} + 
 \dot{q}^k \lh \dd{R^j_2}{q^k} \dd{R^i_1}{\dot{q}^j} - 
 \dd{R^j_1}{q^k} \dd{R^i_2}{\dot{q}^j} \rh = R^i_3, }\\ 
 \\
\dsp{ \dd{R^j_2}{\dot{q}^k} \dd{R^i_1}{\dot{q}^j} - 
 \dd{R^j_1}{\dot{q}^k} \dd{R^i_2}{\dot{q}^j} = 0. }
\ea 
\label{1.13.20}
\ee 
Clearly, the parameter $\eps_3$ of the transformation on the right-hand 
side must be an antisymmetric bilinear combination of the other two 
parameters: 
\be 
\eps_3^{\ag} = f^{\ag}(\eps_1,\eps_2) = - f^{\ag}(\eps_2,\eps_1).
\label{1.13.21}
\ee 

\section{Canonical formalism \label{s1.4}} 

The canonical formalism describes dynamics in terms of phase-space 
coordinates $(q^i,p_i)$ and a Hamiltonian $H(q,p)$, starting form 
an action
\be 
S_{can}[q,p] = \int_1^2 \lh p_i \dot{q}^i - H(q,p) \rh dt.
\label{1.4a.1}
\ee
Variations of the phase-space coordinates change the action to 
first order by 
\be 
\del S_{can} = \int_1^2 dt \left[ \del p_i \lh \dot{q}^i - 
 \dd{H}{p_i} \rh - \del q^i \lh \dot{p}_i + \dd{H}{q^i} \rh 
 + \frac{d}{dt} \lh p_i \del q^i \rh \right].
\label{1.4a.2}
\ee 
The action is stationary under variations vanishing at times 
$(t_1,t_2)$ if Hamilton's equations of motion are satisfied: 
\be 
\dot{p}_i = \dd{H}{q^i}, \hs{2} \dot{q}^i = - \dd{H}{p_i}. 
\label{1.4a.3}
\ee 
This motivates the introduction of the Poisson brackets 
\be 
\left\{ F, G \right\} = \dd{F}{q^i} \dd{G}{p_i} - \dd{F}{p_i} 
 \dd{G}{q^i}, 
\label{1.4a.4}
\ee 
with allow us to write the time derivative of any phase-space 
function $G(q,p)$ as 
\be 
\dot{G} = \dot{q}^i\, \dd{G}{q^i} + \dot{p}_i\, \dd{G}{p_i} 
 = \left\{ G, H \right\}. 
\label{1.4a.5}
\ee 
It follows immediately, that $G$ is a constant of motion if and 
only if 
\be 
\left\{ G, H \right\} = 0, 
\label{1.4a.6}
\ee 
everywhere along the trajectory of the physical system in phase space.
This is guaranteed to be the case if eq.(\ref{1.4a.6}) holds everywhere 
in phase space, but as we discuss below, more subtle situations can arise.

Suppose eq.(\ref{1.4a.6}) is satisfied; then we can construct variations 
of $(q,p)$ defined by 
\be 
\del q^i = \left\{ q^i, G \right\} = \dd{G}{p_i}, \hs{2} 
\del p_i = \left\{ p_i, G \right\} = - \dd{G}{q^i},
\label{1.4a.7}
\ee 
which leave the Hamiltonian invariant:
\be 
\del H = \del q^i\, \dd{H}{q^i} + \del p_i\, \dd{H}{p_i}
 = \dd{G}{p_i} \dd{H}{q^i} - \dd{G}{q^i} \dd{H}{p_i} 
 = \left\{ H, G \right\} = 0. 
\label{1.4a.8}
\ee 
They represent infinitesimal symmetries of the theory provided 
eq.(\ref{1.4a.6}), and hence (\ref{1.4a.8}), is satisfied as an 
identity, irrespective of whether or not the phase-space coordinates 
$(q,p)$ satisfy the equations of motion. To see this, consider the 
variation of the action (\ref{1.4a.2}) with $(\del q, \del p)$ 
given by (\ref{1.4a.7}) and $\del H = 0$ by (\ref{1.4a.8}):
\be 
\del S_{can} = \int_1^2 dt \left[ - \dd{G}{q^i} \dot{q}^i - 
 \dd{G}{p_i} \dot{p}_i + \frac{d}{dt} \lh \dd{G}{p_i} p_i \rh 
 \right] = \int_1^2 dt \frac{d}{dt} \lh \dd{G}{p_i} p_i - G \rh. 
\label{1.4a.9}
\ee 
If we call the quantity inside the parentheses $B(q,p)$, then we 
have rederived eqs.(\ref{1.3.7}) and (\ref{1.3.8}); indeed, 
we then have 
\be 
G = \dd{G}{p_i}\, p_i - B = \del q^i p_i - B, 
\label{1.4a.10}
\ee 
where we know from eq.(\ref{1.4a.5}), that $G$ is a constant of motion 
on classical trajectories (on which Hamilton's equation of motion are 
satisfied). Observe that, whereas in the lagrangean approach we showed
that symmetries imply constants of motion, here we have derived the 
inverse Noether theorem: constants of motion generate symmetries. An 
advantage of this derivation over the lagrangean one is, that we have 
also found explicit expressions for the variations $(\del q, \del p)$. 

A further advantage is, that the infinitesimal group structure 
of the tranformations (the commutator algebra) can be checked directly. 
Indeed, if two symmetry generators $G_{\ag}$ and $G_{\bg}$ both satisfy 
(\ref{1.4a.6}), then the Jacobi identity for Poisson brackets implies 
\be 
\left\{ \left\{ G_{\ag}, G_{\bg} \right\}, H \right\} = 
 \left\{ G_{\ag}, \left\{ G_{\bg}, H \right\} \right\} - 
 \left\{ G_{\bg}, \left\{ G_{\ag}, H \right\} \right\} = 0.
\label{1.4a.11}
\ee 
Hence if the set of generators $\left\{ G_{\ag} \right\}$ is complete,
we must have an identity of the form 
\be 
\left\{ G_{\ag}, G_{\bg} \right\} = P_{\ag\bg} \lh G \rh 
 = - P_{\bg\ag} \lh G \rh,
\label{1.4a.12}
\ee 
where the $P_{\ag\bg}(G)$ are polynomials in the constants of 
motion $G_{\ag}$: 
\be 
P_{\ag\bg}(G) = c_{\ag\bg} + f_{\ag\bg}^{\;\;\;\gam}\, G_{\gam}
 + \frac{1}{2}\, g_{\ag\bg}^{\;\;\;\;\gam\del} G_{\gam} G_{\del} + ....
\label{1.4.12}
\ee 
The coefficients $c_{\ag\bg}$, $f_{\ag\bg}^{\;\;\;\gam}$, 
$g_{\ag\bg}^{\;\;\;\;\gam\del}$, $...$ are constants, having zero 
Poisson brackets with any phase-space function. As such the first 
term $c_{\ag\bg}$ may be called a central charge. 
 
It now follows that the transformation of any phase-space function 
$F(q,p)$, given by 
\be 
\del_{\ag} F = \left\{ F, G_{\ag} \right\},
\label{1.4a.13}
\ee 
satisfies the commutation relation 
\be 
\ba{lll}
\left[ \del_{\ag}, \del_{\bg} \right] F & = & \left\{ \left\{ 
 F, G_{\bg} \right\}, G_{\ag} \right\} - \left\{ \left\{ 
 F, G_{\ag} \right\}, G_{\bg} \right\} = \left\{ F, \left\{ 
 G_{\bg}, G_{\ag} \right\} \right\} \\
 & & \\
 & = & \dsp{ C_{\bg\ag}^{\;\;\;\gam}(G)\; \del_{\gam} F, }
\ea
\label{1.4a.14}
\ee 
where we have introduced the notation 
\be 
C_{\bg\ag}^{\;\;\;\gam}(G) = \dd{P_{\bg\ag}(G)}{G_{\gam}} = 
 f_{\ag\bg}^{\;\;\;\gam} + g_{\ag\bg}^{\;\;\;\;\gam \del} G_{\del} + ... . 
\label{1.4a.15}
\ee 
In particular this holds for the coordinates and momenta $(q,p)$ 
themselves; taking $F$ to be another constraint $G_{\gam}$, we
find from the Jacobi identity for Poisson brackets the 
consistency condition 
\be 
C_{\left[ \ag \bg \rd}^{\;\;\;\;\del}\, P_{\ld \gam \right] \del} = 
 f_{\left[ \ag \bg \rd}^{\;\;\;\;\del}\, c_{\ld \gam \right] \del}
 + \lh f_{\left[ \ag \bg \rd}^{\;\;\;\;\del}\, f_{\ld \gam \right] 
 \del}^{\;\;\;\;\ve} + g_{\left[ \ag\bg \rd}^{\;\;\;\;\del\eps} 
 c_{\ld \gam \right] \del} \rh G_{\ve} + .... = 0.
\label{1.4a.15.1}
\ee  

By the same arguments as in sect.\ \ref{s1.3} (eq.(\ref{1.3.11}
and following) it is established, that whenever the theory generated
by $G_{\ag}$ is a {\em local} symmetry with time-dependent parameters, 
the generator $G_{\ag}$ turns into a constraint: 
\be 
G_{\ag}(q,p) = 0.
\label{1.4a.16}
\ee 
However, compared to the case of rigid symmetries, a subtlety now 
arises: the constraints $G_{\ag} = 0$ define a hypersurface in the 
phase space to which all physical trajectories of the system are 
confined. This implies, that it is sufficient for the constraints 
to commute with the hamiltonian (in the sense of Poisson brackets) 
on the physical hypersurface (i.e., {\em on shell}). Off the hypersurface 
({\em off shell}), the bracket of the hamiltonian with the constraints can 
be anything, as the physical trajectories never enter this part of
phase space. Thus the most general allowed algebraic structure defined 
by the hamiltonian and constraints is
\be 
\left\{ G_{\ag}, G_{\bg} \right\} = P_{\ag\bg}(G), \hs{2} 
 \left\{ H, G_{\ag} \right\} = Z_{\ag}(G),
\label{1.4a.17}
\ee 
where both $P_{\ag\bg}(G)$ and $Z_{\ag}(G)$ are polynomials in the 
constraints with the property that $P_{\ag\bg}(0) = Z_{\ag}(0) = 0$.
This is sufficient to guarantee that in the physical sector of the 
phase space $\{ H, G_{\ag} \}|_{G = 0} = 0$. Note, that in the case of 
local symmetries with generators $G_{\ag}$ defining constraints, the 
central charge in the bracket of the constraints must vanish: 
$c_{\ag\bg} = 0$. This is a genuine restriction on the existence of 
local symmetries. A dynamical system with constraints and hamiltonian 
satisfying eqs.(\ref{1.4a.17}) is said to be {\em first class}. Actually, 
it is quite easy to see that the general first-class algebra of Poisson 
brackets is more appropriate for systems with local symmetries. 
Namely, even if the brackets of the constraints and the hamiltonian 
genuinely vanishes on and off shell, one can always change the 
hamiltonian of the system by adding a polynomial in the constraints:
\be 
H^{\prime} = H + R(G), \hs{2} R(G) = \rg_0 + \rg_1^{\ag} G_{\ag} + 
 \frac{1}{2}\, \rg_2^{\ag\bg} G_{\ag} G_{\bg} + ... 
\label{1.4a.18}
\ee 
This leaves the hamiltonian on the physical shell in phase space 
invariant (up to a constant $\rg_0$), and therefore the physical 
trajectories remain the same. Furthermore, even if $\{ H, G_{\ag} \} 
= 0$, the new hamiltonian satisfies 
\be 
\left\{ H^{\prime}, G_{\ag} \right\} = \left\{ R(G), G_{\ag} \right\}
 = Z^{(R)}_{\ag}(G) \equiv \rg_1^{\bg} P_{\bg\ag}(G) + ...,
\label{1.4a.19}
\ee 
which is of the form (\ref{1.4a.17}). In addition the equations of motion 
for the variables $(q,p)$ are changed by a local symmetry transformation
only, as 
\be 
(\dot{q^i})^{\prime} = \left\{ q^i, H^{\prime} \right\}  = 
 \left\{ q^i, H \right\} + \left\{ q^i, G_{\ag} \right\} \dd{R}{G_{\ag}}
 = \dot{q}^i + \ve^{\ag} \del_{\ag} q^i,
\label{1.4a.20}
\ee 
where $\ve^{\ag}$ are some ---possibly complicated--- local functions 
which may depend on the phase-space coordinates $(q,p)$ themselves.  
A similar observation holds of course for the momenta $p_i$. 
We can actually allow the coefficients $\rg_1^{\ag}, \rg_2^{\ag\bg}, 
...$ to be space-time dependent variables themselves, as this 
does not change the general form of the equations of motion 
(\ref{1.4a.20}), whilst variation of the action w.r.t.\ these 
new variables will only impose the constraints as equations of motion:
\be 
\frac{\del S}{\del \rg^{\ag}_1} = G_{\ag}(q,p) = 0, 
\label{1.45a.21}
\ee 
in agreement with the dynamics already established. 

The same argument shows however,
that the part of the hamiltonian depending on the constraints in not 
unique, and may be changed by terms like $R(G)$. In many cases this 
allows one to get rid of all or part of $h_{\ag}(G)$. 

\section{Quantum dynamics \label{s1.5}} 

In quantum dynamics in the canonical operator formalism, one can follow
largely the same lines of argument as presented for classical theories
in sect.\ \ref{s1.4}. Consider a theory of canonical pairs of operators
$(\hq,\hp)$ with commutation relations 
\be 
\left[ \hq^{\,i}, \hp_j \right] = i \del^i_j, 
\label{1.5.1}
\ee 
and hamiltonian $\hH(\hq,\hp)$ such that 
\be 
i \frac{d\hq^{\,i}}{dt} = \left[ \hq^{\,i},  \hH \right], \hs{2} 
i \frac{d\hp_i}{dt} = \left[ \hp_i,  \hH \right]. 
\label{1.5.2}
\ee 
The $\delta$-symbol on the right-hand side of (\ref{1.5.1}) is 
to be interpreted in a generalized sense: for continuous parameters 
$(i,j)$ it represents a Dirac delta-function rather than a Kronecker 
delta.

In the context of quantum theory, constants of motion become 
operators $\hG$ which commute with the hamiltonian: 
\be 
\left[ \hG, \hH \right] = i \frac{d\hG}{dt} = 0, 
\label{1.5.3}
\ee 
and can therefore be diagonalized on stationary eigenstates. 
We henceforth assume we have at our disposal a complete set 
$\{ \hG_{\ag} \}$ of such constants of motion, in the sense that 
any operator satisfying (\ref{1.5.3})  can be expanded as a polynomial 
in the operators $\hG_{\ag}$. 

In analogy to the classical theory, we define infinitesimal symmetry 
transformations by 
\be 
\del_{\ag} \hq^{i} = -i \left[ \hq^i, \hG_{\ag} \right], \hs{2} 
\del_{\ag} \hp_{i} = -i \left[ \hp_i, \hG_{\ag} \right].
\label{1.5.4}
\ee 
By construction they have the property of leaving the hamiltonian 
invariant: 
\be 
\del_{\ag} \hH = -i \left[ \hH, \hG_{\ag} \right] = 0. 
\label{1.5.5}
\ee 
Therefore the operators $\hG_{\ag}$ are also called symmetry 
generators. It follows by the Jacobi identity, analogous to 
eq.(\ref{1.4a.11}), that the commutator of two such generators 
commutes again with the hamiltonian, and therefore 
\be 
-i \left[ \hG_{\ag}, \hG_{\bg} \right] = P_{\ag\bg}(\hG)
 = c_{\ag\bg} + f_{\ag\bg}^{\;\;\;\gam}\, \hG_{\gam} + .... 
\label{1.5.6}
\ee 
A calculation along the lines of (\ref{1.4a.14}) then shows, that 
for any operator $\hF(\hq,\hp)$ one has 
\be 
\del_\ag \hF = -i \left[ \hF, \hG_{\ag} \right], \hs{2} 
\left[ \del_{\ag}, \del_{\bg} \right] \hF = i f_{\ag\bg}^{\;\;\;\gam}\, 
  \del_{\gam} \hF + ... 
\label{1.5.7}
\ee 
Observe, that compared to the classical theory, in the quantum theory
there is an additional potential source for the appearance of central 
charges in (\ref{1.5.6}), to wit the operator ordering on the right-hand 
side. As a result, even when no central charge is present in the classical 
theory, such central charges can arise in the quantum theory. This is  
a source of anomalous behaviour of symmetries in quantum theory. 

As in the classical theory, local symmetries impose additional 
restrictions; if a symmetry generator $\hG[\eps]$ involves 
time-dependent parameters $\eps^a(t)$, then its evolution
equation (\ref{1.5.3}) is modified to: 
\be 
i \frac{d\hG[\eps]}{dt} = \left[ \hG[\eps], \hH \right] + 
 i\, \dd{\hG[\eps]}{t},
\label{1.5.8}
\ee 
where
\be 
\dd{\hG[\eps]}{t} = \dd{\eps^a}{t}\, \frac{\del \hG[\eps]}{\del \eps^a}. 
\label{1.5.8.1}
\ee 
It follows, that $\hG[\eps]$ can generate symmetries of the hamiltonian 
and be conserved at the same time for arbitrary $\eps^a(t)$ only if
the functional derivative vanishes: 
\be 
\frac{\del \hG[\eps]}{\del \eps^a(t)}  = 0,
\label{1.5.9}
\ee 
which defines a set of operator constraints, the quantum equivalent 
of (\ref{1.3.14}). The important step in this argument is to realize, 
that the transformation properties of the evolution operator should 
be consistent with the Schr\"{o}dinger equation, which can be true only 
if both conditions (symmetry and conservation law) hold. To see this, 
recall that the evolution operator 
\be 
\hU(t,t^{\prime}) = e^{-i (t-t^{\prime}) \hH}, 
\label{1.5.10}
\ee 
is the formal solution of the Schr\"{o}dinger equation 
\be 
\lh i \dd{}{t} - \hH \rh \hU = 0, 
\label{1.5.11}
\ee 
satisfying the initial condition $\hU(t,t) = \hat{1}$. Now under 
a symmetry transformation (\ref{1.5.4}), (\ref{1.5.7}) this equation 
transforms into 
\be 
\ba{lll}
\dsp{ \del \left[ \lh i\dd{}{t} - \hH \rh \hU \right] }& = & \dsp{
 - i \left[ \lh i \dd{}{t} - \hH \rh \hU , \hG[\eps] \right] }\\
 & & \\
 & = & \dsp{ -i \lh i \dd{}{t} -\hH \rh \left[ \hU, \hG[\eps] \right]
 -i \left[ \lh i\dd{}{t} - \hH \rh, \hG[\eps] \right] \hU } 
\ea 
\label{1.5.12}
\ee 
For the transformations to respect the Schr\"{o}dinger equation, 
the left-hand side of this identity must vanish, hence so must the 
right-hand side. But the right-hand side vanishes for arbitrary
$\eps(t)$ if and only if both conditions are met: 
\[
\left[ \hH, \hG[\eps] \right] = 0, \hs{1} \mbox{and} \hs{1} 
\dd{\hG[\eps]}{t} = 0.
\]
This is what we set out to prove. Of course, like in the classical 
hamiltonian formulation, we realize that for generators of local 
symmetries a more general first-class algebra of commutation relations 
is allowed, along the lines of eqs.(\ref{1.4a.17}). Also here, the 
hamiltonian may then be modified by terms involving only the constraints 
and, possibly, corresponding lagrange multipliers. The discussion 
parallels that for the classical case. 

\section{The relativistic particle \label{s1.6}}

In this section and the next we revisit the two examples of 
constrained systems in sect.\ \ref{s1.2} to illustrate the general 
principles of symmetries, conservation laws and constraints above. 
First we consider the relativistic particle. 

The starting point of the analysis is the action (\ref{1.2.8}):
\[
S[x^{\mu};e] = \frac{m}{2}\, \int_1^2\, \lh \frac{1}{e}\, 
 \frac{dx_{\mu}}{d\lb} \frac{dx^{\mu}}{d\lb} - e c^2 \rh d\lb. 
\]
Here $\lb$ plays the role of system time, and the hamiltonian we 
construct is the one generating time-evolution in this sense. 
The canonical momenta are given by 
\be 
p_{\mu} = \frac{\del S}{\del (dx^{\mu}/d\lb)} = \frac{m}{e}\, 
 \frac{dx_{\mu}}{d\lb}, \hs{2} 
p_e = \frac{\del S}{\del (de/d\lb)} = 0. 
\label{1.6.1}
\ee 
The second equation is a constraint on the extended phase space 
spanned by the canonical pairs $(x^{\mu}, p_{\mu}; e, p_e)$. Next 
we perform a legendre transformation to obtain the hamiltonian 
\be 
H = \frac{e}{2m} \lh p^2 + m^2c^2 \rh + p_e\, \frac{de}{d\lb}. 
\label{1.6.2}
\ee 
The last term obviously vanishes upon application of the constraint
$p_e = 0$. The canonical (hamiltonian) action now reads 
\be 
S _{can} = \int_1^2 d\lb\, \lh p_{\mu} \frac{dx^{\mu}}{d\lb} 
 - \frac{e}{2m} \lh p^2 + m^2c^2 \rh \rh. 
\label{1.6.3}
\ee 
Observe, that the dependence on $p_e$ has dropped out, irrespective
of whether we constrain it to vanish or not. The role of the einbein
is now clear: it is a lagrange multiplier imposing the dynamical 
constraint (\ref{1.2.7}): 
\[ 
p^2 + m^2 c^2 = 0. 
\]
Note, that in combination with $p_e = 0$, this constraint implies 
$H = 0$, i.e.\ the hamiltonian consists {\em only} of a polynomial
in the constraints. This is a general feature of systems with 
reparametrization invariance, including for example the theory of 
relativistic strings and general relativity. 

In the example of the relativistic particle, we immediately encounter 
a generic phenomenon: any time we have a constraint on the dynamical 
variables imposed by a lagrange multiplier (here: $e$), its associated 
momentum (here: $p_e$) is constrained to vanish. It has been shown in 
a quite general context, that one may always reformulate hamiltonian
theories with constraints such that all constraints appear with 
lagrange multipliers \ct{fj}; therefore this pairing of constraints 
is a generic feature in hamiltonian dynamics. However, as we have 
already discussed in sect.\ \ref{s1.4}, such lagrange multiplier terms 
do not affect the dynamics, and the multipliers as well as their 
associated momenta can be eliminated from the physical hamiltonian. 

The non-vanishing Poisson brackets of the theory, including the 
lagrange multipliers, are 
\be 
\left\{ x^{\mu}, p_{\nu} \right\} = \del^{\mu}_{\nu}, \hs{2} 
\left\{ e, p_e \right\} = 1.
\label{1.6.4}
\ee 
As follows from the hamiltonian treatment, all equations of motion 
for any quantity $\Fg(x,p;e,p_e)$ can then be obtained from a  
Poisson bracket with the hamiltonian: 
\be 
\frac{d\Fg}{d\lb} = \left\{ \Fg, H \right\}, 
\label{1.6.4.1}
\ee 
although this equation does not imply any non-trivial information 
on the dynamics of the lagrange multipliers. Nevertheless, in this 
formulation of the theory it must be assumed {\em a priori} that 
$(e,p_e)$ are allowed to vary; the dynamics can be projected to 
the hypersurface $p_e = 0$ only after computing Poisson brackets. 
The alternative is to work with a restricted phase space spanned 
only by the physical co-ordinates and momenta $(x^{\mu},p_{\mu})$. 
This is achieved by performing a Legendre transformation only with 
respect to the physical velocities\footnote{This is basically a 
variant of Routh's procedure; see e.g.\ Goldstein \ct{gold}, ch.\ 7.}. 
We first explore the formulation of the theory in the extended 
phase space. 

\nit
All possible symmetries of the theory can be determined by solving 
eq.(\ref{1.4a.6}):
\[
\left\{ G, H \right\} = 0.
\]
Among the solutions we find the generators of the Poincar\'{e} 
group: translations $p_{\mu}$ and Lorentz transformations $M_{\mu\nu} 
= x_{\nu} p_{\mu} - x_{\mu} p_{\nu}$. Indeed, the combination of
generators
\be 
G[\eps] = \eps^{\mu} p_{\mu} + \frac{1}{2}\, \eps^{\mu\nu} M_{\mu\nu}.
\label{1.6.5}
\ee 
with constant $(\eps^{\mu}, \eps^{\mu\nu})$ produces the expected 
infinitesimal transformations 
\be 
\del x^{\mu} = \left\{ x^{\mu}, G[\eps] \right\} = 
 \eps^{\mu} + \eps^{\mu}_{\;\;\nu}\, x^{\nu}, \hs{2} 
\del p_{\mu} = \left\{ p_{\mu}, G[\eps] \right\} = 
 \eps_{\mu}^{\;\;\nu}\, p_{\nu}. 
\label{1.6.6}
\ee 
The commutator algebra of these transformations is well-known to be 
closed: it is the Lie algebra of the Poincar\'{e} group. 
\vs{1} 

\nit
{\bf Exercise 1.4} \nl 
Check that the bracket of $G[\eps]$ and the hamiltonian $H$ vanishes.
Compute the bracket of two Poincar\'{e} transformations $G[\eps_1]$ 
and $G[\eps_2]$. \nl

For the generation of constraints the local reparametrization 
invariance of the theory is the one of interest. The infinitesimal 
form of the transformations (\ref{1.2.10}) is obtained by taking 
$\lb^{\prime} = \lb - \eps(\lb)$, with the result 
\be 
\ba{l}
\dsp{ \del x^{\mu} = x^{\prime\, \mu}(\lb) - x^{\mu}(\lb) 
 = \eps\,\frac{dx^{\mu}}{d\lb}, \hs{2} 
\del p_{\mu} = \eps\, \frac{dp_{\mu}}{d\lb}, }\\
 \\
\dsp{ \del e = e^{\prime}(\lb) - e(\lb) = \frac{d(e \eps)}{d\lb}. }
\ea
\label{1.6.7}
\ee 
Now recall, that $e d\lb = d\tau$ is a reparametrization-invariant 
form. Furthermore, $\eps(\lb)$ is an arbitrary local function of
$\lb$. It follows, that without loss of generality we can consider 
an equivalent set of {\em covariant} transformations with parameter
$\sg = e \eps$: 
\be 
\ba{l} 
\dsp{ \del_{cov}\, x^{\mu} = \frac{\sg}{e}\, \frac{dx^{\mu}}{d\lb}, \hs{2} 
\del_{cov}\, p_{\mu} = \frac{\sg}{e}\, \frac{dp_{\mu}}{d\lb}, }\\
 \\
\dsp{ \del_{cov}\, e = \frac{d\sg}{d\lb}. }
\ea 
\label{1.6.7.1}
\ee
It is straightforward to check, that under these transformations the 
canonical lagrangean (the integrand of (\ref{1.6.3})) transforms into 
a total derivative, and $\del_{cov}\, S_{can} = [B_{cov}]_1^2$, with 
\be
B_{cov}[\sg] = \sg \lh p_{\mu}\, \frac{dx^{\mu}}{ed\lb} - 
 \frac{1}{2m} (p^2 + m^2 c^2) \rh.
\label{1.6.8}
\ee 
Using eq.(\ref{1.4a.10}), we find that the generator of the local 
transformations (\ref{1.6.7}) is given by
\be 
G_{cov}[\sg] = (\del_{cov} x^{\mu}) p_{\mu} + (\del_{cov} e) p_e 
 - B_{cov} = \frac{\sg}{2m}\, \lh p^2 + m^2c^2 \rh + 
 p_e \frac{d\sg}{d\lb}.
\label{1.6.9}
\ee 
It is easily verified, that $dG_{cov}/d\lb = 0$ on physical trajectories
for arbitrary $\sg(\lb)$ if and only if the two earlier constraints are 
satisfied at all times:
\be
p^2 + m^2 c^2 = 0, \hs{2} p_e = 0. 
\label{1.6.constr}
\ee 
It is clear that the Poissonbrackets of these constraints among 
themselves vanish. On the canonical variables, $G_{cov}$ generates the 
transformations 
\be 
\ba{ll} 
\dsp{ \del_G\, x^{\mu} = \left\{ x^{\mu}, G_{cov}[\sg] \right\}
 = \frac{\sg p^{\mu}}{m}, }& \dsp{
\del_G\, p_{\mu} = \left\{ p_{\mu}, G_{cov}[\sg] \right\} = 0, }\\
 \\ 
\dsp{ \del_G\, e = \left\{ e, G_{cov}[\sg] \right\} = \frac{d\sg}{d\lb}, }
 & \dsp{ \del_G\, p_e = \left\{ p_e, G_{cov}[\sg] \right\} = 0. }
\ea 
\label{1.6.10}
\ee 
These transformation rules actually differ from the original ones, 
eq.(\ref{1.6.7.1}). However, all the differences vanish when applying 
the equations of motion: 
\be 
\ba{l} 
\dsp{ \del^{\prime} x^{\mu} = ( \del_{cov} - \del_G) x^{\mu} = 
 \frac{\sg}{m} \lh \frac{m}{e}\, \frac{dx^{\mu}}{d\lb} - p^{\mu} \rh 
 \approx 0, }\\
 \\ 
\dsp{ \del^{\prime} p_{\mu} = (\del_{cov} - \del_G)p_{\mu} = 
 \frac{\sg}{e}\, \frac{dp_{\mu}}{d\lb} \approx 0. }
\ea
\label{1.6.11}
\ee 
The transformations $\del^{\prime}$ are in fact themselves symmetry
transformations of the canonical action, but of a trivial kind: as
they vanish on shell, they do not imply any conservation laws or 
constraints \ct{jwthesis}. Therefore the new transformations 
$\del_G$ are physically equivalent to $\del_{cov}$.  

The upshot of this analysis is, that we can describe the relativistic 
particle by the hamiltonian (\ref{1.6.2}) and the Poisson brackets 
(\ref{1.6.4}), provided we impose on all physical quantities in 
phase space the constraints (\ref{1.6.constr}). 

A few comments are in order. First, the hamiltonian is by construction
the generator of translations in the time coordinate (here: $\lb$); 
therefore after the general exposure in sects.\ \ref{s1.3} and \ref{s1.4}
it should not come as a surprise, that when promoting such translations 
to a local symmetry, the hamiltonian is constrained to vanish. 

Secondly, we briefly discuss the other canonical procedure, 
which takes directly advantage of the the local parametrization 
invariance (\ref{1.2.10}) by using it to fix the einbein; in particular 
the choice $e = 1$ leads to the identification of $\lb$ with proper 
time: $d\tau = e d\lb \rightarrow d\tau = d\lb$. This procedure is 
called gauge fixing. Now the canonical action becomes simply 
\be 
S_{can}|_{e=1} = \int_1^2 d\tau\, \lh p \cdot \dot{x} - \frac{1}{2m}\,
 \lh p^2 + m^2c^2 \rh \rh. 
\label{1.6.12}
\ee 
This is a regular action for a hamiltonian system. It is completely
Lorentz covariant, only the local reparametrization invariance is 
lost. As a result, the constraint $p^2 + m^2c^2 = 0$ can no longer 
be derived from the action; it must now be imposed separately as 
an external condition. Because we have fixed $e$, we do not need to 
introduce its conjugate momentum $p_e$, and we can work in a 
restricted physical phase space spanned by the canonical pairs 
$(x^{\mu}, p_{\mu})$. Thus, a second consistent way to formulate 
classical hamiltonian dynamics for the relativistic particle is to 
use the gauge-fixed hamiltonian and Poisson brackets 
\be 
H_{f} = \frac{1}{2m} \lh p^2 + m^2c^2 \rh, \hs{2} 
 \left\{ x^{\mu}, p_{\nu} \right\} = \del^{\mu}_{\nu},
\label{1.6.13}
\ee
whilst adding the constraint $H_f = 0$ to be satisfied at all
(proper) times. Observe, that the remaining constraint implies 
that one of the momenta $p_{\mu}$ is not independent:
\be 
p_0^2 = \vec{p}^{\,2} + m^2 c^2.
\label{1.6.12.1}
\ee
As this defines a hypersurface in the restricted phase space, the 
dimensionality of the physical phase space is reduced even further. 
To deal with this situation, we can again follow two different 
routes; the first one is to solve the constraint and work in a
reduced phase space. The standard procedure for this is to 
introduce light-cone coordinates $x^{\pm} = (x^0 \pm x^3)/\sqrt{2}$, 
with canonically conjugate momenta $p_{\pm} = (p_0 \pm p_3)/\sqrt{2}$, 
such that
\be 
\left\{ x^{\pm}, p_{\pm} \right\} = 1, \hs{2} 
\left\{ x^{\pm}, p_{\mp} \right\} = 0. 
\label{1.6.12.2}
\ee 
The constraint (\ref{1.6.12.1}) can then be written 
\be 
2p_+ p_- = p_1^2 + p_2^2 + m^2 c^2,
\label{1.6.12.3}
\ee 
which allows us to eliminate the light-cone co-ordinate $x_-$ and its
conjugate momentum $p_- = (p_1^2 + p_2^2 + m^2 c^2)/2p_+$. Of course, 
by this procedure the manifest Lorentz-covariance of the model 
is lost. Therefore one often prefers an alternative route:
to work in the covariant phase space (\ref{1.6.13}), and impose
the constraint on physical phase space functions only after 
solving the dynamical equations. 

\section{The electro-magnetic field \label{s1.7}} 

The second example to be considered here is the electro-magnetic 
field. As our starting point we take the action of exercise 1.2, 
which is the action of eq.(\ref{1.2.18}) modified by a total 
time-derivative, in which the magnetic field has been written 
in terms of the vector potential as $\vec{B}(A) = \vec{\nabla} 
\times \vec{A}$:
\be 
\ba{l}
S_{em}[\fg,\vec{A},\vec{E}] = \dsp{ \int_1^2 dt\, 
 L_{em}(\fg, \vec{A}, \vec{E}), }\\
 \\
L_{em} = \dsp{ \int d^3 x\, \lh  -\frac{1}{2}\, \lh \vec{E}^2 + 
 [\vec{B}(A)]^2 \rh - \fg\, \vec{\nabla} \cdot \vec{E} - \vec{E} 
 \cdot \dd{\vec{A}}{t} \rh }
\ea 
\label{1.7.1}
\ee 
It is clear, that $(\vec{A}, -\vec{E})$ are canonically conjugate; 
by adding the time derivative we have chosen to let $\vec{A}$ play 
the role of co-ordinates, whilst the components of $-\vec{E}$ 
represent the momenta:
\be 
\vec{\pi}_A = -\vec{E} = \frac{\del S_{em}}{\del (\der \vec{A}/\der t)} 
\label{1.7.2}
\ee
Also, like the einbein in the case of the relativistic particle, 
here the scalar potential $\fg = A_0$ plays the role of 
lagrange multiplier to impose the constraint $\vec{\nabla} \cdot 
\vec{E} = 0$; therefore its canonical momentum vanishes: 
\be 
\pi_{\fg} = \frac{\del S_{em}}{\del (\der \fg/ \der t)} = 0. 
\label{1.7.2.1}
\ee 
This is the generic type of constraint for lagrange multipliers,
which we encountered also in the case of the relativistic particle.
Observe, that the lagrangean (\ref{1.7.1}) is already in the 
canonical form, with the hamiltonian given by 
\be 
H_{em} = \int d^3 x\, \lh \frac{1}{2}\, \lh \vec{E}^2 + \vec{B}^2
 \rh + \fg\, \vec{\nabla} \cdot \vec{E} + \pi_{\fg} \dd{\fg}{t} \rh. 
\label{1.7.3}
\ee 
Again, as in the case of the relativistic particle, the last term
can be taken to vanish upon imposing the constraint (\ref{1.7.2.1}),
but in any case it cancels in the canonical action
\be
\ba{lll}
S_{em} & = & \dsp{ \int_1^2 dt \lh \int d^3x \left[ - \vec{E} \cdot 
 \dd{\vec{A}}{t} + \pi_{\fg} \dd{\fg}{t} \right] - 
 H(\vec{E},\vec{A},\pi_{\fg},\fg) \rh }\\
 & & \\
 & = & \dsp{ \int_1^2 dt \lh \int d^3x \left[ - \vec{E} \cdot 
 \dd{\vec{A}}{t} \right] - H(\vec{E},\vec{A},\fg)|_{\pi_{\fg}=0} \rh }
\ea
\label{1.7.3.1}
\ee 
To proceed with the canonical analysis, we have the same choice
as in the case of the particle: to keep the full hamiltonian, 
and include the canonical pair $(\fg, \pi_{\fg})$ in an extended 
phase space; or to use the local gauge invariance to remove 
$\fg$ by fixing it at some particular value. 

In the first case we have to introduce Poisson brackets 
\be 
\left\{ A_i(\vec{x},t), E_j(\vec{y},t) \right\}
 = - \del_{ij}\, \del^3(\vec{x} - \vec{y}), \hs{2} 
\left\{ \fg(\vec{x},t), \pi_{\fg}(\vec{y},t) \right\} 
 = \del^3(\vec{x} - \vec{y}). 
\label{1.7.4}
\ee 
It is straightforward to check, that the Maxwell equations are
reproduced by the brackets with the hamiltonian:
\be 
\dot{\Fg} = \left\{ \Fg, H \right\}, 
\label{1.7.5}
\ee
where $\Fg$ stands for any of the fields $(\vec{A}, \vec{E}, \fg, 
\pi_{\fg})$ above, although in the sector of the scalar potential
the equations are empty of dynamical content.  

Among the quantities commuting with the hamiltonian (in the sense 
of Poisson brackets), the most interesting for our purpose is the 
generator of the gauge transformations
\be 
\del \vec{A} = \vec{\nabla} \Lb, \hs{2} \del \fg = \dd{\Lb}{t},
 \hs{2} \del \vec{E} = \del \vec{B} = 0.
\label{1.7.6}
\ee 
Its construction proceeds according to eq.(\ref{1.4a.10}).
Actually, the action (\ref{1.7.1}) is gauge invariant provided
the gauge parameter vanishes sufficiently fast at spatial infinity, 
as $\del L_{em} = - \int d^3 x\, \vec{\nabla} \cdot ( \vec{E} \der 
\Lb/ \der t)$. Therefore the generator of the gauge transformations 
is 
\be 
\ba{lll}
G[\Lb] & = & \dsp{ \int d^3 x\, \lh - \del \vec{A} \cdot \vec{E} + 
 \del \fg\, \pi_{\fg} \rh }\\
 & & \\
 & = & \dsp{ \int d^3 x\, \lh - \vec{E} \cdot \vec{\nabla} \Lb + 
 \pi_{\fg} \dd{\Lb}{t} \rh\, =\, \int d^3x\, \lh \Lb 
 \vec{\nabla} \cdot \vec{E} + \pi_{\fg} \dd{\Lb}{t} \rh. }
\ea 
\label{1.7.7}
\ee 
The gauge transformations (\ref{1.7.6}) are reproduced by the 
Poisson brackets 
\be 
\del \Fg = \left\{ \Fg, G[\Lb] \right\}.
\label{1.7.6.1}
\ee 
From the result (\ref{1.7.7}) it follows, that conservation of  
$G[\Lb]$ for arbitrary $\Lb(\vec{x},t)$ is due to the constraints 
\be 
\vec{\nabla} \cdot \vec{E} = 0, \hs{2} \pi_{\fg} = 0, 
\label{1.7.8}
\ee 
which are necessary and sufficient. These in turn imply that 
$G[\Lb] = 0$ itself. 

One reason why this treatment might be prefered, is that in
a relativistic notation $\fg = A_0$, $\pi_{\fg} = \pi^0$, the
brackets (\ref{1.7.4}) take the quasi-covariant form 
\be 
\left\{ A_{\mu}(\vec{x},t), \pi^{\nu}(\vec{y},t) \right\} 
 = \del_{\mu}^{\nu}\, \del^3(\vec{x} - \vec{y}), 
\label{1.7.cov.1}
\ee 
and similarly for the generator of the gauge transformations :
\be 
G[\Lb] = - \int d^3x\, \pi^{\mu} \der_{\mu} \Lb.
\label{1.7.cov.2}
\ee  
Of course, the three-dimensional $\del$-function and integral 
show, that the covariance of these equations is not complete. 

The other procedure one can follow, is to use the gauge invariance
to set $\fg = \fg_0$, a constant. Without loss of generality 
this constant can be chosen equal to zero, which just amounts to
fixing the zero of the electric potential. In any case, the term
$\fg\, \vec{\nabla} \cdot \vec{E}$ vanishes from the action and for
the dynamics it suffices to work in the reduced phase space spanned
by $(\vec{A}, \vec{E})$. In particular, the hamiltonian and 
Poisson brackets reduce to  
\be 
H_{red} = \int d^3x\, \frac{1}{2} \lh \vec{E}^2 + \vec{B}^2 \rh, 
\hs{2} \left\{ A_i(\vec{x},t), E_j(\vec{y},t) \right\} = 
 -\del_{ij} \del^3(\vec{x} - \vec{y}).  
\label{1.7.9}
\ee 
The constraint $\vec{\nabla} \cdot \vec{E} = 0$ is no longer a 
consequence of the dynamics, but has to be imposed separately. 
Of course, its bracket with the hamiltonian still vanishes:
$\{H_{red}, \vec{\nabla} \cdot \vec{E}\} = 0$. The constraint 
actually signifies that one of the components of the canonical 
momenta (in fact an infinite set: the longitudinal electric field 
at each point in space) is to vanish; therefore the dimensionality 
of the physical phase space is again reduced by the constraint. As 
the constraint is preserved in time (its Poisson bracket with $H$ 
vanishes), this reduction is consistent. Again, there are two 
options to proceed: solve the constraint and obtain a phase space 
spanned by the physical degrees of freedom only, or keep the 
constraint as a separate condition to be imposed on all solutions 
of the dynamics. The explicit solution in this case consists of 
splitting the electric field in transverse and longitudinal parts 
by projection operators: 
\be 
\vec{E} = \vec{E}_T + \vec{E}_L = 
 \lh 1 - \vec{\nabla} \frac{1}{\Del} \vec{\nabla} \rh \cdot \vec{E}
 + \vec{\nabla} \frac{1}{\Del} \vec{\nabla} \cdot \vec{E},
\label{1.7.10}
\ee 
and similarly for the vector potential. One can  now restrict 
the phase space to the transverse parts of the fields only; this 
is equivalent to requiring $\vec{\nabla} \cdot \vec{E} = 0$ and 
$\vec{\nabla} \cdot \vec{A} = 0$ simultaneously. In practice it is 
much more convenient to use these constraints as such in computing 
physical observables, instead of projecting out the longitudinal 
components explicitly at all intermediate stages. Of course, one 
then has to check that the final result does not depend on any 
arbitrary choice of dynamics attributed to the longitudinal fields.

\section{Yang-Mills theory \label{s1.8}}

Yang-Mills theory is an important extension of Maxwell theory, with 
a very similar canonical structure. The covariant action is a 
direct extension of the electro-magnetic action in exercise 1.2:
\be 
S_{YM} = - \frac{1}{4}\, \int d^4x\, 
 F^a_{\mu\nu} F_a^{\mu\nu},
\label{1.8.1}
\ee 
where $F_{\mu\nu}^a$ is the field strength of the Yang-Mills 
vector potential $A_{\mu}^a$:
\be 
F_{\mu\nu}^a = \der_{\mu} A^a_{\nu} - \der_{\nu} A^a_{\mu} -
 g f_{bc}^{\;\;\;a} A^b_{\mu} A^c_{\nu}. 
\label{1.8.2}
\ee 
Here $g$ is the coupling constant, and the coefficients 
$f_{bc}^{\;\;\;a}$ are the structure constant of a compact Lie 
algebra ${\sf g}$ with (anti-hermitean) generators 
$\left\{ T_a \right\}$: 
\be 
\left[ T_a, T_b \right] = f_{ab}^{\;\;\;c}\, T_c. 
\label{1.8.3}
\ee 
The Yang-Mills action (\ref{1.8.1}) is invariant under (infinitesimal) 
local gauge transformations with parameters $\Lb^a(x)$:
\be 
\del A_{\mu}^a = (D_{\mu} \Lb)^a = \der_{\mu} \Lb^a - 
 g f_{bc}^{\;\;\;a} A_{\mu}^b \Lb^c, 
\label{1.8.4}
\ee 
under which the field strength $F_{\mu\nu}^a$ transforms as 
\be 
\del F_{\mu\nu}^a = gf_{bc}^{\;\;\;a} \Lb^b F_{\mu\nu}^c. 
\label{1.8.5}
\ee 
To obtain a canonical description of the theory, we compute the 
momenta
\be 
\pi_a^{\mu} = \frac{\del S_{YM}}{\del \der_0 A_{\mu}^a} = 
 - F_a^{0 \mu} = \left\{ 
 \ba{rl} 
 - E_{a}^i, & \mu = i = (1,2,3); \\ 
   0, & \mu = 0.
 \ea \rd
\label{1.8.6}
\ee 
Clearly, the last equation is a constraint of the type we 
have encountered before; indeed, the time component of the 
vector field, $A_0^a$, plays the same role of lagrange mutiplier 
for a Gauss-type constraint as the scalar potential $\fg =  A_0$ 
in electro-dynamics, to which the theory reduces in the limit 
$g \rightarrow 0$. This is brought out most clearly in 
the hamiltonian formulation of the theory, with action 
\be 
\ba{l} 
\dsp{ S_{YM} = \int_1^2 dt \lh \int d^3x \left[ - \vec{E}_a \cdot 
 \dd{\vec{A}^a}{t} \right] - H_{YM} \rh, }\\
 \\
\dsp{ H_{YM} = \int d^3x\, \lh \frac{1}{2}\, (\vec{E}_a^{2} + 
 \vec{B}_a^{2}) + A_0^a\, (\vec{D} \cdot \vec{E})_a \rh. }
\ea
\label{1.8.7}
\ee 
Here we have introduced the notation $\vec{B}^a$ for the 
magnetic components of the field strength:
\be 
B_i^a = \frac{1}{2}\, \ve_{ijk} F^a_{jk}.
\label{1.8.7.1}
\ee 
In eqs.(\ref{1.8.7}) we have left out all terms involving the 
time-component of the momentum, since they vanish as a result of 
the constraint $\pi_a^{0} = 0$, eq.(\ref{1.8.6}). Now $A_0^a$ 
appearing only linearly, its variation leads to another constraint 
\be 
(\vec{D} \cdot \vec{E})^a = \vec{\nabla} \cdot \vec{E}^a - 
 g f_{bc}^{\;\;\;a} \vec{A}^b \cdot \vec{E}^c = 0. 
\label{1.8.8}
\ee 
As in the other theories we have encountered so far, the constraints 
come in pairs: one constraint, imposed by a Lagrange multiplier, 
restricts the physical degrees of freedom; the other constraint is 
the vanishing of the momentum associated with the Lagrange multiplier. 

To obtain the equations of motion, we need to specify the 
Poisson brackets:
\be 
\left\{ A_i^a(\vec{x},t), E_{jb}(\vec{y},t) \right\} = 
 - \del_{ij} \del^a_b\, \del^3(\vec{x} - \vec{y}), \hs{1} 
\left\{ A_0^a, (\vec{x},t), \pi^0_{b}(\vec{y},t) \right\} = 
 \del_{ij} \del^a_b\, \del^3(\vec{x} - \vec{y}),
\label{1.8.8.1}
\ee 
or in quasi-covariant notation 
\be 
\left\{ A_\mu^a(\vec{x},t), \pi^{\nu}_{b}(\vec{y},t) \right\} = 
 \del_{\mu}^{\nu} \del^a_b\, \del^3(\vec{x} - \vec{y}).
\label{1.8.8.2}
\ee 
Provided the gauge parameter vanishes sufficiently fast at spatial 
infinity, the canonical action is gauge invariant: 
\be 
\del S_{YM} = - \int_1^2 dt \int d^3x\, \vec{\nabla} \cdot \lh 
 \vec{E}_a \dd{\Lb^a}{t} \rh \simeq 0.
\label{1.8.9}
\ee 
Therefore it is again straightforward to construct the generator 
for the local gauge transformations: 
\be 
\ba{lll}
G[\Lb] & = & \dsp{ \int d^3x\, \lh - \del \vec{A}^a \cdot \vec{E}_a 
 + \del A_0^a\, \pi^0_a \rh }\\  
 & & \\
 & = & \dsp{ \int d^3x\, \pi_a^{\mu} (D_{\mu} \Lb)^a\, \simeq\, 
 \int d^3x\, \lh \Lb_a (\vec{D} \cdot \vec{E})^a + 
 \pi^0_a\, (D_0 \Lb)^a \rh. }
\ea
\label{1.8.10}
\ee 
The new aspect of the gauge generators in the case of Yang-Mills 
theory is, that the constraints satisfy a non-trivial Poisson 
bracket algebra: 
\be 
\left\{ G[\Lb_1], G[\Lb_2] \right\} = G[\Lb_3], 
\label{1.8.11}
\ee 
where the parameter on the right-hand side is defined by 
\be 
\Lb_3 = g f_{bc}^{\;\;\;a}\, \Lb^b_1\, \Lb^c_2. 
\label{1.8.12}
\ee 
We can also write the physical part of the constraint algebra in a 
local form; indeed, let
\be 
G_a(x) = (\vec{D} \cdot \vec{E})_a(x). 
\label{1.8.loc.1}
\ee 
Then a short calculation leads to the result
\be 
\left\{ G_a(\vec{x},t), G_b(\vec{y},t) \right\} = 
 g f_{ab}^{\;\;\;c}\, G_c(\vec{x},t)\, \del^3(\vec{x}-\vec{y}) .
\label{1.8.loc.2}
\ee 
We observe, that the condition $G[\Lb] = 0$ is satisfied for 
arbitrary $\Lb(x)$ if and only if the two local constraints 
hold: 
\be 
(\vec{D} \cdot \vec{E})^a = 0, \hs{2} \pi^0_a = 0. 
\label{1.8.13}
\ee 
This is sufficient to guarantee that $\left\{ G[\Lb], H \right\} = 0$
holds as well. Together with the closure of the algebra of constraints
(\ref{1.8.11}) this guarantees that the constraints $G[\Lb] = 0$ 
are consistent both with the dynamics and among themselves. 

Eq.(\ref{1.8.13}) is the generalization of the transversality 
condition (\ref{1.7.8}) and removes the same number of momenta 
(electric field components) from the physical phase space. Unlike
the case of electrodynamics however, it is non-linear and can not
be solved explicitly. Moreover, the constraint does not determine 
in closed form the conjugate co-ordinate (the combination of gauge 
potentials) to be removed from the physical phase space with it. 
A convenient possibility to impose in classical Yang-Mills theory 
is the transversality condition $\vec{\nabla} \cdot \vec{A}^a = 0$, 
which removes the correct number of components of the vector potential 
and still respects the rigid gauge invariance (with constant parameters 
$\Lb^a$).
\vs{1} 

\nit
{\bf Exercise 1.5} \nl
Prove eqs.(\ref{1.8.11}) and (\ref{1.8.loc.2}).
 
\section{The relativistic string \label{s1.9}}

As the last example in this chapter we consider the massless 
relativistic (bosonic) string, as described by the Polyakov action 
\be 
S_{str} = \int d^2\xi\, \lh - \frac{1}{2}\, \sqrt{-g} g^{ab}\, 
 \der_a X^{\mu} \der_b X_{\mu} \rh, 
\label{1.9.1}
\ee 
where $\xi^a = (\xi^0,\xi^1) = (\tau, \sg)$ are co-ordinates 
parametrizing the two-dimensional world sheet swept out by the 
string, $g_{ab}$ is a metric on the world sheet, with $g$ its 
determinant, and $X^{\mu}(\xi)$ are the co-ordinates of the string 
in the $D$-dimensional embedding space-time (the target space), 
which for simplicity we take to be flat (Minkowskian). As a 
generally covariant two-dimensional field theory, the action is 
manifestly invariant under reparametrizations of the world sheet:
\be 
X^{\prime}_{\mu}(\xi^{\prime}) = X_{\mu}(\xi), \hs{2} 
g^{\prime}_{ab}(\xi^{\prime}) = g_{cd}(\xi)\, 
 \dd{\xi^c}{\xi^{\prime\,a}} \dd{\xi^d}{\xi^{\prime\, b}}. 
\label{1.9.1.1}
\ee 
The canonical momenta are 
\be 
\Pi_{\mu} = \frac{\del S_{str}}{\del \der_0 X^{\mu}} 
 = - \sqrt{-g}\, \der^{\,0} X_{\mu}, \hs{2}
\pi_{ab} = \frac{\del S_{str}}{\del \der_0 g^{ab}} = 0. 
\label{1.9.2}
\ee 
The latter equation brings out, that the inverse metric $g^{ab}$, 
or rather the combination $h^{ab} = \sqrt{-g} g^{ab}$, acts as a 
set of lagrange multipliers, imposing the vanishing of the 
symmetric energy-momentum tensor:
\be 
T_{ab} = \frac{2}{\sqrt{-g}}\, \frac{\del S_{str}}{\del g^{ab}} 
 = - \der_a X^{\mu} \der_b X_{\mu} + \frac{1}{2}\, g_{ab} g^{cd}
 \der_c X^{\mu} \der_d X_{\mu} = 0. 
\label{1.9.3}
\ee 
Such a constraint arises because of the local reparametrization 
invariance of the action. Note however, that the energy-momentum
tensor is traceless: 
\be 
T_a^{\;\;a} = g^{ab}\, T_{ab} = 0. 
\label{1.9.4}
\ee 
and as a result it has only two independent components. The origin 
of this reduction of the number of constraints is the local Weyl 
invariance of the action (\ref{1.9.1})
\be 
g_{ab}(\xi) \rightarrow \bar{g}_{ab}(\xi) = e^{\Lb(\xi)}\, g_{ab}(\xi),
 \hs{2} X^{\mu}(\xi) \rightarrow \bar{X}^{\mu}(\xi) = X^{\mu}(\xi),
\label{1.9.5}
\ee 
which leaves $h^{ab}$ invariant: $\bar{h}^{ab} = h^{ab}$. Indeed, 
$h^{ab}$ itself also has only two independent components, as
the negative of its determinant is unity: $-h = -\det h^{ab} = 1$. 

The hamiltonian is obtained by Legendre transformation, and taking 
into account $\pi^{ab} = 0$ it reads 
\be 
\ba{lll}
H & = & \dsp{ \frac{1}{2}\, \int d\sg \lh \sqrt{-g}\, 
 \lh - g^{00} [\der_0 X]^2 + g^{11} [\der_1 X]^2 \rh\, 
 +\, \pi^{ab}\, \der_0\, g_{ab} \rh }\\
 & & \\ 
 & = & \dsp{ \int d\sg\, \lh T^{\,0}_{\;\;0} + \pi^{ab}\,
 \der_0\, g_{ab} \rh.}
\ea
\label{1.9.6}
\ee 
The Poisson brackets are 
\be 
\ba{l}
\dsp{ \left\{ X^{\mu}(\tau,\sg), \Pi_{\nu}(\tau,\sg^{\prime})
 \right\}\, =\, \del^{\mu}_{\nu}\, \del(\sg - \sg^{\prime}), }\\
 \\
\dsp{ \left\{ g_{ab}(\tau,\sg), \pi^{cd}(\tau,\sg^{\prime}) 
 \right\}\, =\, \frac{1}{2}\, \lh \del_a^c \del_b^d + \del_a^d
 \del_b^c \rh\, \del(\sg - \sg^{\prime}). }
\ea
\label{1.9.6.1}
\ee 
The constraints (\ref{1.9.3}) are most conveniently expressed in the 
hybrid forms (using relations $g = g_{00} g_{11} - g_{01}^2$ and 
$g_{11} = gg^{00}$):
\be 
\ba{rll}
g T^{00} & = & \dsp{ - T_{11}\, =\, 
 \frac{1}{2}\, \lh \Pi^2 + [\der_1 X]^2 \rh\, =\, 0, }\\ 
 & & \\ 
\sqrt{-g}\, T^{\,0}_{\;\;1} & = & \dsp{ \Pi \cdot \der_1 X = 0. } 
\ea 
\label{1.9.7}
\ee 
These results imply, that the hamiltonian (\ref{1.9.6}) actually
vanishes, as in the case of the relativistic particle. The 
reason is also the same: reparametrization invariance, now on 
a two-dimensional world sheet rather than on a one-dimensional 
world line. 

The infinitesimal form of the transformations (\ref{1.9.1.1}) 
with $\xi^{\prime} = \xi - \Lb(\xi)$ is
\be 
\ba{lll}
\del X^{\mu}(\xi) & = & \dsp{ X^{\prime\,\mu}(\xi) - X^{\mu}(\xi) 
 = \Lb^a \der_a X^{\mu} = \frac{1}{g g^{00}}\, \lh 
 \sqrt{-g}\, \Lb^0 \Pi^{\mu} + \Lb_1 \der_{\sg} X^{\mu} \rh, }\\
 & & \\ 
\del g_{ab}(\xi) & = & \dsp{ (\der_a \Lb^c) g_{cb} + (\der_b \Lb^c) 
 g_{ac} + \Lb^c \der_c g_{ab} = D_a \Lb_b + D_b \Lb_a, }
\ea
\label{1.9.8}
\ee 
where we use the covariant derivative $D_a \Lb_b = \der_a \Lb_b 
- \Gam_{ab}^{\;\;\;c} \Lb_c$. 
The generator of these transformations as constructed by our 
standard procedure now becomes 
\be 
\ba{lll}
G[\Lb] & = & \dsp{ \int d\sg \lh \Lb^a \der_a X \cdot \Pi + 
 \frac{1}{2}\, \Lb^0 \sqrt{-g}\, g^{ab} \der_a X \cdot \der_b X 
 + \pi^{ab} (D_a \Lb_b + D_b \Lb_a) \rh }\\
 & & \\
 & = & \dsp{ \int d\sg\, \lh  - \sqrt{-g}\, \Lb^a T^{\,0}_{\;\;a} 
 + 2 \pi^{ab} D_a \Lb_b \rh. }
\ea
\label{1.9.9}
\ee 
which has to vanish inorder to represent a canonical symmetry: 
the constraint $G[\Lb] = 0$ summarizes all constraints introduced 
above. The brackets of $G[\Lb]$ now take the form 
\be  
\left\{ X^{\mu}, G[\Lb] \right\} = \Lb^a \der_a X^{\mu} = \del X^{\mu}, 
\hs{2}
\left\{ g_{ab}, G[\Lb] \right\} = D_a \Lb_b + D_b \Lb_a = \del g_{ab}, 
\label{1.9.10}
\ee 
and in particular
\be 
\left\{ G_[\Lb_1], G[\Lb_2] \right\} = G_[\Lb_3], \hs{2} 
\Lb_3^a = \Lb_{\left[ 1 \rd}^b \der_b \Lb_{\ld 2 \right]}^a. 
\label{1.9.11}
\ee 
It takes quite a long and difficult calculation to check this 
result. 

Most practioners of string theory prefer to work in the restricted 
phase space, in which the metric $g_{ab}$ is not a dynamical variable, 
and there is no need to introduce its conjugate momentum $\pi^{ab}$. 
Instead, $g_{ab}$ is chosen to have a convenient value by exploiting 
the reparametrization invariance (\ref{1.9.1.1}) or (\ref{1.9.8}):
\be 
g_{ab} = \rg\, \eta_{ab} = \rg\, \lh \ba{cc} -1 & 0 \\
                                            0 & 1 \ea \rh. 
\label{1.9.12}
\ee 
Because of the Weyl invariance (\ref{1.9.5}) $\rg$ never appears 
explicitly in any physical quantity, so it does not have to be 
fixed itself. In particular, the hamiltonian becomes 
\be 
H_{red} = \frac{1}{2}\, \int d\sg \lh [\der_0 X]^2 + [\der_1 X]^2 \rh 
 = \frac{1}{2}\, \int d\sg \lh \Pi^2 + [\der_{\sg} X]^2 \rh, 
\label{1.9.13}
\ee 
whilst the constrained gauge generators (\ref{1.9.9}) become 
\be 
G_{red}[\Lb] = \int d\sg \lh \frac{1}{2}\, \Lb^0 \lh \Pi^2 + 
 [\der_{\sg} X]^2 \rh + \Lb^1 \Pi \cdot \der_{\sg} X \rh.
\label{1.9.14}
\ee 
Remarkably, these generators still satisfy a closed bracket 
algebra: 
\be 
\left\{ G_{red}[\Lb_1], G_{red}[\Lb_2] \right\} = G_{red}[\Lb_3], 
\label{1.9.15}
\ee 
but the structure constants have changed, as becomes evident from
the expression for $\Lb_3$: 
\be
\ba{l} 
\Lb_3^0 = \Lb^1_{\left[ 1 \rd} \der_{\sg} \Lb^0_{\ld 2 \right]} 
 + \Lb^0_{\left[ 1 \rd} \der_{\sg} \Lb^1_{\ld 2 \right]}, \\
 \\
\Lb_3^1 = \Lb^0_{\left[ 1 \rd} \der_{\sg} \Lb^0_{\ld 2 \right]}
 + \Lb^1_{\left[ 1 \rd} \der_{\sg} \Lb^1_{\ld 2 \right]}
\ea 
\label{1.9.16}
\ee 
The condition for $G_{red}[\Lb]$ to generate a symmetry of the 
hamiltonian $H_{red}$ (and hence to be conserved), is again 
$G_{red}[\Lb] = 0$. Observe, that these expressions reduce to 
those of (\ref{1.9.11}) when the $\Lb^a$ satisfy
\be 
\der_{\sg} \Lb^1 = \der_{\tau} \Lb^0, \hs{2} \der_{\sg} \Lb^0 = 
 \der_{\tau} \Lb^1. 
\label{1.9.17}
\ee 
In terms of the light-cone co-ordinates $u = \tau - \sg$ or $v = 
\tau + \sg$ this can be written: 
\be 
\der_u (\Lb^1 + \Lb^0) = 0, \hs{2} 
\der_v (\Lb^1 - \Lb^0) =0. 
\label{1.9.18}
\ee 
As a result, the algebras are identical for parameters living
on only one branch of the (two-dimensional) light-cone: 
\be 
\Lb^0(u,v) = \Lb_+(v) - \Lb_-(u),   \hs{2} 
\Lb^1(u,v) = \Lb_+(v) + \Lb_-(u), 
\label{1.9.19}
\ee 
with $\Lb_{\pm} = (\Lb_1 \pm \Lb_0)/2$. 
\vs{1} 

\nit
{\bf Exercise 1.6} \nl 
a.\ Compute the commutator of two infinitesimal transformations 
(\ref{1.9.8}) and show it results in a similar transformation 
with parameter $\Lb_3$ of eq.(\ref{1.9.11}). \nl
b.\ Prove equations (\ref{1.9.15}) and (\ref{1.9.16}).

%% file: brs2.tex
\chapter{Canonical BRST construction \label{ch2}} 

Many interesting physical theories incorporate constraints 
arising from a local gauge symmetry, which forces certain 
components of the momenta to vanish in the physical phase 
space. For reparametrization-invariant systems (like the 
relativistic particle or the relativistic string) these 
constraints are quadratic in the momenta, whereas in abelian 
or non-abelian gauge theories of Maxwell-Yang-Mills type they 
are linear in the momenta (i.e., in the electric components 
of the field strength). 
 
There are several ways to deal with such constraints. The most 
obvious one is to solve them and formulate the theory purely 
in terms of physical degrees of freedom. However, this is 
possible only in the simplest cases, like the relativistic 
particle or an unbroken abelian gauge theory (electrodynamics). 
And even then, there can arise complications such as non-local 
interactions. Therefore in most cases and for most applications 
an alternative strategy is more fruitful; this prefered strategy 
is to keep (some) unphysical degrees of freedom in the theory in 
such a way that desirable properties of the description, like
locality, and rotation or Lorentz-invariance, can be preserved 
at intermediate stages of calculations. In this chapter we discuss
methods for dealing with such a situation, when unphysical degrees
of freedom are taken along in the analysis of the dynamics. 

The central idea of the BRST construction is to identify the 
solutions of the constraints with the cohomology classes of a 
certain nilpotent operator, the BRST operator $\Og$. To construct 
this operator we introduce a new class of variables, the ghost 
variables. For the theories we have discussed in chapter \ref{ch1}, 
which do not involve fermion fields in essential way (at least 
from the point of view of constraints), the ghosts are anticommuting 
variables: odd elements of a Grassmann algebra. However, theories 
with more general types of gauge symmetries involving fermionic 
degrees of freedom, like supersymmetry or Siegel's $\kg$-invariance 
in the theory of superparticles and superstrings, or theories 
with reducible gauge symmetries, require commuting ghost 
variables as well. Nevertheless, to bring out the central 
ideas of the BRST construction as clearly as possible, here
we discuss theories with bosonic symmetries only.

\section{Grassmann variables \label{s2.1}}

The BRST construction involves anticommuting variables, which 
are odd elements of a Grassmann algebra. The theory of such 
variables plays an important role in quantum field theory, 
most prominently in the description of fermion fields as they 
naturally describe systems satisfying the Pauli exclusion 
principle. For these reasons we briefly review the basic elements 
of the theory of anticommuting variables at this point. For more 
detailed expositions we refer to the references \ct{ber,dwitt}.

A Grassmann algebra of rank $n$ is the set of polynomials constructed 
from elements $\left\{ e, \thg_1, ... , \thg_n \right\}$ with 
the properties 
\be 
e^2 = e, \hs{2} e \thg_i = \thg_i e = \thg_i, \hs{2} 
\thg_i \thg_j + \thg_j \thg_i = 0. 
\label{2.1.1}
\ee 
Thus $e$ is the identity element, which will often not be written 
out explicitly. The elements $\thg_i$ are nilpotent: $\thg_i^2 = 0$,  
whilst for $i \neq j$ the elements $\thg_i$ and $\thg_j$ anticommute. 
As a result, a general element of the algebra consists of $2^n$ terms 
and takes the form 
\be 
g = \ag e + \sum_{i=1}^n\, \ag^i\, \thg_i + \sum_{(i,j) = 1}^n\, 
 \frac{1}{2!}\, \ag^{ij}\, \thg_i \thg_j + ... + \tilde{\ag}\, 
 \thg_1 ... \thg_n,
\label{2.1.2}
\ee 
where the coefficients $\ag^{i_1..i_p}$ are completely antisymmetric 
in the indices. The elements $\left\{ \thg_i \right \}$ are called the 
generators of the algebra. An obvious example of a Grassmann algebra 
is the algebra of differential forms on an $n$-dimensional manifold. 

On the Grassmann algebra we can define a co-algebra of polynomials 
in elements $\left\{\bar{\thg}^1, ... , \bar{\thg}^n \right\}$,
which together with the unit element $e$ is a Grassmann algebra by
itself, but which in addition has the property
\be 
[ \bar{\thg}^i, \thg_j ]_+ = \bar{\thg^i}\, \thg_j + 
 \thg_j\, \bar{\thg}^i = \del^i_j\, e.
\label{2.1.3}
\ee 
This algebra can be interpreted as the algebra of derivations on 
the Grassmann algebra spanned by $(e,\thg_i)$. 

By the property (\ref{2.1.3}) the complete set of elements 
$\left\{ e; \thg_i; \bar{\thg}^i \right\}$ is actually turned into 
a Clifford algebra, which has a (basically unique) representation 
in terms of Dirac matrices in $2n$-dimensional space. The relation 
can be established by considering the following complex linear 
combinations of Grassmann generators:
\be 
\Gam_i = \gam_{i} = \bar{\thg}^i + \thg_i, \hs{2} 
\tilde{\Gam}_i = \gam_{i+n} = i \lh \bar{\thg}^i - \thg_i \rh, 
 \hs{2} i = 1, ... ,n.
\label{2.1.4}
\ee 
By construction these elements satisfy the relation
\be 
[ \gam_a, \gam_b ]_+ = 2\, \del_{ab}\, e, \hs{3}
(a,b) = 1, ... ,2n, 
\label{2.1.5}
\ee 
but actually the subsets $\left\{ \Gam_i \right\}$ and $\left\{ 
\tilde{\Gam}_i \right\}$ define two mutually anti-commuting 
Clifford algebras of rank $n$: 
\be 
[ \Gam_i, \Gam_j ]_+ = [ \tilde{\Gam}_i, \tilde{\Gam}_j ]_+ 
 = 2\, \del_{ij}, \hs{2} 
[ \Gam_i, \tilde{\Gam}_j ]_+ = 0. 
\label{2.1.6}
\ee 
Of course, the construction can be turned around to construct a
Grassmann algebra of rank $n$ and its co-algebra of derivations 
out of a Clifford algebra of rank $2n$. 

In field theory applications we are mostly interested in Grassmann
algebras of infinite rank, not only $n \rightarrow \infty$, but 
particularly also the continuous case 
\be 
[ \bar{\thg}(t), \thg(s) ]_+ = \del(t - s),
\label{2.1.7}
\ee 
where $(s,t)$ are real-valued arguments. Obviously, a Grassmann
{\em variable} $\xi$ is a quantity taking values in a set of 
linear Grassmann forms $\sum_i \ag^i \thg_i$ or its continuous 
generalization $\int_t \ag(t)\, \thg(t)$. Similarly, one can 
define derivative operators $\der / \der \xi$ as linear operators 
mapping Grassmann forms of rank $p$ into forms of rank $p-1$, by
\be 
\dd{}{\xi}\, \xi = 1 - \xi \dd{}{\xi},
\label{2.1.8}
\ee 
and its generalization for systems of multi-Grassmann variables.
These derivative operators can be constructed as linear forms 
in $\bar{\thg}^i$ or $\bar{\thg}(t)$.

In addition to differentiation one can also define Grassmann 
integration. In fact, Grassmann integration is defined as 
identical with Grassmann differentiation. For a single Grassmann
variable, let $f(\xi) = f_0 + \xi f_1$; then one defines
\be 
\int d\xi\, f(\xi) = f_1.
\label{2.1.9}
\ee 
This definition satisfies all standard properties of indefinite 
integrals:
\begin{enumerate} 
\item linearity: 
\be 
\int d\xi\, \left[ \ag f(\xi) + \bg g(\xi) \right] = 
 \ag \int d\xi\, f(\xi) + \bg \int d\xi\, g(\xi);
\label{2.1.10}
\ee 
\item translation invariance: 
\be 
\int d\xi\, f(\xi + \eta) = \int d\xi\, f(\xi); 
\label{2.1.11}
\ee 
\item fundamental theorem of calculus (Gauss-Stokes): 
\be
\int d\xi \dd{f}{\xi} = 0;
\label{2.1.12}
\ee 
\item reality: for {\em real} functions $f(\xi)$ (i.e.\ 
$f_{0,1} \in {\bf R}$)
\be 
\int d\xi f(\xi) = f_1 \in {\bf R}.
\label{2.1.13}
\ee 
\end{enumerate} 
A particularly useful result is the evaluation of Gaussian 
Grassmann integrals. First we observed, that 
\be 
\int \left[ d\xi_1 ... d\xi_n \right] \xi_{\ag_1} ... \xi_{\ag_n}
 = \ve_{\ag_1 ... \ag_n}. 
\label{2.1.14}
\ee 
From this it follows, that a general Gaussian Grassmann integral is 
\be 
\int \left[ d\xi_1 ... d\xi_n \right] \exp \lh \frac{1}{2}\,
 \xi_{\ag} A_{\ag\bg}\, \xi_{\bg} \rh = \pm \sqrt{ |\det A| }. 
\label{2.1.15}
\ee 
This is quite obvious after bringing $A$ into block-diagonal form: 
\be 
A = \lh \ba{ccc} 
     \ba{cc} 0 & \og_1 \\ - \og_1 & 0 \ea & & 0 \\
       & \ba{cc} 0 & \og_2 \\ - \og_2 & 0 \ea & \\
      0 & & \ba{cc} \cdot & \\  & \cdot \ea 
 \ea \rh.
\label{2.1.16}
\ee 
There are then two possibilities: \nl
(i) If the dimensionality of the matrix $A$ is even ($(\ag, \bg) 
= 1, ... , 2r$) and none of the characteristic values $\og_i$
vanishes, then every $2 \times 2$ block gives a contribution 
$2 \og_i$ to the exponential:
\be 
\exp\lh \frac{1}{2}\, \xi_{\ag} A_{\ag \bg}\, \xi_{\bg} \rh 
= \exp \lh \sum_{i = 1}^r\, \og_i\, \xi_{2i - 1} \xi_{2i} \rh 
 = 1 + ... + \prod_{i = 1}^r (\og_i\, \xi_{2i-1} \xi_{2i}).
\label{2.1.17}
\ee 
The final result is then established by performing the Grassmann
integrations, which leaves a non-zero contribution only from the 
last term, reading
\be 
\prod_{i=1}^r \og_i = \pm \sqrt{ |\det A| },
\label{2.1.18}
\ee 
the sign depending on the number of negative characteristic 
values $\og_i$. \nl
(ii) If the dimensionality of $A$ is odd, the last block is 
one-dimensional representing a zero-mode; then the integral 
vanishes, as does the determinant. Of course, the same is true 
for even-dimensional $A$ if one of the values $\og_i$ vanishes.

Another useful result is, that one can define a Grassmann-valued
delta-function:
\be 
\del(\xi - \xi^{\prime}) = - \del(\xi^{\prime} - \xi) 
 = \xi - \xi^{\prime},
\label{2.1.19}
\ee 
with the properties
\be 
\int d\xi\, \del(\xi - \xi^{\prime}) = 1, \hs{2} 
\int d\xi\, \del(\xi - \xi^{\prime}) f(\xi) = f(\xi^{\prime}). 
\label{2.1.20}
\ee 
The proof follows simply by writing out the integrants and using
the fundamental rule of integration (\ref{2.1.9}). 

\section{Classical BRST transformations \label{s2.2}} 

Consider again a general dynamical system subject to a set of 
constraints $G_{\ag} = 0$, as defined in eqs.(\ref{1.3.11}) or 
(\ref{1.4a.10}). We take the algebra of constraints to be 
first-class, as in eq.(\ref{1.4a.17}): 
\be 
\left\{ G_{\ag}, G_{\bg} \right\} = P_{\ag\bg}(G), \hs{2}
\left\{ G_{\ag}, H \right\} = Z_{\ag}(G).
\label{2.2.1}
\ee 
Here $P(G)$ and $Z(G)$ are polynomial expressions in the constraints, 
such that $P(0) = Z(0) = 0$; in particular this implies that the 
constant terms vanish: $c_{\ag\bg} = 0$. 

The BRST construction starts with the introduction of canonical
pairs of Grassmann degrees of freedom $(c^{\ag}, b_{\bg})$, one 
for each constraint $G_{\ag}$, with Poisson brackets 
\be 
\left\{ c^{\ag},  b_{\bg} \right\} = 
 \left\{ b_{\bg}, c^{\ag} \right\} = -i \del^{\ag}_{\bg}, 
\label{2.2.2}
\ee 
These anti-commuting variables are known as ghosts; the complete 
Poisson brackets on the extended phase space are given by
\be  
\left\{ A, B \right\} = \dd{A}{q^i} \dd{B}{p_i} - \dd{A}{p_i} 
 \dd{B}{q^i} + i (-1)^A \lh \dd{A}{c^{\ag}} \dd{B}{b_{\ag}} 
 + \dd{A}{b_{\ag}} \dd{B}{c^{\ag}} \rh,
\label{2.2.3}
\ee 
where $(-1)^A$ denotes the Grassmann parity of $A$: $+1$, if $A$
is Grassmann-even (commuting), and $-1$ if $A$ is Grassmann-odd
(anti-commuting). 

With the help of these ghost degrees of freedom one defines the 
BRST charge $\Og$, which has Grassmann parity $(-1)^{\Og} = -1$, as 
\be 
\Og = c^{\ag} \lh G_{\ag} + M_{\ag} \rh,
\label{2.2.4}
\ee 
where $M_{\ag}$ is Grassmann-even and of the form
\be 
\ba{lll}
M_{\ag} & = & \dsp{ \sum_{n \geq 1}\, \frac{i^n}{2 n!}\, c^{\ag_1} ... 
 c^{\ag_{n}} M_{\ag\ag_1...\ag_{n}}^{\;\;\bg_1...\bg_{n}} b_{\bg_1} 
 ... b_{\bg_{n}} }\\
 & & \\
 & = & \dsp{ \frac{i}{2}\, c^{\ag_1} M_{\ag\ag_1}^{\;\;\bg_1} b_{\bg_1} - 
 \frac{1}{4}\, c^{\ag_1} c^{\ag_2} M_{\ag\ag_1\ag_2}^{\;\;\bg_1\bg_2} 
 b_{\bg_1} b_{\bg_2} + ... }
\ea
\label{2.2.5}
\ee 
The quantities $M_{\ag\ag_1...\ag_p}^{\;\;\bg_1...\bg_p}$ are functions 
of the classical phase-space variables via the constraints $G_{\ag}$, 
and are defined such that
\be 
\left\{ \Og, \Og \right\} = 0. 
\label{2.2.6}
\ee 
As $\Og$ is Grassmann-odd, this is a non-trivial property, from 
which the BRST charge can be constructed inductively:
\be 
\ba{lll}
\left\{ \Og, \Og \right\} & = & c^{\ag} c^{\bg} \lh P_{\ag\bg}
 + M_{\ag\bg}^{\;\;\gam} G_{\gam} \rh \\
 & & \\ 
 & & +\, i c^{\ag} c^{\bg} c^{\gam} \lh \left\{ G_{\ag}, 
 M_{\bg\gam}^{\;\;\del} \right\} - M_{\ag\bg}^{\;\;\,\ve} 
 M_{\gam\eps}^{\;\;\del} + M_{\ag\bg\gam}^{\;\;\del\,\ve}\, 
 G_{\ve} \rh b_{\del} + ...
\ea
\label{2.2.7}
\ee 
This vanishes if and only if 
\be 
\ba{l}
M_{\ag\bg}^{\;\;\gam} G_{\gam} = - P_{\ag\bg}, \\
 \\
M_{\ag\bg\gam}^{\;\;\del\,\ve} G_{\ve} = \left\{ 
 M_{\left[\ag\bg\rd}^{\;\;\;\del}, G_{\ld\gam\right]} \right\} + 
 M_{\left[\ag\bg\rd}^{\;\;\,\ve} M_{\ld \gam \right]\ve}^{\;\;\,\del}, \\
 \\
 ...
\ea
\label{2.2.8}
\ee 
Observe, that the first relation can only be satisfied under the 
condition $c_{\ag\bg} = 0$, with the solution
\be 
M_{\ag\bg}^{\;\;\gam} = f_{\ag\bg}^{\;\;\;\gam} + \frac{1}{2}\, 
 g_{\ag\bg}^{\;\;\;\;\gam\del} G_{\del} + ...
\label{2.2.8.1}
\ee 
The same condition guarantees that the 
second relation can be solved: the bracket on the right-hand side is 
\be 
\left\{ M_{\ag\bg}^{\;\;\del}, G_{\gam} \right\} = 
 \dd{M_{\ag\bg}^{\;\;\del}}{G_{\ve}}\, P_{\ve\gam} = \frac{1}{2}\,
 g_{\ag\bg}^{\;\;\;\;\del\ve} f_{\ve\gam}^{\;\;\sg} G_{\sg} + ...
\label{2.2.9}
\ee 
whilst the Jacobi identity (\ref{1.4a.15.1}) implies that 
\be
f_{\left[\ag\bg\rd}^{\;\;\;\ve} f_{\ld\gam\right]\ve}^{\;\;\;\del} = 0, 
\label{2.2.10}
\ee 
and therefore $M_{\left[\ag\bg\rd}^{\;\;\,\ve} 
M_{\ld\gam\right]\ve}^{\;\;\,\del} = {\cal O}[G_{\sg}]$. This allows
to determine $M_{\ag\bg\gam}^{\;\;\del\ve}$. Any higher-order terms can 
be calculated similarly. In practice $P_{\ag\bg}$ and $M_{\ag}$ usually 
contain only a small number of terms.  

Next we observe, that we can extend the classical hamiltonian $H = H_0$ 
with ghost terms such that 
\be 
H_c = H_0 + \sum_{n \geq 1}\, \frac{i^n}{n!}\, c^{\ag_1} ... c^{\ag_n}\, 
 h_{\ag_1 ... \ag_n}^{(n)\, \bg_1 ... \bg_n}(G)\, b_{\bg_1} ... b_{\bg_n}, 
 \hs{2} \left\{ \Og, H_c \right\} = 0. 
\label{2.2.11}
\ee 
Observe, that on the physical hypersurface in the phase space this 
hamiltonian coincides with the original classical hamiltonian modulo
terms which do not affect the time-evolution of the classical 
phase-space variables $(q,p)$. We illustrate the procedure by 
constructing the first term: 
\be 
\ba{lll}
\left\{ \Og, H_c \right\} & = & \dsp{ \left\{ c^{\ag} G_{\ag}, H_0 
 \right\} + \frac{i}{2}\, \left\{ c^{\ag} G_{\ag}, c^{\gam} 
 h_{\gam}^{(1)\, \bg} b_{\bg} \right\} + \frac{i}{2} \left\{ 
 c^{\ag_1} c^{\ag_2} M_{\ag_1 \ag_2}^{\bg} b_{\bg}, 
 H_0 \right\} + ... }\\ 
 & & \\
 & = & \dsp{ c^{\ag} \lh Z_{\ag} - h_{\ag}^{(1)\, \bg} G_{\bg} \rh
 + ... } 
\ea 
\label{2.2.11.1}
\ee 
Hence the bracket vanishes if the hamiltonian is extended by ghost
terms such that 
\be 
h_{\ag}^{(1)\, \bg}(G)\, G_{\bg} = Z_{\ag}(G), \hs{2} ... 
\label{2.2.11.2}
\ee 
This equation is guaranteed to have a solution by the condition 
$Z(0) = 0$. 

As the BRST charge commutes with the ghost-extended hamiltonian, we 
can use it to generate ghost-dependent symmetry transformations of the 
classical phase-space variables: the BRST transformations
\be 
\ba{lll}
\del_{\Og}\, q^i & = & \dsp{ - \left\{ \Og, q^i \right\} = 
 \dd{\Og}{p_i} = c^{\ag}\, \dd{G_{\ag}}{p_i} +\, \mbox{ghost extensions}, }\\
 \\
\del_{\Og}\, p_i & = & \dsp{ - \left\{ \Og, p_i \right\} = -
 \dd{\Og}{q_i} =  c^{\ag}\, \dd{G_{\ag}}{q^i} +\, \mbox{ghost extensions}. }
\ea
\label{2.2.12}
\ee 
These BRST transformations are just the gauge transformations with the 
parameters $\eps^{\ag}$ replaced by the ghost variables $c^{\ag}$, plus
(possibly) some ghost-dependent extension. 

Similarly, one can define BRST transformations of the ghosts: 
\be 
\ba{lll}
\del_{\Og}\, c^{\ag} & = & \dsp{ - \left\{ \Og, c^{\ag} \right\} 
 = i \dd{\Og}{b_{\ag}} =  - \frac{1}{2}\, c^{\bg} c^{\gam} 
 M_{\bg\gam}^{\;\;\ag} + ..., }\\
 & & \\
\del_{\Og}\, b_{\ag} & = & \dsp{ - \left\{ \Og, b_{\ag} \right\} 
 = i \dd{\Og}{c^{\ag}} = iG_{\ag} - c^{\bg} M_{\ag\bg}^{\;\;\gam}\, 
 b_{\gam} + ...} 
\ea 
\label{2.2.13}
\ee 
An important property of these transformations is their nilpotence: 
\be 
\del_{\Og}^2 = 0.
\label{2.2.14}
\ee 
This follows most directly from the Jacobi identity for the Poisson 
brackets of the BRST charge with any phase-space function $A$: 
\be 
\del_{\Og}^2\, A = \left\{ \Og, \left\{ \Og, A \right\} \right\} = 
 - \frac{1}{2}\, \left\{ A, \left\{ \Og, \Og \right\} \right\} = 0. 
\label{2.2.15}
\ee 
Thus the BRST variation $\del_{\Og}$ behaves like an exterior  
derivative. Next we observe, that gauge invariant physical 
quantities $F$ have the properties 
\be 
\left\{ F, c^{\ag} \right\} = i\dd{F}{b_{\ag}} = 0, \hs{2} 
\left\{ F, b_{\ag} \right\} = i \dd{F}{c^{\ag}} = 0, \hs{2}
\left\{ F, G_{\ag} \right\} = \del_{\ag} F = 0. 
\label{2.2.16}
\ee 
As a result, such physical quantities must be BRST invariant:
\be 
\del_{\Og}\, F = - \left\{ \Og, F \right\} = 0. 
\label{2.2.17}
\ee 
In the terminology of algebraic geometry, such a function $F$ is 
called BRST closed. Now because of the nilpotence, there are trivial 
solutions to this condition, of the form
\be 
F_0 = \del_{\Og}\, F_1 = - \left\{ \Og, F_1 \right\}. 
\label{2.2.18}
\ee 
These solutions are called BRST exact; they always depend on the 
ghosts $(c^{\ag}, b_{\ag})$, and can not be physically relevant. 
We conclude, that true physical quantities must be BRST closed, 
but not BRST exact. Such non-trivial solutions of the BRST 
condition (\ref{2.2.17}) define the BRST cohomology, which is the 
set
\be 
\cH(\del_{\Og}) = \frac{\mbox{Ker} (\del_{\Og})}{\mbox{Im} (\del_{\Og})}. 
\label{2.2.19}
\ee 
We will make this more precise later on. 

\section{Examples \label{s2.3}} 

As an application of the above construction, we now present the 
classical BRST charges and transformations for the gauge systems 
discussed in chapter \ref{ch1}. \nl

\nit
1.\ {\em Relativistic particle.} We consider the gauge-fixed 
version of the relativistic particle. Taking $c = 1$, the only 
constraint is 
\be 
H_0 = \frac{1}{2m} (p^2 + m^2) = 0,
\label{2.3.1}
\ee
and hence in this case $P_{\ag\bg} = 0$. We only introduce one pair 
of ghost variables, and define 
\be 
\Og = \frac{c}{2m} (p^2 + m^2).
\label{2.3.2}
\ee 
It is trivially nilpotent, and the BRST transformations of the 
phase space variables read
\be 
\ba{ll}
\dsp{ \del_{\Og} x^{\mu} = \left\{ x^{\mu}, \Og \right\} 
 = \frac{cp^{\mu}}{m}, } & 
\dsp{ \del_{\Og} p_{\mu} = \left\{ p_{\mu}, \Og \right\} = 0,}\\
 & \\
\dsp{ \del_{\Og} c = -\left\{ c, \Og \right\} = 0, }& 
\dsp{ \del_{\Og} b = -\left\{ b, \Og \right\} 
 = \frac{i}{2m}\, (p^2 +  m^2) \approx 0. } 
\ea 
\label{2.3.3}
\ee  
The $b$-ghost transforms into the constraint, hence it vanishes  
on the physical hypersurface in the phase space. It is straightforward 
to verify that $\del_{\Og}^2 = 0$. \nl

\nit
2.\ {\em Electrodynamics.} In the gauge fixed Maxwell's electrodynamics 
there is again only a single constraint, and a single pair of ghost fields 
to be introduced. We define the BRST charge
\be 
\Og = \int d^3x\, c \vec{\nabla} \cdot \vec{E}.
\label{2.3.4}
\ee 
The classical BRST transformations are just ghost-dependend 
gauge transformations: 
\be 
\ba{ll}
\del_{\Og} \vec{A} = \left\{ \vec{A}, \Og \right\} = \vec{\nabla} c, &
\del_{\Og} \vec{E} = \left\{ \vec{E}, \Og \right\} = 0, \\
 & \\
\del_{\Og} c = -\left\{ c, \Og \right\} = 0, & 
\del_{\Og} b = -\left\{ b, \Og \right\} = i \vec{\nabla} \cdot \vec{E}
 \approx 0. 
\ea 
\label{2.3.5}
\ee 

\nit
3.\ {\em Yang-Mills theory.} One of the simplest non-trivial systems 
of constraints is that of Yang-Mills theory, in which the constraints
define a local Lie algebra (\ref{1.8.loc.2}). The BRST charge becomes 
\be
\Og = \int d^3x\, \lh c^a G_a - \frac{ig}{2}\, c^a c^b f_{ab}^{\;\;\;c} 
 b_c \rh,
\label{2.3.6}
\ee 
with $G_a = (\vec{D} \cdot \vec{E})_a$. It is now non-trivial that the 
bracket of $\Og$ with itself vanishes; it is true because of the closure 
of the Lie algebra, and the Jacobi identity for the structure constants. 

The classical BRST transformations of the fields become 
\be 
\ba{ll}
\del_{\Og} \vec{A}^a = \left\{ \vec{A}^a, \Og \right\} = (\vec{D} c)^a,
 & \del_{\Og} \vec{E}_a = \left\{ \vec{E}_a, \Og \right\} 
 = g f_{ab}^{\;\;\;c} c^b \vec{E}_c, \\
 & \\
\dsp{ \del_{\Og} c^a = -\left\{ c^a, \Og \right\} = \frac{g}{2}\, 
 f_{bc}^{\;\;\;a}\, c^b c^c, }& 
\del_{\Og} b_a = - \left\{ b_a, \Og \right\} = i\, G_a + 
 g f_{ab}^{\;\;\;c}\, c^b\, b_c. 
\ea 
\label{2.3.7}
\ee 
Again, it can be checked by explicit calculation that $\del_{\Og}^2 = 0$ 
for all variations (\ref{2.3.7}). \nl

\nit
{\bf Exercise} Show that 
\[
\del_{\Og}\, G_a = g f_{ab}^{\;\;\;c} c^b\, G_c.
\]
From this, prove that $\del_{\Og}^2\, b_a = 0$.\nl 

\nit
4.\ {\em Relativistic string.} Finally, we discuss the free relativistic 
string. We take the reduced constraints (\ref{1.9.14}), satisfying the 
algebra (\ref{1.9.15}), (\ref{1.9.16}). The the BRST charge takes the 
form 
\be 
\ba{lll}
\Og  & = & \dsp{ \int d\sg \left[ \frac{1}{2}\, c^0 \lh \Pi^2 + 
 [\der_{\sg} X]^2 \rh + c^1 \Pi \cdot \der_{\sg} X \rd }\\
 & & \\
 & & \dsp{ \ld -\, i \lh c^1 \der_{\sg} c^0 + c^0 \der_{\sg} c^1 \rh b_0
      - i \lh c^0 \der_{\sg} c^0 + c^1 \der_{\sg} c^1 \rh b_1 \right]. }
\ea 
\label{2.3.8}
\ee 
The BRST transformations generated by the Poisson brackets of this 
charge read 
\be 
\ba{ll} 
\del_{\Og} X^{\mu} = \left\{ X^{\mu}, \Og \right\} = c^0 \Pi^{\mu}  
 + c^1 \der_{\sg} X^{\mu} \approx c^a \der_a X^{\mu}, & \\
 & \\
\del_{\Og} \Pi_{\mu} = \left\{ \Pi_{\mu}, \Og \right\} = 
 \der_{\sg} \lh c^0 \der_{\sg} X_{\mu} + c^1 \Pi_{\mu} \rh
 \approx \der_{\sg} \lh \ve^{ab} c_a \der_b X^{\mu} \rh, & \\
 & \\ 
\del_{\Og} c^0 = -\left\{ c^0, \Og \right\} = c^1 \der_{\sg} c^0
 + c^0 \der_{\sg} c^1, \\ 
 & \\
\del_{\Og} c^1 = -\left\{ c^0, \Og \right\} = c^0 \der_{\sg} c^0
 + c^1 \der_{\sg} c^1, \\
 & \\
\del_{\Og} b_0 = -\left\{ b_0, \Og \right\} = \frac{i}{2}\, 
 \lh \Pi^2 + [\der_{\sg} X]^2 \rh + c^1 \der_{\sg} b_0 + 
 c^0 \der_{\sg} b_1 + 2 \der_{\sg} c^1 \, b_0 + 
 2\, \der_{\sg} c^0\, b_1, \\
 & \\
\del_{\Og} b_1 = -\left\{ b_1, \Og \right\} = i\, \Pi \cdot 
 \der_{\sg} X + c^0 \der_{\sg} b_0 + c^1 \der_{\sg} b_1 + 
 2 \der_{\sg} c^0\, b_0 + 2\, \der_{\sg} c^1\, b_1. 
\ea 
\label{2.3.9}
\ee 
A tedious calculation shows, that these transformations are nilpotent 
indeed: $\del_{\Og}^2 = 0$. 

\section{Quantum BRST cohomology \label{s2.4}} 

The construction of a quantum theory for constrained systems poses 
the following problem: to have a local and/or covariant description 
of the quantum system, it is advantageous to work in an extended 
Hilbert space of states, with unphysical components, like gauge and 
ghost degrees of freedom. Therefore we need first of all a way to 
characterize physical states within this extended Hilbert space, 
and secondly a way to construct a unitary evolution operator 
which does not mix physical and unphysical components. In this 
section we show, that the BRST construction can solve both these 
problems \ct{spiegel,fs,hwm}.

We begin with a quantum system subject to constraints $G_{\ag}$;
we impose these constraints on the physical states:
\be 
G_{\ag} |\Psi \rangle = 0,
\label{2.4.0} 
\ee 
implying that physical states are gauge-invariant. In the quantum 
theory the generators of constraints are operators, which satisfy 
the commutation relations (\ref{1.5.7}):
\be 
-i \left[ G_{\ag}, G_{\bg} \right] = P_{\ag\bg}(G), 
\label{2.4.0.1}
\ee 
where we omit the hat on operators for ease of notation.

Next we introduce corresponding ghost field operators 
$(c_{\ag}, b_{\bg})$ with equal-time anti-commutation relations 
\be 
\left[ c^{\ag}, b_{\bg} \right]_+ = c^{\ag} b_{\bg} + \bg_{\bg} c^{\ag} 
 = \del^{\ag}_{\bg}. 
\label{2.4.1}
\ee 
(For simplicity, the time-dependence in the notation has been suppressed). 
In the ghost-extended Hilbert space we now construct a BRST operator 
\be 
\Og = c^{\ag} \lh G_{\ag} + \sum_{n \geq 1}\, \frac{i^n}{2n!}\, 
 c^{\ag_1} ... c^{\ag_n} M_{\ag\ag_1...\ag_n}^{\;\;\bg_1...\bg_n} 
 b_{\bg_1} ... b_{\bg_n} \rh, 
\label{2.4.2}
\ee 
which is required to satisfy the anti-commutation relation
\be 
\left[ \Og, \Og \right]_+ = 2 \Og^2 = 0. 
\label{2.4.3}
\ee 
In words, the BRST operator is nilpotent. Working out the square
of the BRST operator, we get
\be 
\ba{lll} 
\Og^2 & = & \dsp{ \frac{i}{2}\, c^{\ag} c^{\bg} \lh -i \left[ 
 G_{\ag}, G_{\bg} \right] + M_{\ag\bg}^{\;\;\gam} G_{\gam} \rh }\\
 & & \\
 & & \dsp{ -\, \frac{1}{2} c^{\ag} c^{\bg} c^{\gam} \lh -i \left[ 
 G_{\ag}, M_{\bg\gam}^{\;\;\del} \right] + M_{\ag\bg}^{\;\;\ve} 
 M_{\gam\ve}^{\;\;\del} + M_{\ag\bg\gam}^{\;\;\del\ve} G_{\ve} \rh 
 b_{\del} + ... }
\ea 
\label{2.4.4}
\ee 
As a consequence, the coefficients $M_{\ag}$ are defined as the 
solutions of the set of equations 
\be 
\ba{l} 
i \left[ G_{\ag}, G_{\bg} \right] = - P_{\ag\bg} 
 = M_{\ag\bg}^{\gam} G_{\gam}, \\
 \\
 i \left[ G_{[\ag}, M_{\bg\gam]}^{\;\;\del} \right] +
 M_{[\ag\bg}^{\;\;\,\ve} M_{\gam]\ve}^{\;\;\del} =
 M_{\ag\bg\gam}^{\;\;\del\ve} G_{\ve} \\
 \\
 ...
\ea
\label{2.4.5}
\ee 
These are operator versions of the classical equations (\ref{2.2.8}).
As in the classical case, their solution requires the absence of a
central charge: $c_{\ag\bg} = 0$. 

Observe, that the Jacobi identity for the generators $G_{\ag}$ 
implies some restrictions on the higher terms in the expansion of
$\Og$: 
\be 
\ba{l}
0 = \left[ G_{\ag}, \left[ G_{\bg}, G_{\gam} \right] \right] + 
\mbox{(terms cyclic in $[\ag\bg\gam]$)} = -3i \left[ G_{[\ag}, 
 M_{\bg\gam]}^{\;\;\del} G_{\del} \right] \\ 
 \\
 \hspace{3em} \dsp{ = -3 \lh i \left[ G_{[\ag}, M_{\bg\gam]}^{\;\;\del} 
 \right] + M_{[\ag\bg}^{\;\;\,\ve} M_{\ag]\,\ve}^{\;\;\,\del} \rh 
 G_{\del} = - \frac{3i}{2}\, M_{\ag\bg\gam}^{\;\;\del\ve} 
 M_{\del\ve}^{\;\;\sg} G_{\sg}. }
\ea 
\label{2.4.6}
\ee 
The equality on the first line follows from the first equation 
(\ref{2.4.5}), the last equality from the second one. 

To describe the states in the extended Hilbert space, we introduce 
a ghost-state module, a basis for the ghost states consisting of
monomials in the ghost operators $c^{\ag}$: 
\be 
|[\ag_1 \ag_2 ... \ag_p]\rangle_{gh} = \frac{1}{p!}\, c^{\ag_1} 
 c^{\ag_2} ... c^{\ag_p} |0\rangle_{gh},
\label{2.4.7}
\ee
with $|0\rangle_{gh}$ the ghost vacuum state annihilated by all 
$b_{\bg}$. By construction these states are completely anti-symmetric 
in the indices $[\ag_1 \ag_2 ... \ag_p]$, i.e.\ the ghosts satisfy 
Fermi-Dirac statistics, even though they do not carry spin. This 
confirms their unphysical nature. As a result of this choice of basis, 
we can decompose an arbitrary state in components with different ghost
number (= rank of the ghost polynomial):
\be 
|\Psi\rangle = |\Psi^{(0)}\rangle + c^{\ag} |\Psi^{(1)}_{\ag} \rangle 
 + \frac{1}{2}\, c^{\ag} c^{\bg} |\Psi^{(2)}_{\ag\bg} \rangle + ...
\label{2.4.8}
\ee 
where the states $|\Psi^{(n)}_{\ag_1 ... \ag_n}\rangle$ corresponding
to ghost number $n$ are of the form $|\psi^{(n)}_{\ag_1 ... \ag_n}(q) 
\rangle \times |0\rangle_{gh}$, with $|\psi^{(n)}_{\ag_1 ... \ag_n}(q) 
\rangle$ states of zero-ghost number, depending only on the degrees of 
freedom of the constrained (gauge) system; therefore we have 
\be 
b_{\bg} |\Psi^{(n)}_{\ag_1 ... \ag_n} \rangle = 0.
\label{2.4.9}
\ee 
To do the ghost-counting, it is convenient to introduce the ghost-number 
operator 
\be 
N_g = \sum_{\ag}\, c^{\ag} b_{\ag}, \hs{2} 
\left[ N_g, c^{\ag} \right] = c^{\ag}, \hs{2}
\left[ N_g, b_{\ag} \right] = - b_{\ag}, 
\label{2.4.9.1}
\ee
where as usual the summation over $\ag$ has to be interpreted in a 
generalized sense (it includes integration over space when appropriate).
It follows, that the BRST operator has ghost number $+1$:
\be 
\left[ N_g, \Og \right] = \Og.
\label{2.4.9.2}
\ee 
Now consider a BRST-invariant state: 
\be 
\Og |\Psi \rangle  = 0.
\label{2.4.10}
\ee 
Substitution of the ghost-expansions of $\Og$ and $|\Psi\rangle$
gives 
\be 
\ba{lll}
\Og |\Psi\rangle & = & \dsp{ c^{\ag} G_{\ag} |\Psi^{(0)}\rangle 
 + \frac{1}{2}\, c^{\ag} c^{\bg} \lh G_{\ag} |\Psi^{(1)}_{\bg}\rangle 
 - G_{\bg} |\Psi^{(1)}_{\ag}\rangle + i M_{\ag\bg}^{\;\;\gam} 
 |\Psi^{(1)}_{\gam} \rangle \rh }\\
 & & \\
 & & \dsp{ +\, \frac{1}{2}\, c^{\ag} c^{\bg} c^{\gam} \lh G_{\ag} 
 |\Psi^{(2)}_{\bg\gam} \rangle - i M_{\ag\bg}^{\;\;\del} 
 |\Psi^{(2)}_{\gam\del} \rangle + \frac{1}{2}\, 
 M_{\ag\bg\gam}^{\;\;\del\ve} |\Psi^{(2)}_{\del\ve}\rangle \rh + ...
}
\ea 
\label{2.4.11}
\ee 
Its vanishing then implies 
\be 
\ba{l}
G_{\ag} |\Psi^{(0)}\rangle = 0, \\
 \\ 
G_{\ag} |\Psi^{(1)}_{\bg}\rangle - G_{\bg} |\Psi^{(1)}_{\ag}\rangle 
 + i M_{\ag\bg}^{\;\;\gam} |\Psi^{(1)}_{\gam} \rangle = 0, \\
 \\
\dsp{ G_{[\ag} |\Psi^{(2)}_{\bg\gam]}\rangle - i M_{[\ag\bg}^{\;\;\del} 
 |\Psi^{(2)}_{\gam]\del}\rangle + \frac{1}{2}\, 
 M_{\ag\bg\gam}^{\;\;\del\ve} |\Psi^{(2)}_{\del\ve}\rangle = 0, }\\
 \\
 ...
\ea 
\label{2.4.12}
\ee 
These conditions admit solutions of the form
\be 
\ba{l} 
|\Psi^{(1)}_{\ag}\rangle = G_{\ag} |\chi^{(0)}\rangle, \\
 \\
|\Psi^{(2)}_{\ag\bg}\rangle = G_{\ag} |\chi^{(1)}_{\bg}\rangle 
 - G_{\bg} |\chi^{(1)}_{\ag}\rangle + i M_{\ag\bg}^{\;\;\gam} 
 |\chi^{(1)}_{\gam}\rangle, \\ 
 \\
 ...   
\ea 
\label{2.4.13}
\ee 
where the states $|\chi^{(n)}\rangle$ have zero ghost number:
$b_{\ag} |\chi^{(n)}\rangle = 0$. 
Substitution of these expressions into eq.(\ref{2.4.8}) gives 
\be 
\ba{lll}
|\Psi\rangle & = & \dsp{ |\Psi^{(0)}\rangle + c^{\ag} G_{\ag} 
 |\chi^{(0)}\rangle + c^{\ag} c^{\bg} G_{\ag} |\chi^{(1)}_{\bg}\rangle 
 + \frac{i}{2}\, c^{\ag} c^{\bg} M_{\ag\bg}^{\;\;\gam} 
 |\chi^{(1)}_{\gam}\rangle }\\
 & & \\
 & = & |\Psi^{(0)}\rangle + \Og \lh |\chi^{(0)}\rangle + 
 c^{\ag} |\chi^{(1)}_{\ag}\rangle + ... \rh \\
 & & \\
 & = & |\Psi^{(0)}\rangle + \Og\, |\chi \rangle.
\ea 
\label{2.4.14}
\ee 
The second term is trivially BRST invariant because of the nilpotence 
of the BRST operator: $\Og^2 = 0$. Assuming that $\Og$ is hermitean, 
it follows, that $|\Psi\rangle$ is normalized if and only if 
$|\Psi^{(0)}\rangle$ is: 
\be 
\langle \Psi| \Psi \rangle = \langle \Psi^{(0)} | \Psi^{(0)} \rangle 
 + 2\, \mbox{Re}\, \langle \chi| \Og | \Psi^{(0)} \rangle 
 + \langle \chi | \Og^2 | \chi \rangle 
 = \langle \Psi^{(0)} | \Psi^{(0)} \rangle.
\label{2.4.14.1}
\ee 
We conclude, that the class of normalizable BRST-invariant states 
includes the set of states which can be decomposed into a normalizable 
gauge-invariant state $|\Psi^{(0)}\rangle$ at ghost number zero, plus 
a trivially invariant zero-norm state $\Og |\chi\rangle$. These states 
are members of the BRST cohomology, the classes of states which are
BRST invariant (BRST closed) modulo states in the image of $\Og$
(BRST-exact states):
\be 
\cH(\Og) = \frac{\mbox{Ker}\, \Og}{\mbox{Im}\, \Og}. 
\label{2.4.15}
\ee

\section{BRST-Hodge decomposition of states \label{s2.4a}} 

We have shown by explicit construction, that physical states can be 
identified with the BRST-cohomology classes of which the lowest, 
non-trivial, component has zero ghost-number. However, our analysis 
does not show to what extent these solutions are unique. In this 
section we present a general discussion of BRST cohomology to 
establish conditions for the existence of a direct correspondence 
between physical states and BRST cohomology classes \ct{raz-rybkin,kvh}. 

We assume that the BRST operator is self-adjoint w.r.t.\ the 
physical inner product. An immediate consequence is, that the 
ghost-extended Hilbert space of states contains zero-norm states.
Let 
\be
|\Lb\rangle = \Og |\chi\rangle.
\label{2.4a.1}
\ee 
These states are all orthogonal to each other, including themselves, 
and thus they have zero-norm indeed:
\be 
\langle \Lb^{\prime} | \Lb \rangle = 
 \langle \chi^{\prime} |\Og^2| \chi \rangle = 0 \hs{1} 
 \Rightarrow \hs{1} \langle \Lb | \Lb \rangle = 0.
\label{2.4a.2}
\ee 
Moreover, these states are orthogonal to all normalizable BRST-invariant
states:
\be 
\Og |\Psi\rangle = 0 \hs{1} \Rightarrow \hs{1}
 \langle \Lb|\Psi \rangle = 0. 
\label{2.4a.3}
\ee 
Clearly, the BRST-exact states can not be physical. On the other 
hand, BRST-closed states are defined only modulo BRST-exact states.
We prove, that if on the extended Hilbert space $\cH_{ext}$ there 
exists a non-degenerate inner product ({\em not} the physical inner 
product), which is also non-degenerate when restricted to the subspace 
Im $\Og$ of BRST-exact states, then all physical states must be 
members of the BRST cohomology.  

A non-degenerate inner product $(\, ,\, )$ on $\cH_{ext}$ is an 
inner product with the property, that 
\be 
(\fg, \chi) = 0, \hs{1} \forall \fg, \hs{1} \Leftrightarrow \hs{1} 
\chi = 0. 
\label{2.4a.4}
\ee 
If the restriction of this inner product to Im $\Og$ is non-degenerate
as well, then 
\be 
(\Og \fg, \Og \chi) = 0, \hs{1} \forall \fg, \hs{1} \Leftrightarrow
 \hs{1} \Og \chi = 0. 
\label{2.4a.5}
\ee 
As there are no non-trivial zero-norm states w.r.t.\ this inner 
product, the BRST operator can not be self-adjoint; its adjoint, 
denoted by $^*\Og$ then defines a second nilpotent operator:
\be 
(\Og \fg, \chi) = (\fg, {^*\Og} \chi) \hs{2} \Rightarrow \hs{2} 
(\Og^2 \fg, \chi) = (\fg, {^*\Og}^2 \chi) = 0, \hs{1} \forall \fg.
\label{2.4a.6}
\ee 
The non-degeneracy of the inner product implies that ${^*\Og}^2 = 0$. 
The adjoint ${^*\Og}$ is called the co-BRST operator. Note, that 
from eq.(\ref{2.4a.5}) one infers
\be 
(\fg, {^*\Og}\, \Og \chi) = 0, \hs{1} \forall \fg, \hs{1} \Leftrightarrow 
\hs{1} {^*\Og}\, \Og \chi = 0 \hs{1} \Leftrightarrow \hs{1} \Og \chi = 0. 
\label{2.4a.6.1}
\ee 

\let\picnaturalsize=N 
\def\picsize{3.0in} 
\def\picfilename{Hext.eps} 
\ifx\nopictures Y\else{\ifx\epsfloaded Y\else\input epsf \fi 
\let\epsfloaded=Y 
\centerline{\ifx\picnaturalsize N\epsfxsize \picsize\fi \epsfbox{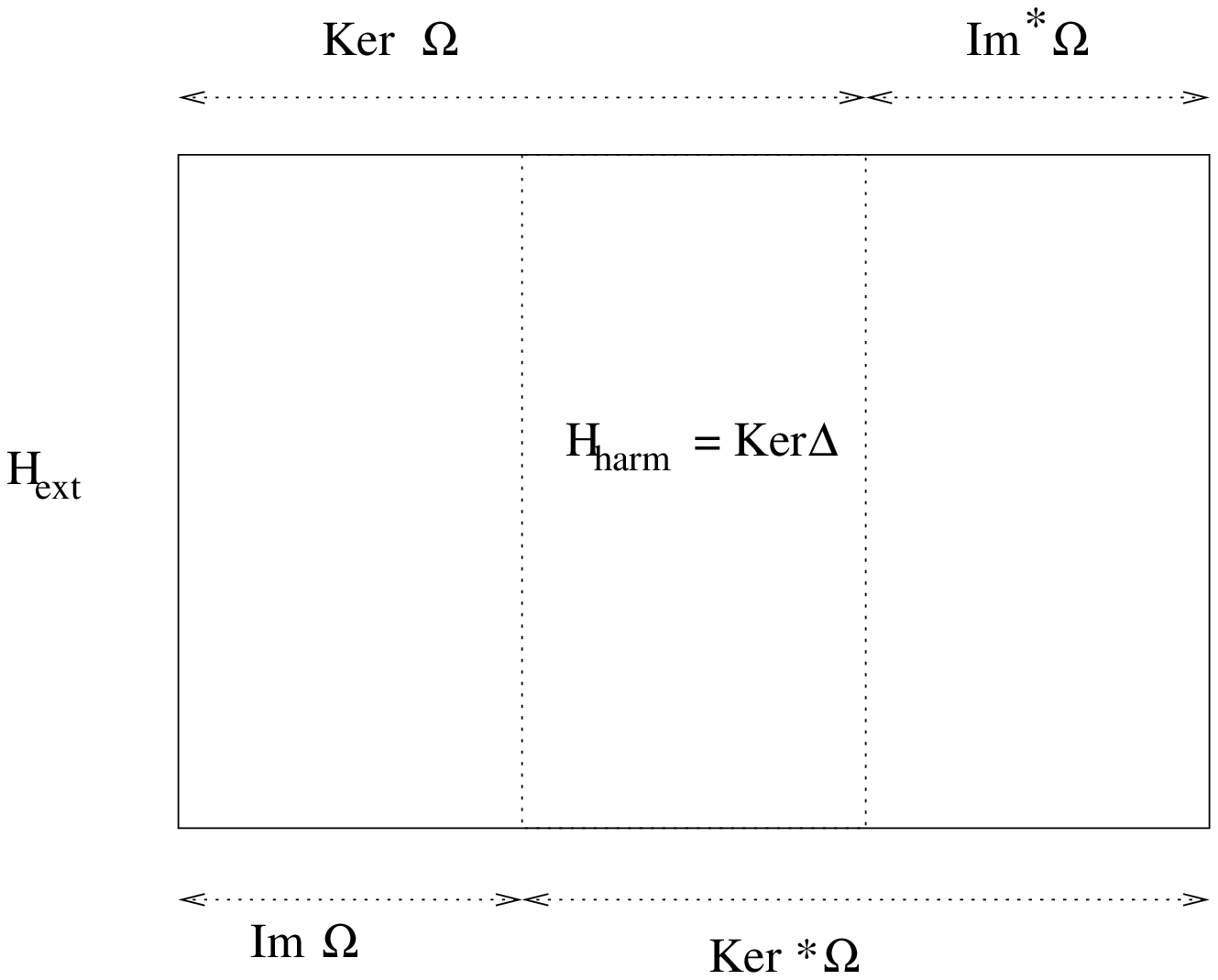}}}\fi 
\begin{center}
{\footnotesize{Fig.\ 1: Decomposition of the extended Hilbert space}}
\end{center} 

\nit
It follows immediately, that any BRST-closed vector $\Og \psi = 0$ is 
determined uniquely by requiring it to be co-closed as well. Indeed, 
let $^*\Og \psi = 0$; then   
\be 
{^*\Og}( \psi + \Og \chi) = 0 \hs{1} \Leftrightarrow \hs{1} 
{^*\Og}\, \Og \chi = 0 \hs{1} \Leftrightarrow \hs{1} \Og \chi = 0. 
\label{2.4a.7}
\ee 
Thus, if we regard the BRST transformations as gauge transformations 
on states in the extended Hilbert space generated by $\Og$, then 
$^*\Og$ represents a gauge-fixing operator determining a single 
particular state out of the complete BRST orbit. States which
are both closed and co-closed are called (BRST) {\em harmonic}.

Denoting the subspace of harmonic states by $\cH_{harm}$, we can 
now prove the following theorem: the extended Hilbert space $\cH_{ext}$ 
can be decomposed exactly into three subspaces (Fig.\ 1): 
\be 
\cH_{ext} = \cH_{harm} + \mbox{Im}\, \Og + \mbox{Im}\, ^*\Og. 
\label{2.4a.8}
\ee 
Equivalently, any vector in $\cH_{ext}$ can be decomposed as
\be 
\psi = \og + \Og \chi + {^*\Og} \fg, \hs{2} \mbox{where} \hs{1}
\Og \og = {^*\Og} \og  = 0. 
\label{2.4a.9}
\ee 
We sketch the proof. Denote the space of zero modes of the BRST operator
(the BRST-closed vectors) by Ker $\Og$, and the zero modes of the 
co-BRST operator (co-closed vectors) by Ker ${^*\Og}$. Then
\be 
\psi \in \mbox{Ker}\, \Og \hs{1} \Leftrightarrow \hs{1} 
(\Og \psi, \fg) = 0, \hs{1} \forall \fg, \hs{1} \Leftrightarrow 
\hs{1} (\psi, {^*\Og} \fg) = 0, \hs{1} \forall \fg.
\label{2.4a.10}
\ee 
$\psi$ being orthogonal to all vectors in Im ${^*\Og}$, it follows
that 
\be 
\mbox{Ker}\, \Og = \lh \mbox{Im}\, {^*\Og} \rh^{\perp},
\label{2.4a.11}
\ee 
the orthoplement of Im ${^*\Og}$. Similarly we prove 
\be 
\mbox{Ker}\, {^*\Og} = \lh \mbox{Im}\, \Og \rh^{\perp}.
\label{2.4a.12}
\ee 
Therefore any vector which is not in Im $\Og$ {\em and} not in 
Im ${^*\Og}$ must belong to the orthoplement of both, i.e.\ to 
Ker ${^*\Og}$ and Ker $\Og$ simultaneously; such a vector is 
therefore harmonic. 

Now as the BRST-operator and the co-BRST operator are both nilpotent,
\be 
\mbox{Im}\, \Og \subset \mbox{Ker}\, \Og = 
 \lh \mbox{Im}\, {^*\Og} \rh^{\perp}, \hs{2} 
\mbox{Im}\, {^*\Og} \subset \mbox{Ker}\, {^*\Og} = 
 \lh \mbox{Im}\, {\Og} \rh^{\perp}.
\label{2.4a.13}
\ee 
Therefore Im $\Og$ and Im ${^*\Og}$ have no elements in common 
(recall that the null-vector is not in the space of states). 
Obviously, they also have no elements in common with their own
orthoplements (because of the non-degeneracy of the inner product), 
and in particular with $\cH_{harm}$, which is the set of common 
states in both orthoplements. This proves the theorem. 

We can define a BRST-laplacian $\Del_{BRST}$ as the semi positive
definite self-adjoint operator
\be 
\Del_{BRST} = (\Og + {^*\Og})^2 = {^*\Og}\, \Og + \Og\, {^*\Og}, 
\label{2.4a.14}
\ee 
which commutes with both $\Og$ and ${^*\Og}$. Consider its 
zero-modes $\og$:
\be 
\Del_{BRST}\, \og = 0 \hs{1} \Leftrightarrow \hs{1}  
{^*\Og}\, \Og\, \og + \Og\, {^*\Og}\, \og = 0. 
\label{2.4a.15}
\ee 
The left-hand side of the last expression is a sum of a vector in 
Im $\Og$ and one in Im ${^*\Og}$; as these subspaces are orthogonal
w.r.t.\ the non-degenerate inner product, it follows that 
\be 
{^*\Og}\, \Og\, \og = 0 \hs{1} \wedge \hs{1} \Og\, {^*\Og}\, \og\, = 0,
\label{2.4a.16}
\ee
separately. This in turn implies $\Og \og = 0$ and ${^*\Og} \og = 0$, 
and $\og$ must be a harmonic state: 
\be 
\Del_{BRST}\, \og = 0 \hs{1} \Leftrightarrow \hs{1} 
\og \in \cH_{harm}; 
\label{2.4a.17}
\ee 
hence Ker $\Del_{BRST} = \cH_{harm}$. The BRST-Hodge decomposition 
theorem can therefore be expressed as 
\be 
\cH_{ext} = \mbox{Ker}\, \Del_{BRST} + \mbox{Im}\, \Og
 + \mbox{Im} {^*\Og}. 
\label{2.4a.18}
\ee 
The BRST-laplacian allows us to discuss the representation theory 
of BRST-transformations. First of all, the BRST-laplacian commutes
with the BRST- and co-BRST operators $\Og$ and ${^*\Og}$:
\be 
\left[ \Del_{BRST}, \Og \right] = 0, \hs{2} 
\left[ \Del_{BRST}, {^*\Og} \right] = 0. 
\label{2.4a.19}
\ee 
As a result, BRST-multiplets can be characterized by the eigenvalues 
of $\Del_{BRST}$: the action of $\Og$ or ${^*\Og}$ does not change 
this eigenvalue. Basically we must then distinguish between zero-modes 
and non-zero modes of the BRST-laplacian. The zero-modes, the harmonic 
states, are BRST-singlets:
\[
\Og | \og \rangle = 0, \hs{2} {^*\Og} | \og \rangle = 0. 
\label{2.4a.20}
\]
In contrast, the non-zero modes occur in pairs of BRST- and co-BRST-exact 
states:
\be 
\Del_{BRST} |\fg_{\pm} \rangle = \lb^2 |\fg_{\pm} \rangle \hs{1} 
\Rightarrow \hs{1} \Og |\fg_+ \rangle = \lb\, |\fg_- \rangle, \hs{1} 
 {^*\Og} |\fg_- \rangle = \lb\, | \fg_+ \rangle. 
\label{2.4a.21}
\ee 
Eq.\ (\ref{2.4a.2}) guarantees that $|\fg_{\pm} \rangle$ have zero 
(physical) norm; we can however rescale these states such that  
\be 
\langle \fg_- | \fg_+ \rangle = \langle \fg_+ | \fg_- \rangle = 1. 
\label{2.4a.22}
\ee 
It follows, that the linear combinations 
\be 
|\chi_{\pm} \rangle = \frac{1}{\sqrt{2}}\, \lh |\fg_+ \rangle \pm 
 | \fg_- \rangle \rh
\label{2.4a.23}
\ee 
define a pair of positive- or negative-norm states:
\be 
\langle \chi_{\pm} | \chi_{\pm} \rangle = \pm 1, \hs{2} 
\langle \chi_{\mp} | \chi_{\pm} \rangle = 0. 
\label{2.4a.24}
\ee 
They are eigenstates of the operator $\Og + {^*\Og}$ with eigenvalues
$(\lb, -\lb)$:
\be 
(\Og + {^*\Og}) |\chi_{\pm}\rangle = \pm \lb |\chi_{\pm}\rangle.
\label{2.4a.25}
\ee  
As physical states must have positive norm, all BRST-doublets must 
be unphysical, and only BRST-singlets (harmonic) states can represent 
physical states. Conversely, if all harmonic states are to be 
physical, only the components of the BRST-doublets are allowed to 
have non-positive norm. Observe, however, that this condition can be 
violated if the inner product $(\, ,\, )$ becomes degenerate on the 
subspace Im $\Og$; in that case the harmonic gauge does not remove 
all freedom to make BRST-transformations and zero-norm states can 
survive in the subspace of harmonic states. 

\section{BRST operator cohomology \label{s2.5}} 

The BRST construction replaces a complete set of constraints, 
imposed by the generators of gauge transformations, by a single 
condition: BRST invariance. However, the normalizable solutions 
of the BRST condition (\ref{2.4.10}): 
\[
\Og |\Psi\rangle = 0, \hs{2} \langle \Psi|\Psi \rangle = 1,
\]
are not unique: from any solution one can construct an infinite 
set of other solutions 
\be 
|\Psi^{\prime}\rangle = |\Psi\rangle + \Og |\chi\rangle, \hs{2} 
\langle \Psi^{\prime}|\Psi^{\prime}\rangle = 1,
\label{2.5.1}
\ee 
provided the BRST operator is self-adjoint w.r.t.\ the physical
inner product. Under the conditions discussed in sect.\ \ref{s2.4a} 
the normalizable part of the state vector is unique, hence the 
transformed state is not physically different from the original one, a
nd we actually identify a single physical state with the complete class 
of solutions (\ref{2.5.1}). 
As observed before, in this respect the quantum theory in the extended 
Hilbert space behaves much like an abelian gauge theory, with the BRST 
transformations acting as gauge transformations. 

Keeping this in mind, it is clearly necessary that the action of 
dynamical observables of the theory on physical states is invariant 
under BRST transformations: an observable ${\cal O}$ maps physical 
states to physical states; therefore if $|\Psi\rangle$ is a physical 
state, then 
\be 
\Og {\cal O} |\Psi\rangle = \left[ \Og, {\cal O} \right] |\Psi\rangle 
 = 0. 
\label{2.5.4}
\ee 
Again, the solution of this condition for any given observable is 
not unique: for an observable with ghost number $N_g = 0$, and any 
operator $\Fg$ with ghost number $N_g = -1$, 
\be 
{\cal O}^{\prime} = {\cal O} + \left[ \Og, \Fg \right]_+
\label{2.5.5}
\ee 
also satisfies condition (\ref{2.5.4}). The proof follows directly
from the Jacobi identity:
\be 
\left[ \Og, \left[ \Og, \Fg \right]_+ \right] = 
 \left[ \Og^2, \Fg \right] = 0. 
\label{2.5.6}
\ee 
This holds in particular for the hamiltonian; indeed, the time-evolution 
of states in the unphysical sector (the gauge and ghost fields) is not 
determined a priori, and can be chosen by an appropriate BRST extension
of the hamiltonian: 
\be
H_{ext} = H_{phys} + \left[ \Og, \Fg \right]_+.
\label{2.5.7}
\ee 
Here $H_{phys}$ is the hamiltonian of the physical degrees of freedom. 
The BRST-exact extension $[ \Og, \Fg ]_+$ acts only on the unphysical 
sector, and can be used to define the dynamics of the gauge- and ghost 
degrees of freedom. 

\section{Lie-algebra cohomology \label{s2.7}}

We illustrate the BRST construction with a simple example: a system 
of constraints defining an ordinary $n$-dimensional compact Lie-algebra 
\ct{jwvh2}. The Lie algebra is taken to be a direct sum of semi-simple 
and abelian $u(1)$ algebras, of the form
\be 
\left[ G_a, G_b \right] = if_{ab}^{\;\;\;c} G_c, \hs{3} 
 (a,b,c) = 1,...,n, 
\label{2.7.1}
\ee 
where the generators $G_a$ are hermitean, and the $f_{ab}^{\;\;\;c} 
= - f_{ba}^{\;\;\;c}$ are real structure constants. We assume the 
generators normalized such that the Killing metric is unity:
\be 
- \frac{1}{2}\, f_{ac}^{\;\;\;d} f_{bd}^{\;\;\;c} = \del_{ab}. 
\label{2.7.2}
\ee 
Then $f_{abc} = f_{ab}^{\;\;\;d} \del_{dc}$ is completely anti-symmetric. 
We introduce ghost operators $(c^a, b_b)$ with canonical anti-commutation
relations (\ref{2.4.1}):
\[
\left[ c^a, b_b \right]_+ = \del^a_b, \hs{2} \left[ c^a, c^b \right]_+
 = \left[ b_a, b_b \right]_+ = 0.
\]
This implies, that in the `co-ordinate representation', in which the
ghosts $c^a$ are represented by Grassmann variables, the $b_a$ can be 
represented by a Grassmann derivative: 
\be 
b_a = \dd{}{c^a}. 
\label{2.7.2.1}
\ee 
The nilpotent BRST operator takes the simple form 
\be 
\Og = c^a G_a - \frac{i}{2}\, c^a c^b f_{ab}^{\;\;\;c} b_c, \hs{2} 
\Og^2 = 0.
\label{2.7.3}
\ee 
We define a ghost-extended state space with elements 
\be 
\psi[c] = \sum_{k = 0}^n\, \frac{1}{k!}\,  c^{a_1} ... c^{a_k}\, 
 \psi^{(k)}_{a_1...a_k}.
\label{2.7.4}
\ee 
The coefficients $\psi^{(k)}_{a_1..a_k}$ of ghost number $k$ carry 
completely anti-symmetric product representations of the Lie algebra. 
 
On the state space we introduce an indefinite inner product, with 
respect to which the ghosts $c^a$ and $b_a$ are self-adjoint; this 
is realized by the Berezin integral over the ghost variables
\be 
\langle \fg, \psi \rangle = \int [dc^n ... dc^1]\, \fg^{\dagger}\, 
\psi = \frac{1}{n!}\, \ve^{a_1 ... a_n} \sum_{k=0}^n\, \lh 
\ba{c} n \\ k \ea \rh\, \fg^{(n-k)\, *}_{a_{n-k} ... a_1}\,
\psi^{(k)}_{a_{n-k+1} ... a_n}.
\label{2.7.5}
\ee 
In components, the action of the ghosts is given by 
\be 
(c^a \psi)^{(k)}_{a_1...a_k} = \del^a_{a_1} \psi^{(k-1)}_{a_2 a_3...a_k}
 - \del^a_{a_2} \psi^{(k-1)}_{a_1 a_3...a_k} + ... + (-1)^{k-1}
 \del^a_{a_k} \psi^{(k-1)}_{a_1 a_2...a_{k-1}},
\label{2.7.6}
\ee 
and similarly 
\be 
(b_a \psi)^{(k)}_{a_1...a_k} = \lh \dd{\psi}{c^a} \rh^{(k)}_{a_1...a_k} 
 = \psi^{(k+1)}_{aa_1...a_k}.
\label{2.7.7}
\ee 
It is now easy to check, that the ghost operators are self-adjoint 
w.r.t.\ the inner product (\ref{2.7.5}):
\be 
\langle \fg, c^a \psi \rangle = \langle c^a \fg, \psi \rangle, \hs{2} 
\langle \fg, b_a \psi \rangle = \langle b_a \fg, \psi \rangle. 
\label{2.7.8}
\ee 
It follows directly, that the BRST operator (\ref{2.7.3}) is self-adjoint 
as well:
\be 
\langle \fg,  \Og \psi \rangle = \langle \Og \fg, \psi \rangle. 
\label{2.7.9}
\ee 
Now we can introduce a second inner product, which is positive definite 
and therefore manifestly non-degenerate:
\be 
(\fg, \psi) = \sum_{k=0}^n\, \frac{1}{k!}\, \lh \fg^{(k)\, *}\rh^{a_1...a_k}
 \psi^{(k)}_{a_1...a_k}. 
\label{2.7.10}
\ee 
It is related to the first indefinite inner product by Hodge duality:
define the Hodge $*$-operator by
\be 
{^*\psi}^{(k)\, a_1...a_k} = \frac{1}{(n-k)!}\, 
 \ve^{a_1...a_ka_{k+1}...a_n}\, \psi^{(n-k)}_{a_{k+1}...a_n}.
\label{2.7.11}
\ee 
Furthermore, define the ghost permutation operator $\cP$ as the operator
which reverses the order of the ghosts in $\psi[c]$; equivalently:
\be 
(\cP \psi)^{(k)}_{a_1...a_k} = \psi^{(k)}_{a_k...a_1}.
\label{2.7.12}
\ee 
Then the two inner products are related by
\be 
(\fg, \psi) = \langle \cP\, {^*\fg}, \psi \rangle. 
\label{2.7.13}
\ee 
An important property of the non-degenerate inner product is, that the 
ghosts $c^a$ and $b_a$ are adjoint to one another:
\be 
(\fg, c^a \psi) = (b_a \fg, \psi). 
\label{2.7.14}
\ee 
Then the adjoint of the BRST operator is given by the co-BRST operator 
\be 
{^*\Og} = b_a G^a - \frac{i}{2}\, c^c\, f^{ab}_{\;\;\;c}\, b_a b_b.
\label{2.7.15}
\ee 
Here raising and lowering indices on the generators and structure 
constants is done with the help of the Killing metric ($\del_{ab}$
in our normalization). It is easy to check, that ${^*\Og}^2 = 0$, as 
expected. 

The harmonic states are both BRST- and co-BRST-closed:
$\Og \psi = {^*\Og} \psi = 0$. They are zero-modes of the 
BRST-laplacian:
\be 
\Del_{BRST} = {^*\Og}\, \Og + \Og {^*\Og} = \lh {^*\Og} + \Og \rh^2,
\label{2.7.16}
\ee 
as follows from the observation that 
\be
( \psi, \Del_{BRST}\, \psi ) = (\Og \psi, \Og \psi) + ({^*\Og}\psi,
 {^*\Og} \psi) = 0 \hs{1} \Leftrightarrow \hs{1} 
 \Og \psi = {^*\Og} \psi = 0. 
\label{2.7.17}
\ee 
For the case at hand, these conditions become 
\be 
G_a \psi = 0, \hs{2} \Sg_a \psi = 0,
\label{2.7.18}
\ee 
where $\Sg_a$ is defined as 
\be 
\Sg_a = \Sg_a^{\dagger} = -i f_{ab}^{\;\;\;c} c^{\,b} b_c.
\label{2.7.19}
\ee
From the Jacobi identity it is quite easy to verify that $\Sg_a$ 
defines a representation of the Lie-algebra:
\be 
\left[ \Sg_a, \Sg_b \right] = i f_{ab}^{\;\;\;c} \Sg_c, \hs{2} 
\left[ G_a, \Sg_b \right] = 0.
\label{2.7.20}
\ee 
The conditions (\ref{2.7.18}) are proven as follows. Substitute the 
explicit expressions for $\Og$ and ${^*\Og}$ into eq.(\ref{2.7.16})
for $\Del_{BRST}$. After some algebra one then finds
\be 
\Del_{BRST} = G^2 + G \cdot \Sg + \frac{1}{2}\, \Sg^2
 = \frac{1}{2}\, G^2 + \frac{1}{2}\, (G + \Sg)^2.
\label{2.7.21}
\ee 
This being a sum of squares, any zero mode must satisfy (\ref{2.7.18}). 
Q.E.D.

Looking for solutions, we observe that in components the second
condition reads
\be 
(\Sg_a \psi)^{(k)}_{a_1...a_k} = -i f_{a[a_1}^{\;\;\;\;b} 
 \psi^{(k)}_{a_2...a_k]b} = 0. 
\label{2.7.22}
\ee 
It acts trivially on states of ghost number $k = 0$; hence bona fide 
solutions are the gauge-invariant states of zero ghost number:
\be 
\psi = \psi^{(0)}, \hs{3} G_a \psi^{(0)} = 0.
\label{2.7.23}
\ee 
However, other solutions with non-zero ghost number exist.
A general solution is for example 
\be 
\psi = \frac{1}{3!}\, f_{abc}\, c^a c^b c^c\, \chi, \hs{2} 
 G_a \chi = 0. 
\label{2.7.24}
\ee 
The 3-ghost state $\psi^{(3)}_{abc} = f_{abc} \chi$ indeed satisfies 
(\ref{2.7.22}) as a result of the Jacobi identity. The states $\chi$
are obviously in one-to-one correspondence with the states $\psi^{(0)}$.
Hence in general there exist several copies of the space of physical 
states in the BRST cohomology, at different ghost number. We infer,
that in addition to requiring physical states to belong to the BRST 
cohomology, it is also necessary to fix the ghost number for the 
definition of physical states to be unique.

%% file: brs2a.tex
\chapter{Action formalism \label{ch2a}}

The canonical construction of the BRST cohomology we have 
described, can be given a basis in the action formulation,
either in lagrangean or hamiltonian form. The latter one 
relates most directly to the canonical bracket formulation. 
It is then straightforward to switch to a gauge-fixed lagrangean 
formulation. Once we have the lagrangean formulation, a covariant 
approach to gauge-fixing and quantization can be developed. In 
this chapter these constructions are presented, and the relations
between various formulations are discussed. 
\vs{1}

\section{BRST invariance from Hamilton's principle \label{s2a.1}}

We have observed in section \ref{s2.5}, that the effective 
hamiltonian in the ghost-extended phase space is defined only 
modulo BRST-exact terms: 
\be 
H_{eff} = H_c+ i \left\{ \Og, \Psi \right\} = H_c - i \del_{\Og} \Psi,
\label{2a.1.1}
\ee 
where $\Psi$ is a function of the phase space variables with ghost number 
$N_g(\Psi) = -1$. Moreover, the ghosts $(c,b)$ are canonically conjugate: 
\[ 
\left\{ c^{\ag}, b_{\bg} \right\} = -i \del^{\ag}_{\bg}. 
\] 
Thus we are lead to construct a pseudo-classical action of the form 
\be 
S_{eff} = \int dt\, \lh p_i \dot{q}^i + i b_{\ag} \dot{c}^{\ag} 
 - H_{eff} \rh. 
\label{2a.1.2}
\ee 
That this is indeed the correct action for our purposes follows from 
the ghost equations of motion obtained from this action, reading 
\be 
\dot{c}^{\ag} = - i\, \dd{H_{eff}}{b_{\ag}}. \hs{2} 
 \dot{b}_{\ag} = - i\, \dd{H_{eff}}{c^{\ag}}.
\label{2a.1.3}
\ee 
These equations are in full agreement with the definition of the 
extended Poisson brackets (\ref{2.2.3}):
\be 
\dot{c}^{\ag} = - \left\{ H_{eff}, c^{\ag} \right\}, \hs{2} 
\dot{b}_{\ag} = - \left\{ H_{eff}, b_{\ag} \right\}.
\label{2a.1.4}
\ee 
As $H_c$ is BRST invariant, $H_{eff}$ is BRST-invariant as well: the 
BRST variations are nilpotent and therefore $\del_{\Og}^2 \Fg = 0$. 
It is then easy to show, that the action $S_{eff}$ is BRST-symmetric, 
and the conserved Noether charge is the BRST charge as defined 
previously:
\be 
\ba{lll}
\del_{\Og} S_{eff} & = & \dsp{ \int dt \left[ \lh \del_{\Og} p_i \dot q^i 
 - \del_{\Og} q^i \dot{p}_i + i \del_{\Og} b_{\ag} \dot{c}^{\ag} 
 + i \del_{\Og} c^{\ag} \dot{b}_{\ag} - \del_{\Og} H_{eff} \rh \rd }\\
 & & \\
 & & \dsp{ \hs{2.5} +\, \frac{d}{dt} (p_i \del_{\Og} q^i - 
 i b_{\ag} \del_{\Og} c^{\ag}) ] }\\
 & & \\\
 & = & \dsp{ \int dt \frac{d}{dt} \lh p_i \del_{\Og} q^i - i b_{\ag} 
 \del_{\Og} c^{\ag}  - \Og \rh. }
\ea
\label{2a.1.5}
\ee 
To obtain the last equality we have used eqs.(\ref{2.2.12}) and 
(\ref{2.2.13}), which can be summarized 
\[
\ba{l}
\dsp{ \del_{\Og} q^i = \dd{\Og}{p_i}, \hs{2} 
\del_{\Og} p_i = - \dd{\Og}{q^i}, }\\
 \\
\dsp{ \del_{\Og} c^{\ag} = i \dd{\Og}{b_{\ag}}, \hs{2} 
\del_{\Og} b_{\ag} = i \dd{\Og}{c^{\ag}}. }
\ea 
\]
The action is therefore invariant up to a total time-derivative, and 
by comparison with eq.(\ref{1.4a.9}) we conclude, that $\Og$ is the 
conserved Noether charge. 
 
\section{Examples \label{s2a.2}} 

1.\ {\em The relativistic particle.} 
A simple example of the procedure presented above is the relativistic
particle \ct{hf}. The canonical hamiltonian $H_0$ is constrained to vanish 
itself. As a result, the effective hamiltonian is a pure BRST term:
\be 
H_{eff} = i \left\{ \Og, \Psi \right\}.
\label{2a.2.1}
\ee 
A simple choice for the gauge fermion is $\Psi = b$, which has the 
correct ghost number $N_g = -1$. With this choice, and the BRST 
generator  $\Og$ of eq.(\ref{2.3.2}), the effective hamiltonian is 
\be 
H_{eff} = i \left\{ \frac{c}{2m}\, (p^2 + m^2), b \right\}
 = \frac{1}{2m}\, \lh p^2 + m^2 \rh. 
\label{2a.2.1a}
\ee 
Then the effective action becomes 
\be 
S_{eff} = \int d\tau \lh p \cdot \dot{x} + i b \dot{c} - 
 \frac{1}{2m}\, (p^2 + m^2) \rh. 
\label{2a.2.2}
\ee 
This action is invariant under the BRST transformations (\ref{2.3.3}) :
\[ 
\ba{ll}
\dsp{ \del_{\Og} x^{\mu} = \left\{ x^{\mu}, \Og \right\} 
 = \frac{cp^{\mu}}{m}, } & 
\dsp{ \del_{\Og} p_{\mu} = \left\{ p_{\mu}, \Og \right\} = 0,}\\
 & \\
\dsp{ \del_{\Og} c = -\left\{ c, \Og \right\} = 0, }& 
\dsp{ \del_{\Og} b = -\left\{ b, \Og \right\} 
 = \frac{i}{2m}\, (p^2 +  m^2), } 
\ea 
\] 
up to a total proper-time derivative:
\be 
\del_{\Og} S_{eff} = \int d\tau \frac{d}{d\tau} \left[ c 
 \lh \frac{p^2 - m^2}{2m} \rh \right]. 
\label{2a.2.3}
\ee 
Implementing the Noether construction, the conserved charge 
resulting from the BRST transformations is 
\be 
\Og = p \cdot \del_{\Og} x  + i b\, \del_{\Og} c - \frac{c}{2m}\, 
 (p^2 - m^2) = \frac{c}{2m}\, (p^2 + m^2).
\label{2a.2.4}
\ee 
Thus we have reobtained the BRST charge from the action (\ref{2a.2.2}) 
and the transformations (\ref{2.3.3}), confirming that together with 
the BRST-cohomology principle, they correctly describe the dynamics of 
the relativistic particle. 

From the hamiltonian formulation (\ref{2a.2.2}) it is straightforward 
to construct a lagrangean one by using the hamilton equation $p^{\mu} 
= m \dot{x}^{\mu}$ to eliminate the momenta as independent variables;
the result is 
\be 
S_{eff} \simeq \int d\tau \lh \frac{m}{2}\, (\dot{x}^2  - 1) +
 i b \dot{c} \rh. 
\label{2a.2.4.1}
\ee 
\vs{1} 

\nit 
2.\ {\em Maxwell-Yang-Mills theory.} 
The BRST generator of the Maxwell-Yang-Mills theory in the temporal 
gauge has been given in (\ref{2.3.6}):
\[ 
\Og = \int d^3x\, \lh c^a G_a - \frac{ig}{2}\, f_{ab}^{\;\;\;c}
 c^a c^b b_c \rh,
\]
with $G_a = (\vec{D} \cdot \vec{E})_a$. The BRST-invariant 
effective hamiltonian takes the form 
\be 
H_{eff} = \frac{1}{2} \lh \vec{E}_a^2 + \vec{B}_a^2 \rh + 
 i \left\{ \Og, \Psi \right\}.
\label{2a.2.5}
\ee 
A simple choice of the gauge fermion: $\Psi = \lb^a b_a$, with 
$\lb_a$ some constants, then gives a effective action
\be 
S_{eff} = \int d^4x\, \left[ - \vec{E} \cdot \dd{\vec{A}}{t}
 + i b_a \dot{c}^a - \frac{1}{2} \lh \vec{E}_a^2 + \vec{B}_a^2 \rh 
 - \lb^a (\vec{D} \cdot \vec{E})_a + ig \lb^a f_{ab}^{\;\;\;c} c^b b_c
 \right].
\label{2a.2.6}
\ee 
The choice $\lb^a = 0$ would in effect turn the ghosts into free fields. 
However, if we eliminate the electric fields $\vec{E}^a$ as independend
degrees of freedom by the substitution $E_i^a = F_{i0}^a = \der_i A_0^a 
- \der_0 A_i^a - g f_{bc}^{\;\;\;a} A_i^b A_0^c$, and recalling the 
classical hamiltonian (\ref{1.8.7}), we observe that we might actually 
interpret $\lb^a$ as a constant scalar potential $A_0^a = \lb^a$, in a 
BRST-extended relativistic action 
\be 
S_{eff} = \int d^4x\, \left[ -\frac{1}{4}\, (F_{\mu\nu}^a)^2 
 + i b_a (D_0 c)^a \right]_{A_0^a = \lb^a},
\label{2a.2.7}
\ee 
where $(D_0 c)^a = \der_0 c^a - g f_{bc}^{\;\;\;a} A_0^b c^c$. 
The action is invariant under the classical BRST transformations 
(\ref{2.3.7}):
\[ 
\ba{ll}
\del_{\Og} \vec{A}^a = (\vec{D} c)^a,
 & \del_{\Og} \vec{E}_a = g f_{ab}^{\;\;\;c} c^b \vec{E}_c, \\
 & \\
\dsp{ \del_{\Og} c^a = \frac{g}{2}\, f_{bc}^{\;\;\;a}\, c^b c^c, }& 
\del_{\Og} b_a = i\, G_a + g f_{ab}^{\;\;\;c}\, c^b\, b_c, 
\ea 
\]
with the above BRST generator (\ref{2.3.6}) as the conserved Noether 
charge. All of the above applies to Maxwell electrodynamics as well,
except that in an abelian theory there is only a single vector
field, and all structure constants vanish: $f_{ab}^{\;\;\;c} = 0$.

\section{Lagrangean BRST formalism \label{s2a.3}}

From the hamiltonian formulation of BRST-invariant dynamical 
systems it is straightforward to develop an equivalent lagrangean 
formalism, by eliminating the momenta $p_i$ as independent degrees 
of freedom. This proceeds as usual by solving Hamilton's equation 
\[ 
\dot{q}^i = \dd{H}{p_i},
\] 
for the momenta in terms of the velocities, and performing the inverse 
Legendre transformation. We have already seen how this works for the 
examples of the relativistic particle and the Maxwell-Yang-Mills theory. 
As the lagrangean is a scalar function under space-time transformations, 
it is better suited for the development of a manifestly covariant 
formulation of gauge-fixed BRST-extended dynamics of theories with 
local symmetries, including Maxwell-Yang-Mills theory and the 
relativistic particle as well as string theory and general relativity. 

The procedure follows quite naturally the steps outlined in the 
previous sections (\ref{s2a.1} and \ref{s2a.2}): \nl 
a.\ Start from a gauge-invariant lagrangean $L_0(q,\dot{q})$. \nl 
b.\ For each gauge degree of freedom (each gauge parameter), introduce 
a ghost variable $c^a$; by definition these ghost variables carry 
ghost number $N_g[c^a]= +1$. Construct BRST transformations 
$\del_{\Og} X$ for the extended configuration-space variables 
$X = (q^i, c^a)$, satisfying the requirement that they leave $L_0$ 
invariant (possibly modulo a total derivative), and are nilpotent: 
$\del_{\Og}^2 X = 0$. \nl 
c.\ Add a trivially BRST-invariant set of terms to the action, of 
the form $\del_{\Og} \Psi$ for some anti-commuting function $\Psi$ 
(the gauge fermion). \nl 
The last step is to result in an effective lagrangean $L_{eff}$ 
with net ghost number $N_g[L_{eff}] = 0$. To achieve this, the 
gauge fermion  must have ghost number $N_g[\Psi] = -1$. However, 
so far we only have introduced dynamical variables with non-negative 
ghost number: $N_g[q^i,c^a] = (0, +1)$. To solve this problem we 
introduce anti-commuting anti-ghosts $b_a$, with ghost number 
$N_g[b_a] = -1$. The BRST-transforms of these variables must then 
be commuting objects $\ag_a$, with ghost number $N_g[\ag] = 0$. In 
order for the BRST-transformations to be nilpotent, we require
\be 
\del_{\Og}\, b_a = i \ag_a, \hs{2} \del_{\Og}\, \ag_a = 0,
\label{2a.3.1}
\ee 
which indeed trivially satisfy $\del_{\Og}^2 = 0$. The examples of 
the previous section illustrate this procedure. 
\vs{1} 

\nit
1.\ {\em Relativistic particle.}  
The starting point for the description of the relativistic particle 
was the reparametrization-invariant action (\ref{1.2.8}). We identify
the integrand as the lagrangean $L_0$. Next we introduce the 
Grassmann-odd ghost variable $c(\lb)$, and define the BRST 
transformations 
\be 
\del_{\Og}\, x^{\mu} = c\, \frac{dx^{\mu}}{d\lb}, \hs{2} 
\del_{\Og}\, e = \frac{d(ce)}{d\lb}, \hs{2}
\del_{\Og}\, c = c\, \frac{dc}{d\lb}.
\label{2a.3.2}
\ee 
As $c^2 = 0$, these transformations are nilpotent indeed. In addition, 
introduce the anti-ghost representation $(b, \ag)$ with the
transformation rules (\ref{2a.3.1}). We can now construct a gauge
fermion. We make the choice 
\be 
\Psi(b,e) = b(e - 1) \hs{1} \Rightarrow \hs{1} 
\del_{\Og}\, \Psi = i \ag (e - 1) - b\, \frac{d(ce)}{d\lb}.
\label{2a.3.3}
\ee 
As a result, the effective lagrangean (in natural units) becomes 
\be 
L_{eff} = L_0 - i \del_{\Og} \Psi = \frac{m}{2e}\, \frac{dx_{\mu}}{d\lb} 
 \frac{dx^{\mu}}{d\lb} - \frac{e m}{2} + \ag (e - 1) + i b\, 
 \frac{d(ce)}{d\lb} .
\label{2a.3.4}
\ee 
Observing that the variable $\ag$ plays the role of a lagrange multiplier,
fixing the einbein to its canonical value $e = 1$ such that $d\lb = 
d\tau$, this lagrangean is seen to reproduce the action (\ref{2a.2.4.1}): 
\[ 
S_{eff} = \int d\tau L_{eff} \simeq \int d\tau \lh \frac{m}{2}\, 
 (\dot{x}^2  - 1) + i b\, \dot{c} \rh. 
\]

\nit
2.\ {\em Maxwell-Yang-Mills theory.}
The covariant classical action of the Maxwell-Yang-Mills theory was 
presented in eq.(\ref{1.8.1}): 
\[
S_0 = - \frac{1}{4}\, \int d^4x\, \lh F_{\mu\nu}^a \rh^2.
\]
Introducing the ghost fields $c^a$, we can define nilpotent BRST 
transformations 
\be 
\del_{\Og}\, A_{\mu}^a = \lh D_{\mu} c \rh^a, \hs{2} 
\del_{\Og}\, c^a = \frac{g}{2}\, f_{bc}^{\;\;\;a} c^b c^c.
\label{2a.3.5}
\ee 
Next we add the anti-ghost BRST multiplets $(b_a, \ag_a)$, with the 
transformation rules (\ref{2a.3.1}). Choose the gauge fermion
\be 
\Psi(A_0^a,b_a) = b_a (A_0^a - \lb^a) \hs{1} \Rightarrow \hs{1} 
\del_{\Og}\, \Psi = i \ag_a (A_0^a - \lb^a) - b_a (D_0 c)^a,
\label{2a.3.6}
\ee 
where $\lb^a$ are some constants (possibly zero). Adding this to the
classical action gives 
\be 
S_{eff} = \int d^4x\, \left[ - \frac{1}{4}\, (F_{\mu\nu}^a)^2 + 
 \ag_a (A_0^a - \lb^a) + i b_a (D_0 c)^a \right].
\label{2a.3.7}
\ee 
Again, the fields $\ag_a$ act as lagrange multipliers, fixing 
the electric potentials to the constant values $\lb^a$.  
After substitution of these values, the action reduces to 
the form (\ref{2a.2.7}). 
\vs{1} 

\nit 
We have thus demonstrated that the lagrangean and canonical procedures
lead to equivalent results; however, we stress that in both cases the
procedure involves the choice of a gauge fermion $\Psi$, restricted by 
the requirement that it has ghost number $N_g[\Psi] = -1$. 

The advantage of the lagrangean formalism is, that it is easier to 
formulate the theory with different choices of the gauge fermion. In 
particular, it is possible to make choices of gauge which manifestly 
respect the Lorentz-invariance of Minkoswki space. This is not an 
issue for the study of the relativistic particle, but it is an issue 
in the case of Maxwell-Yang-Mills theory, which we have constructed 
so far only in the temporal gauge $A_0^a =$ constant. 

We now show how to construct a covariant gauge-fixed and BRST-invariant 
effective lagrangean for Maxwell-Yang-Mills theory, using the same 
procedure. In stead of (\ref{2a.3.6}), we choose the gauge fermion 
\be 
\Psi = b_a \lh \der \cdot A^a - \frac{\lb}{2}\, \ag^a \rh \hs{1}
 \Rightarrow \hs{1} \del_{\Og} \Psi = i \ag_a\, \der \cdot A^a 
 - \frac{i \lb}{2}\, \ag_a^2 - b_a\, \der \cdot (D c)^a.
\label{2a.3.8}
\ee 
Here the parameter $\lb$ is a arbitrary real number, which can be
used to obtain a convenient form of the propagator in perturbation
theory. The effective action obtained with this choice of gauge-fixing 
fermion is, after a partial integration:
\be 
S_{eff} = \int d^4x\, \left[ - \frac{1}{4}\, (F_{\mu\nu}^a)^2 + 
 \ag_a\, \der \cdot A^a - \frac{\lb}{2}\, \ag^2_a - i \der b_a 
 \cdot (D c)^a \right].
\label{2a.3.9}
\ee
As we have introduced quadratic terms in the bosonic variables
$\ag_a$, they now behave more like auxiliary fields, rather than
lagrange multipliers. Their variational equations lead to the 
result 
\be 
\ag^a = \frac{1}{\lb}\, \der \cdot A^a.
\label{2a.3.10}
\ee 
Eliminating the auxiliary fields by this equation, the effective action 
becomes 
\be
S_{eff} = \int d^4x\, \left[ - \frac{1}{4}\, (F_{\mu\nu}^a)^2 + 
 \frac{1}{2\lb}\, (\der \cdot A^a)^2 - i \der b_a \cdot (D c)^a \right].
\label{2a.3.11}
\ee
This is the standard form of the Yang-Mills action used in covariant 
perturbation theory. 
Observe, that the elimination of the auxiliary field $\ag_a$ also
changes the BRST-transformation of the anti-ghost $b_a$ to:
\be 
\del_{\Og}\, b^a = \frac{i}{\lb}\, \der \cdot A^a 
 \hs{1} \Rightarrow \hs{1}  
\del_{\Og}^2\, b^a = \frac{i}{\lb}\, \der \cdot (Dc)^a \simeq 0.
\label{2a.3.12}
\ee 
The transformation is now nilpotent only after using the ghost 
field equation. 

The BRST-Noether charge can be computed from the action (\ref{2a.3.11})
by the standard procedure, and leads to the expression 
\be 
\Og = \int d^3x\, \lh \pi_a^{\mu} (D_{\mu} c)^a - \frac{ig}{2}\, 
 f_{ab}^{\;\;\;c} c^a c^b \gam_c \rh, 
\label{2a.3.13}
\ee 
where $\pi_a^{\mu}$ is the canonical momentum of the vector potential 
$A_{\mu}^a$, and $(\bg^a, \gam_a)$ denote the canonical momenta of
the ghost fields $(b_a, c^a)$:
\be 
\ba{ll}
\dsp{ \pi_a^i = \dd{\cL_{eff}}{\dot{A}^a_i}\, = - F_a^{0i} = - E_a^i, }& 
 \dsp{ \pi_a^0 = \dd{\cL_{eff}}{\dot{A}^a_0}\, =
 - \frac{1}{\lb}\, \der \cdot A_{a}, }\\
 & \\
\dsp{ \bg^a = i \dd{\cL_{eff}}{\dot{b_a}}\, = - (D_0 c)^a, }& 
\dsp{ \gam_a = i \dd{\cL_{eff}}{\dot{c}^a}\, = \der_0 b_a. }
\ea 
\label{2a.3.14}
\ee 
Each ghost field $(b_a, c^a)$ now has its own conjugate momentum,
because the ghost terms in the action (\ref{2a.3.11}) are quadratic 
in derivatives, rather than linear as before. Note also, that 
a factor $i$ has been absorbed in the ghost momenta to make them 
real; this leads to the standard Poisson brackets 
\be 
\left\{ c^a(\vec{x};t), \gam_b(\vec{y};t) \right\} = 
 -i \del^a_b \del^3(\vec{x} - \vec{y}),  \hs{2} 
\left\{ b_a(\vec{x};t), \bg^b(\vec{y};t) \right\} = 
 -i \del_a^b \del^3(\vec{x} - \vec{y}).
\label{2a.3.15}
\ee 
As our calculation shows, all explicit dependence on $(b_a, \bg^a)$ has 
dropped out of the expression (\ref{2a.3.13}) for the BRST charge. 

The parameter $\lb$ is still a free parameter, and in actual 
calculations it is often useful to check partial gauge-independence 
of physical results, like cross sections, by establishing that 
they do not depend on this parameter. What needs to be shown more 
generally is, that physical results do not depend on the choice 
of gauge fermion. This follows formally from the BRST cohomology 
being independent of the choice of gauge fermion. Indeed, from the 
expression (\ref{2a.3.13}) for $\Og$ we observe that it is of the 
same form as the one we have used previously in the temporal gauge, 
even though now $\pi^0_a$ no longer vanishes identically. In the 
quantum theory this implies, that the BRST-cohomology classes at 
ghost number zero correspond to gauge-invariant states, in which
\be 
\lh \vec{D} \cdot \vec{E} \rh^a = 0,  \hs{2} 
\der \cdot A^a = 0. 
\label{2a.3.16}
\ee 
The second equation implies, that the time-evolution of the 0-component 
of the vector potential is fixed completely by the initial conditions
and the evolution of the spatial components $\vec{A}^a$. In particular,
$A_0^a = \lb^a =$ constant is a consistent solution if by a gauge 
transformation we take the spatial components to satisfy $\vec{\nabla}
\cdot \vec{A}^a = 0$. 

In actual computations, especially in perturbation theory, the 
matter is more subtle however: the theory needs to be renormalized,
and this implies that the action and BRST-transformation rules 
have to be adjusted to the introduction of counter terms. To prove
the gauge independence of the renormalized theory it must be shown,
that the renormalized action still possesses a BRST-invariance, 
and the cohomology classes at ghost-number zero satisfy the 
renormalized conditions (\ref{2a.3.16}). In four-dimensional 
space-time this can indeed be done for the pure Maxwell-Yang-Mills 
theory, as there exists a manifestly BRST-invariant regularization 
scheme (dimensional regularization) in which the theory defined by 
the action (\ref{2a.3.11}) is renormalizable by power counting. 
The result can be extended to gauge theories interacting with 
scalars and spin-1/2 fermions, except for the case in which 
the Yang-Mills fields interact with chiral fermions in anomalous 
representations of the gauge group.  

\section{The master equation \label{s2a.4}}

Consider a BRST-invariant action $S_{eff}[\Fg^A] = S_0 + \int dt\, 
(i \del_{\Og} \Psi)$, where the variables $\Fg^A = (q^i, c^a, b_a, \ag_a)$ 
parametrize the extended configuration space of the system, and 
$\Psi$ is the gauge fermion, which is Grassmann-odd and has ghost 
number $N_g[\Psi] = -1$. Now by construction 
\be 
\del_{\Og} \Psi = \del_{\Og} \Fg^A\, \dd{\Psi}{\Fg^A}, 
\label{2a.4.1}
\ee 
and therefore we can write the effective action also as 
\be 
S_{eff}[\Fg^A] = S_0 + i \int dt\, \left[ \del_{\Og} \Fg^A\, \Fg^*_A 
 \right]_{\Fg^*_A = \dd{\Psi}{\Fg^A}}.
\label{2a.4.2}
\ee 
This way of writing considers the action as a functional on a doubled 
configuration space, parametrized by variables $(\Fg^A, \Fg^*_A)$,
the first set $\Fg^A$ being called the {\em fields}, and the second set 
$\Fg^*_A$ called the {\em anti-fields}. In the generalized action 
\be 
S^*[\Fg^A, \Fg^*_A] = S_0 + i \int dt\, \del_{\Og} \Fg^A\, \Fg^*_A, 
\label{2a.4.3}
\ee 
the anti-fields play the role of sources for the BRST-variations of the
fields $\Fg^A$; the effective action $S_{eff}$ is the restriction to the 
hypersurface $\Sg[\Psi]:\, \Fg^*_A = \der \Psi/\der \Fg^A$. We observe, 
that by construction the antifields have Grassmann parity opposite to 
that of the corresponding fields, and ghost number $N_g[\Fg^*_A] = 
- (N_g[\Fg^A] + 1)$. 

In the doubled configuration space the BRST variations of the fields 
can be written as 
\be 
i \del_{\Og} \Fg^A = (-1)^{A} \frac{\del S^*}{\del \Fg^*_A},
\label{2a.4.4}
\ee 
where $(-1)^A$ is the Grassmann parity of the field $\Fg^A$, whilst 
$-(-1)^A = (-1)^{A+1}$ is the Grassmann parity of the anti-field 
$\Fg^*_A$. We now define the {\em anti-bracket} of two functionals
$F(\Fg^A, \Fg_A^*)$ and $G(\Fg^A, \Fg_A^*)$ on the large 
configuration space by 
\be 
(F, G) = (-1)^{F + G + FG}\, (G, F) = 
 (-1)^{A(F+1)}\, \lh \frac{\del F}{\del \Fg^A} 
 \frac{\del G}{\del \Fg_A^*} + (-1)^F\, \frac{\del F}{\del \Fg^*_A} 
 \frac{\del G}{\del \Fg^A} \rh. 
\label{2a.4.5}
\ee 
These brackets are symmetric in $F$ and $G$ if both are Grassmann-even 
(bosonic), and anti-symmetric in all other cases. Sometimes one 
introduces the notion of {\em right derivative}:
\be 
\frac{F\stackrel{\leftarrow}{\del}}{\del \Fg^A} \equiv 
 (-1)^{A(F + 1)}\, \frac{\del F}{\del \Fg^A}. 
\label{2a.4.6}
\ee 
Then the anti-brackets take the simple form 
\be 
(F, G) = \frac{F\stackrel{\leftarrow}{\del}}{\del \Fg^A}\, 
 \frac{\stackrel{\rightarrow}{\del} G}{\del \Fg_A^*} - 
 \frac{F\stackrel{\leftarrow}{\del}}{\del 
 \Fg^*_A}\, \frac{\stackrel{\rightarrow}{\del} G}{\del \Fg^A},
\label{2a.4.7}
\ee 
where the derivatives with a right arrow denote the standard {\em left}
derivatives. In terms of the anti-brackets, the BRST transformations 
(\ref{2a.4.4}) can be written in the form 
\be 
i \del_{\Og}\, \Fg^A = (S^*, \Fg^A). 
\label{2a.4.8}
\ee 
In analogy, we can define 
\be 
i \del_{\Og}\, \Fg^*_A = (S^*, \Fg^*_A) = (-1)^A \frac{\del S^*}{\del 
 \Fg^A}. 
\label{2a.4.8.1}
\ee  
Then the BRST transformation of any functional $Y(\Fg^A, \Fg^*_A)$
is given by 
\be 
i \del_{\Og} Y = (S^*, Y). 
\label{2a.4.8.2}
\ee 
In particular, the BRST-invariance of the action $S^*$ can be expressed 
as
\be 
(S^*, S^*) = 0. 
\label{2a.4.9}
\ee  
This equation is known as the {\em master equation}. The formalism presented
here was initiated in the work by Zinn-Justin \ct{zj} and Batalin and
Vilkovisky \ct{bv}.

Next we observe, that on the physical hypersurface $\Sg[\Psi]$ the BRST 
transformations of the antifields are given by the classical field equations; 
indeed, introducing an anti-commuting parameter $\mu$ for infinitesimal 
BRST transformations
\be 
i \mu\, \del_{\Og} \Fg^*_A = \frac{\del S^*}{\del \Fg^A}\, \mu \hs{1}
 \stackrel{\Sg[\Psi]}{\longrightarrow}\hs{1} 
 \frac{\del S_{eff}}{\del \Fg^A}\, \mu \simeq 0, 
\label{2a.4.10}
\ee 
where the last equality holds only for solutions of the classical 
field equations. Because of this result, it is customary to redefine 
the BRST transformations of the antifields such that they vanish:
\be 
\del_{\Og} \Fg^*_A = 0,
\label{2a.4.11}
\ee
instead of (\ref{2a.4.8.1}). As the BRST transformations are nilpotent, 
this is consistent with the identification $\Fg^*_A = 
\der \Psi/\der \Fg^A$ in the action; indeed, it now follows that 
\be 
\del_{\Og} \lh \del_{\Og} \Fg^A\, \Fg^*_A \rh = 0,
\label{2a.4.12}
\ee 
which holds before the identification as a result of (\ref{2a.4.11}), 
and after the identification because it reduces to $\del_{\Og}^2 \Psi 
= 0$. Note, that the condition for BRST invariance of the action now 
becomes
\be 
i \del_{\Og} S^* = \frac{1}{2}\, (S^*, S^*) = 0,
\label{2a.4.13}
\ee
which still implies the master equation (\ref{2a.4.9}).

\section{Path-integral quantization \label{s2a.5}}

The construction of BRST-invariant actions $S_{eff} = S^*[\Fg_A^*
 = \der \Psi/ \der \Fg^A]$ and the anti-bracket formalism is 
especially useful in the context of path-integral quantization. 
The path integral provides a representation of the matrix elements 
of the evolution operator in the configuration space:
\be 
\langle q_f, T/2 | e^{-i T H} | q_i, -T/2 \rangle = 
 \int_{q_i}^{q_f} Dq(t)\, e^{i \int_{-T/2}^{T/2} L(q,\dot{q}) dt}.  
\label{2a.5.1}
\ee 
In field theory one usually considers the vacuum-to-vacuum amplitude 
in the presence of sources, which is a generating functional for
time-ordered vacuum Green's functions: 
\be 
Z[J] = \int D\Fg\, e^{iS[\Fg] + i \int J \Fg},
\label{2a.5.2}
\ee 
such that 
\be 
\langle 0| T(\Fg_1 ... \Fg_k) | 0 \rangle = 
 \left. \frac{\del^k Z[J]}{\del J_1 ... \del J_k} \right|_{J=0}. 
\label{2a.5.3}
\ee 
The corresponding generating functional $W[J]$ for the connected Green's 
functions is related to $Z[J]$ by 
\be 
Z[J] = e^{i\, W[J]}. 
\label{2a.5.4}
\ee 
For theories with gauge invariances, the evolution operator is 
constructed from the BRST-invariant hamiltonian; then the action to 
be used is the in the path integral (\ref{2a.5.2}) is the BRST invariant 
action:
\be 
Z[J] = e^{i\, W[J]} = \int D\Fg^A\, \left. 
 e^{i\, S^*[\Fg^A, \Fg_A^*] + i \int J_A \Fg^A} \right|_{\Fg^*_A 
 = \der \Psi/\der \Fg^A}, 
\label{2a.5.5}
\ee 
where the sources $J_A$ for the fields are supposed to be BRST invariant
themselves. For the complete generating functional to be BRST invariant, 
it is not sufficient that only the action $S^*$ is BRST invariant, as
guaranteed by the master equation (\ref{2a.4.9}): the functional 
integration measure must be BRST invariant as well. Under an infinitesimal 
BRST transformation $\mu \del_{\Og} \Fg^A$ the measure changes by 
a graded jacobian (superdeterminant) \ct{ber,dwitt} 
\be 
{\cal J} = \mbox{SDet} \lh \del^A_B + \mu (-1)^B\, 
 \frac{\del (\del_{\Og} \Fg^A)}{\del \Fg^B} \rh \approx 1 + \mu\, 
 \mbox{Tr} \frac{\del (\del_{\Og} \Fg^A)}{\del \Fg^B}.
\label{2a.5.6}
\ee 
We now define 
\be 
\frac{\del (i \del_{\Og} \Fg^A)}{\del \Fg^A} = 
 (-1)^A\, \frac{\del^2 S^*}{\del \Fg^A \del \Fg_A^*}
 \equiv \bar{\Del} S^*. 
\label{2a.5.7}
\ee 
The operator $\bar{\Del}$ defined by  
\be
\bar{\Del}  = (-1)^{A}\, \frac{\del^2}{\del \Fg^A \del \Fg_A^*}. 
\label{2a.5.7.1}
\ee 
is a laplacian on the field/anti-field configuration space, with the
property $\bar{\Del}^2 = 0$. 
The condition of invariance of the measure requires the BRST jacobian 
(\ref{2a.5.6}) to be unity:
\be 
{\cal J} = 1 - i \mu\, \bar{\Del} S^* = 1,
\label{2a.5.7.2}
\ee 
which reduces to the vanishing of the laplacian of $S^*$: 
\be 
\bar{\Del} S^* = 0. 
\label{2a.5.8}
\ee 
The two conditions (\ref{2a.4.9}) and (\ref{2a.5.8}) imply the 
BRST invariance of the path integral (\ref{2a.5.5}). Actually, a
somewhat more general situation is possible, in which neither the 
action nor the functional measure are invariant independently, only 
the combined functional integral. Let the action generating 
the BRST transformations be denoted by $W^*[\Fg^A, \Fg^*_A]$:
\be 
i \del_{\Og} \Fg^A = (W^*, \Fg^A), \hs{2} 
i \del_{\Og} \Fg^*_A = 0. 
\label{2a.5.9}
\ee 
As a result the graded jacobian for a transformation with parameter 
$\mu$ is 
\be 
\mbox{SDet} \lh \del^A_B + \mu (-1)^B\, \frac{\del(\del_{\Og} \Fg^A)}
 {\del \Fg^B} \rh \approx 1 - i \mu\, \bar{\Del} W^*.
\label{2a.5.10}
\ee 
Then the functional $W^*$ itself needs to satisfy the generalized 
master equation 
\be 
\frac{1}{2}\, (W^*, W^*) = i \bar{\Del} W^*,
\label{2a.5.11}
\ee 
for the path-integral to be BRST invariant. This equation can be 
neatly summarized in the form
\be 
\bar{\Del}\, e^{iW^*} = 0. 
\label{2a.5.12}
\ee 
Solutions of this equation restricted to the hypersurface $\Fg^*_A 
= \der \Psi/\der \Fg^A$ are acceptable actions for the construction 
of BRST-invariant path integrals. 

A geometrical interpretation of the field/anti-field construction 
and the master equation has been discussed in refs.\ct{witten2,schwarz,aoy}.

%% file: brs3.tex
\chapter{Applications of BRST methods \label{c3}}

In the final chapter of these lecture notes, we turn to some application
of BRST-methods other than the perturbative quantization of gauge 
theories. We deal with two topics; the first is the construction of
BRST field theories, presented in the context of the scalar point 
particle. This is the simplest case \ct{huef,jwvh3}; for more complicated 
ones, like the superparticle \ct{siegel2,aklvh} or the string 
\ct{siegel2,witten,siegel}, we refer to the literature. 

The second application concerns the classification of anomalies in 
gauge theories of the Yang-Mills type. Much progress has been made
in this field in recent years \ct{henn3}, of which a summary is 
presented here. 

\section{BRST Field theory \label{s3.1}} 

The examples of the relativistic particle and string show, that
in theories with local reparametrization invariance the hamiltonian
is one of the generators of gauge symmetries, and as such is 
constrained to vanish. The same phenomenon also occurs in 
general relativity, leading to the well-known Wheeler-deWitt
equation. In such case the {\em full} dynamics of the system is
actually contained in the BRST cohomology. This opens up the 
possibility for constructing quantum field theories for particles
\ct{siegel,huef,jwvh3}, or strings \ct{siegel,siegel2,witten}, in 
a BRST formulation, in which the usual BRST operator becomes the 
kinetic operator for the fields. This formulation has some formal 
similarities with the Dirac equation for spin-1/2 fields. 

As our starting point we consider the BRST-operator for the 
relativistic quantum scalar particle, which for free particles,
after some rescaling, reads 
\be 
\Og = c (p^2 + m^2), \hs{3} \Og^2 = 0. 
\label{3.1.1}
\ee 
It acts on fields $\Psi(x,c) = \psi_0(x) + c \psi_1(x)$, with the 
result 
\be 
\Og \Psi(x,c) = c (p^2 + m^2)\, \psi_0(x).
\label{3.1.2}
\ee 
As in the case of Lie-algebra cohomology (\ref{2.7.10}), we introduce 
the non-degenerate (positive definite) inner product
\be 
(\Fg, \Psi) = \int d^dx\, \lh \fg^*_0 \psi_0 + \fg_1^* \psi_1 \rh.
\label{3.1.3}
\ee 
With respect to this inner product the ghosts $(b,c)$ are 
mutually adjoint: 
\be 
(\Fg, c \Psi) = (b \Fg, \Psi) \hs{1} \leftrightarrow \hs{1} 
 b = c^{\dagger}.
\label{3.1.4}
\ee 
Then the BRST operator $\Og$ is not self-adjoint, but rather  
\be 
\Og^{\dagger} = b (p^2 + m^2), \hs{3} \Og^{\dagger\, 2} = 0.
\label{3.1.5}
\ee 
Quite generally, we can construct actions for quantum scalar fields 
coupled to external sources $J$ of the form
\be 
S_G[J] = \frac{1}{2}\, \lh \Psi, G\, \Og \Psi \rh - \lh \Psi, J \rh, 
\label{3.1.6}
\ee 
where the operator $G$ is chosen such that 
\be
G \Og = (G \Og)^{\dagger} = \Og^{\dagger} G^{\dagger}.
\label{3.1.7}
\ee 
This guarantees that the action is real. From the action we then derive 
the field equation 
\be 
G \Og\, \Psi = \Og^{\dagger} G^{\dagger}\, \Psi = J. 
\label{3.1.8}
\ee 
Its consistency requires the co-BRST invariance of the source:
\be 
\Og^{\dagger} J = 0. 
\label{3.1.9}
\ee 
This reflects the invariance of the action and the field equation 
under BRST transformations 
\be 
\Psi \rightarrow \Psi^{\prime} = \Psi + \Og \chi.
\label{3.1.10}
\ee 
In order to solve the field equation we therefore have to impose 
a gauge condition, selecting a particular element of the equivalence 
class of solutions (\ref{3.1.10}). 

A particularly convenient condition is 
\be 
\Og G^{\dagger}\, \Psi = 0. 
\label{3.1.11}
\ee 
In this gauge, the field equation can be rewritten in the form 
\be 
\Del G^{\dagger}\, \Psi = \lh \Og^{\dagger} \Og + 
 \Og \Og^{\dagger} \rh G^{\dagger} \Psi = \Og\, J. 
\label{3.1.12}
\ee 
Here $\Del$ is the BRST laplacean, which can be inverted using 
a standard analytic continuation in the complex plane, to give 
\be 
G^{\dagger}\, \Psi = \frac{1}{\Del}\, \Og\, J. 
\label{3.1.13}
\ee 
We interpret the operator $\Del^{-1} \Og$ on the right-hand side 
as the (tree-level) propagator of the field.

We now implement the general scheme (\ref{3.1.6})-(\ref{3.1.13}) 
by choosing the inner product (\ref{3.1.3}), and $G = b$. 
Then
\be 
G \Og = bc (p^2 + m^2) = \Og^{\dagger} G^{\dagger}, 
\label{3.1.14}
\ee 
and therefore 
\be
\frac{1}{2}\, ( \Psi, G\, \Og \Psi) = \frac{1}{2}\,
 \int d^dx\, \psi_0^* (p^2 + m^2) \psi_0,
\label{3.1.15}
\ee 
which is the standard action for a free scalar field\footnote{Of course,
there is no loss of generality here if we restrict the coefficients 
$\psi_a$ to be real.}. 

The laplacean for the BRST operators (\ref{3.1.2}), (\ref{3.1.5}) 
is 
\be 
\Del = \Og\, \Og^{\dagger} + \Og^{\dagger} \Og = 
 (p^2 + m^2)^2,
\label{3.1.16}
\ee 
which is manifestly non-negative, but might give rise to propagators 
with double poles, or negative residues, indicating the appearance 
of ghost states. However, in the expression (\ref{3.1.13}) for the 
propagator, one of the poles is canceled by the zero of the 
BRST operator; in the present context the equation reads 
\be 
c \psi_0 = \frac{1}{(p^2 + m^2)^2}\, c (p^2 + m^2)\, J_0.
\label{3.1.17}
\ee
This leads to the desired result 
\be 
\psi_0 = \frac{1}{p^2 + m^2}\, J_0, 
\label{3.1.18}
\ee 
and we recover the standard scalar field theory indeed. It is not 
very difficult to extend the theory to particles in external 
gravitational or electromagnetic fields\footnote{See the discussion
in \ct{jwvh3}, which uses however a less elegant implementation 
of the action.}, or to spinning particles \ct{huef,jwvh4}.  

However, a different and more difficult problem is the inclusion
of self interactions \ct{huef}. This question has been addressed mostly 
in the context of string theory \ct{siegel,witten}. As it is expected to 
depend on spin, no unique prescription has been constructed for point 
particles to date. 

\section{Anomalies and BRST cohomology \label{s3.2}}

In the preceding chapters we have seen how local gauge symmetries are 
encoded in the BRST-transformations. First, the BRST-transformations 
of the classical variables correspond to ghost-dependent gauge 
transformations. Second, the closure of the algebra of the gauge 
transformations (and the Poisson brackets or commutators of the 
constraints), as well as the corresponding Jacobi-identities, are 
part of the condition that the BRST transformations are nilpotent. 

It is important to stress, as we observed earlier, that the closure 
of the classical gauge algebra does not necessarily guarantee the 
closure of the gauge algebra in the quantum theory, because it may 
be spoiled by anomalies. Equivalently, in the presence of anomalies 
there is no nilpotent quantum BRST operator, and no local action 
satisfying the master equation (\ref{2a.5.12}). A particular case 
in point is that of a Yang-Mills field coupled to chiral fermions, 
as in the electro-weak standard model. In the following we consider 
chiral gauge theories in some detail. 

The action of chiral fermions coupled to an abelian or non-abelian 
gauge field reads
\be 
S_F[A] = \int d^4x\, \bar{\psi}_L \sDer \psi_L.
\label{3.2.1}
\ee 
Here $D_{\mu} \psi_L = \der_{\mu} \psi_L -  g A_{\mu}^a T_a \psi_L$, 
with $T_a$ the generators of the gauge group in the representation 
according to which the spinors $\psi_L$ transform. In the 
path-integral formulation of quantum field theory the fermions 
make the following contribution to the effective action for the 
gauge fields:
\be 
e^{iW[A]} = \int D\bar{\psi}_L D\psi_L\, e^{iS_F[A]}.
\label{3.2.2}
\ee 
An infinitesimal local gauge transformation with parameter $\Lb^a$ 
changes the effective action $W[A]$ by 
\be 
\del(\Lb) W[A] = \int d^4x\, (D_{\mu} \Lb)^a\,
 \frac{\del W[A]}{\del A^a_{\mu}}\, = - \int d^4x\, \Lb^a 
 \lh \der_{\mu} \frac{\del}{\del A^a_{\mu}} - g f_{ab}^{\;\;c}
 A_{\mu}^b \frac{\del}{\del A_{\mu}^c} \rh W[A],
\label{3.2.3}
\ee 
assuming boundary terms to vanish. By construction, the fermion action 
$S_F[A]$ itself is gauge invariant, but this is generally not true for 
the fermionic functional integration measure. If the measure is not 
invariant: 
\be 
\ba{l}
\dsp{ \del(\Lb) W[A] = - \int d^4x\, \Lb^a \Gam_a[A] \neq 0, }\\
 \\ 
\dsp{ \Gam_a[A] = \cD_a W[A] \equiv \lh \der_{\mu} 
 \frac{\del}{\del A^a_{\mu}} - g f_{ab}^{\;\;c} A_{\mu}^b 
 \frac{\del}{\del A_{\mu}^c} \rh W[A]. }
\ea 
\label{3.2.4}
\ee 
Even though the action $W[A]$ may not be invariant, its variation 
should still be covariant and satisfy the condition
\be 
\cD_a \Gam_b[A] - \cD_b \Gam_a[A] = \left[ \cD_a, \cD_b \right] W[A]
 = g f_{ab}^{\;\;c} \cD_c W[A] = g f_{ab}^{\;\;c} \Gam_c[A].
\label{3.2.5}
\ee 
This consistency condition was first derived by Wess and Zumino 
\ct{wz}, and its solutions determine the functional form of the 
anomalous variation $\Gam_a[A]$ of the effective action $W[A]$. 
It can be derived from the BRST cohomology of the gauge theory 
\ct{zumino,zwz,henn3}.

To make the connection, observe that the Wess-Zumino consistency 
condition (\ref{3.2.5}) can be rewritten after contraction with 
ghosts as follows:
\be 
\ba{lll}
0 & = & \dsp{ \int d^4x\, c^a c^b \lh \cD_a \Gam_b[A] - \cD_b \Gam_a[A] - 
 g f_{ab}^{\;\;c} \Gam_c[A] \rh }\\ 
 & & \\
 & = & \dsp{ 2 \int d^4x\, c^a c^b\, \lh \cD_a \Gam_b - \frac{g}{2}\, 
 f_{ab}^{\;\;c}\, \Gam_c \rh = - 2\, \del_{\Og} \int d^4x c^a \Gam_a, }
\ea
\label{3.2.15}
\ee 
provided we can ignore boundary terms. The integrand is a 4-form of ghost 
number $+1$:
\be
I^1_4 = d^4x\, c^a \Gam_a[A] = \frac{1}{4!}\, \ve_{\mu\nu\kg\lb}\, 
 dx^{\mu} \wedge dx^{\nu} \wedge dx^{\kg} \wedge dx^{\lb}\, c^a \Gam_a[A].
\label{3.2.16}
\ee 
The Wess-Zumino consistency condition (\ref{3.2.15}) then implies 
that non-trivial solutions of this condition must be of the form 
\be 
\del_{\Og} I_4^1 = d I^2_3,
\label{3.2.17}
\ee 
where $I^2_3$ is a 3-form of ghost number $+2$, vanishing on any
boundary of the space-time $\cM$. 

Now we make a very interesting and useful observation: the BRST 
construction can be mapped to a standard cohomology problem on a 
principle fibre bundle with local structure $\cM \times G$, where 
$\cM$ is the space-time and $G$ is the gauge group viewed as a 
manifold \ct{bc-r}. 
First note, that the gauge field is a function of both the co-ordinates 
$x^{\mu}$ on the space-time manifold $\cM$, and of the parameters $\Lb^a$ 
on the group manifold $G$. We denote the combined set of these co-ordinates 
by $\xi = (x,\Lb)$. To make the dependence on space-time and gauge group 
explicit, we introduce the Lie-algebra valued 1-form 
\be 
A(x) = dx^{\mu} A_{\mu}^a(x) T_a,
\label{3.2.5.1}
\ee 
with $T_a$ a generator of the gauge group, and $A_{\mu}^a(x)$ the 
gauge field at the point $x$ in the space-time manifold $\cM$. Starting 
from $A$, all gauge-equivalent configurations are obtained by local gauge 
transformations, generated by group elements $a(\xi)$ according to
\be 
\cA(\xi) = - \frac{1}{g}\, a^{-1}(\xi)\, d a(\xi) 
 + a^{-1}(\xi)\, A(x)\, a(\xi), \hs{2} A(x) = \cA(x,0)
\label{3.2.6}
\ee 
where $d$ is the ordinary differential operator on the space-time $\cM$:
\be 
d a(x,\Lb) = dx^{\mu} \dd{a}{x^{\mu}}(x,\Lb).
\label{3.2.6.1}
\ee  
Furthermore, the parametrization of the group is chosen such that 
$a(x,0) = 1$, the identity element. Then, if $a(\xi)$ is close to 
the identity:
\be
a(\xi) = e^{\,g \Lb(x) \cdot T} \approx 1 + g\, \Lb^a(x) T_a + 
 \cO(g^2 \Lb^2), 
\label{3.2.7}
\ee 
and eq.(\ref{3.2.6}) represents the infinitesimally transformed 
gauge field 1-form (\ref{1.8.4}). In the following we interpret 
$\cA(\xi)$ as a particular 1-form living on the fibre bundle with 
local structure $\cM \times G$. 

A general one-form $\bN$ on the bundle can be decomposed as
\be 
\bN(\xi) = d \xi^i N_i = d x^{\mu} N_{\mu} + d\Lb^a N_a. 
\label{3.2.8}
\ee 
Correspondingly, we introduce the differential operators 
\be 
d = dx^{\mu} \dd{}{x^{\mu}}, \hs{2} 
s = d\Lb^a \dd{}{\Lb^a}, \hs{2} \bd = d + s,
\label{3.2.9}
\ee 
with the properties
\be 
d^2 = 0, \hs{2} s^2 = 0, \hs{2} 
\bd^2 = ds + sd = 0. 
\label{3.2.10}
\ee 
Next define the left-invariant 1-forms on the group $C(\xi)$ by 
\be 
C = a^{-1}\, s a, \hs{2} c(x) = C(x,0).
\label{3.2.11}
\ee 
By construction, using $sa^{-1} = - a^{-1} sa\, a^{-1}$, these 
forms satisfy
\be 
sC = - C^2 . 
\label{3.2.12}
\ee 
The action of of the group differential $s$ on the one-form $A$
is 
\be 
s \cA = \frac{1}{g}\, DC = \frac{1}{g} \lh dC- g [ \cA, C ]_+ \rh.
\label{3.2.13}
\ee 
Finally, the field strength $\cF(\xi)$ for the gauge field $\cA$ is 
defined as the 2-form 
\be 
\cF = d \cA - g \cA^2 = a^{-1} F\, a, 
 \hs{2} F(x) = \cF(x,0). 
\label{3.2.13.1}
\ee 
The action of $s$ on $\cF$ is given by 
\be 
s \cF = [ \cF, C ].
\label{3.2.13.2}
\ee 
Clearly, the above system of equations are in one-to-one correspondence 
with the BRST transformations of the Yang-Mills fields, described by 
the Lie-algebra valued one-form $A = dx^{\mu} A_{\mu}^a T_a$, and the 
ghosts described by the Lie-algebra valued grassmann variable $c = c^a 
T_a$, upon the identification $-gs|_{\Lb=0} \rightarrow \del_{\Og}$:
\be 
\ba{l}
-gs \cA|_{\Lb=0} \rightarrow \del_{\Og} A = - dx^{\mu}\, 
 (D_{\mu} c)^a T_a = - Dc, \\
 \\
\dsp{ -gs C|_{\Lb=0} \rightarrow \del_{\Og} c = \frac{g}{2}\, 
 f_{ab}^{\;\;c} c^a c^b\, T_c = \frac{g}{2}\, c^a c^b\, [T_a, T_b] 
 = g c^2.}\\
 \\
\dsp{ -gs \cF|_{\Lb=0} \rightarrow \del_{\Og} F = - \frac{g}{2}\, 
 dx^{\mu} \wedge dx^{\nu} f_{ab}^{\;\;c} F_{\mu\nu}^a c^b\, T_c = 
 - g [ F, c], }
\ea
\label{3.2.14}
\ee 
provided we take the BRST variational derivative $\del_{\Og}$ and 
the ghosts $c$ to anti-commute with the differential operator $d$: 
\be 
d \del_{\Og} + \del_{\Og} d = 0, \hs{2} 
 d c + c d = dx^{\mu} (\der_{\mu} c).  
\label{3.2.14.1}
\ee 
Returning to the Wess-Zumino consistency condition (\ref{3.2.17}), we
now see that it can be restated as a cohomology problem on the principle 
fibre bundle on which the 1-form $\cA$ lives. This is achieved by 
mapping the 4-form of ghost number $+1$ to a particular 5-form on the 
bundle, which is a local 4-form on $\cM$ and a 1-form on $G$; similarly 
one maps the 3-form of ghost number $+2$ to another 5-form which is a 
local 3-form on $\cM$ and a 2-form on $G$:
\be 
I^1_4 \rightarrow \og^1_4, \hs{2} 
I^2_3 \rightarrow \og^2_3,
\label{3.2.18}
\ee
where the two 5-forms must be related by 
\be 
-gs \og^1_4 = d \og^2_3.
\label{3.2.19}
\ee 
We now show how to solve this equation as part of a whole chain of
equations known as the {\em descent equations}. The starting point 
is a set of invariant polynomials known as the Chern characters 
of order $n$. They are constructed in terms of the field-strength
2-form:
\be 
F = dA - g A^2 = \frac{1}{2}\, dx^{\mu} \wedge dx^{\nu} 
 F^a_{\mu\nu}\, T_a, 
\label{3.2.20}
\ee 
which satisfies the Bianchi identity 
\be 
D F = dF - g \left[ A, F \right] = 0.
\label{3.2.21}
\ee 
The two-form $F$ transforms covariantly under gauge transformations 
(\ref{3.2.6}): 
\be 
F \rightarrow a^{-1} F a = \cF.
\label{3.2.22}
\ee 
It follows that the Chern character of order $n$, defined by 
\be 
Ch_n[A] = \mbox{Tr}\, F^n = \mbox{Tr}\, \cF^n, 
\label{3.2.23}
\ee 
is an invariant $2n$-form: $Ch_n[A] = Ch_n[\cA]$. It is also closed, 
as a result of the Bianchi identity: 
\be 
d\, Ch_n[A] = n \mbox{Tr}\, [ (DF)F^{n-1} ] = 0.
\label{3.2.24}
\ee 
The solution of this equation is given by the exact $2n$-forms:
\be 
Ch_n[A] = d \og^0_{2n-1}[A]. 
\label{3.2.25}
\ee 
Note, that the exact $2n$-form on the right-hand side lies entirely 
in the local space-time part $\cM$ of the bundle, because this is 
manifestly true for the left-hand side. 

\nit
Proof of the result (\ref{3.2.25}) is to be given; for the time being 
we take it for granted and continue our argument. 
First we define a generalized connection on the bundle by 
\be 
\bA(\xi) \equiv - \frac{1}{g}\, a^{-1}(\xi)\, \bd a(\xi) + a^{-1}(\xi) 
 A(x) a(\xi) = - \frac{1}{g}\, C(\xi) + \cA(\xi). 
\label{3.2.26}
\ee 
It follows, that the corresponding field strength on the bundle is 
\be 
\ba{lll}
\bF & = & \dsp{ \bd \bA - g \bA^2 = \lh d + s \rh \lh \cA - 
 \frac{1}{g}\, C \rh - g \lh \cA - \frac{1}{g}\, C \rh^2 }\\ 
 & & \\
 & = & d \cA - g \cA^2 = \cF.
\ea
\label{3.2.27}
\ee 
To go from the first to the second line we have used eq.(\ref{3.2.13}). 
This result is sometimes refered to as {\em the Russian formula} 
\ct{stora}. The result implies, that the components of the generalized
field-strength in the directions of the group manifold all vanish.

It is now obvious, that 
\be 
Ch_n[\bA] = \mbox{Tr}\, \bF^n = Ch_n[A];
\label{3.2.28}
\ee
moreover $\bF$ satisfies the Bianchi identity
\be 
\bD \bF = \bd \bF - g [\bA, \bF] = 0. 
\label{3.2.29}
\ee 
Again, this leads us to infer that 
\be 
\bd Ch_n[\bA] = 0 \hs{1} \Rightarrow \hs{1} 
 Ch_n[\bA] = \bd \og^0_{2n-1}[\bA] = d\og^0_{2n-1}[A], 
\label{3.2.30}
\ee 
where the last equality follows from eqs.(\ref{3.2.28}) and 
(\ref{3.2.25}). The middle step, which states that the $(2n-1)$-form
of which $Ch_n[\bA]$ is the total exterior derivative has the same 
functional form in terms of $\bA$, as the one of which it is the 
exterior space-time derviative has in terms of $A$, will be 
justified shortly. 

We first conclude the derivation of the chain of descent equations, 
which follow from the last result by expansion in terms of $C$: 
\be 
\ba{lll}
d \og^0_{2n-1}[A] & = & (d + s)\, \og^0_{2n-1}[\cA - C/g] \\
 & & \\
 & = & \dsp{ (d + s)\, \lh \og^0_{2n-1}[\cA] + \frac{1}{g}\, 
 \og^1_{2n-2}[\cA,C] + ... + \frac{1}{g^{2n-1}}\, \og^{2n-1}_0[\cA,C] \rh. }
\ea  
\label{3.2.31}
\ee 
Comparing terms of the same degree, we find 
\be 
\ba{l}
d \og^0_{2n-1}[A] = d \og^0_{2n-1}[\cA], \\
 \\
- gs \og^0_{2n-1}[\cA] = d \og^1_{2n-2}[\cA,C], \\ 
 \\ 
- gs \og^1_{2n-2}[\cA,C] = d \og^2_{2n-3}[\cA,C], \\
 \\
 ... \\
 \\
- gs\og^{2n-1}_0 [\cA,C] = 0. 
\ea 
\label{3.2.32}
\ee
Obviously, this result carries over to the BRST differentials: with 
$I^0_n[A] = \og_n^0[A]$, one obtains 
\be 
\del_{\Og} I_m^k[A,c] = d I_{m-1}^{k+1}[A,c], \hs{2} m + k = 2n-1,\;\; 
 k = 0,1,2,...,2n-1.
\label{3.2.33}
\ee 
The first line just states the gauge independence of the Chern character. 
Taking $n = 3$, we find that the third line is the Wess-Zumino 
consistency condition (\ref{3.2.19}):
\[ 
\del_{\Og} \, I^1_4[A,c]  = d I^2_{3}[A,c].  
\]

\nit
{\bf Proofs and solutions} \nl

\nit
We now show how to derive the result (\ref{3.2.25}); this will provide
us at the same time with the tools to solve the Wess-Zumino consistency 
condition. Consider an arbitrary gauge field configuration described 
by the Lie-algebra valued 1-form $A$. From this we define a whole 
family of gauge fields 
\be 
A_t = tA, \hs{2} t \in [0,1].
\label{3.3.1}
\ee 
It follows, that 
\be 
F_t \equiv F[A_t] = t dA - g t^2 A^2 = t F[A] -g (t^2 - t) A^2.
\label{3.3.2}
\ee 
This field strength 2-form satisfied the appropriate Bianchi identity:
\be 
D_t F_t = d F_t - g [ A_t, F ] = 0. 
\label{3.3.3}
\ee 
In addition, one easily derives 
\be 
\frac{dF_t}{dt}\, = dA - [A_t,A]_+ = D_t A,
\label{3.3.4}
\ee 
where the anti-commutator of the 1-forms implies a {\em commutator}
of the Lie-algebra elements. Now we can compute the Chern character 
\be 
\ba{lll}
Ch_n[A] & = & \dsp{ \int_0^1 dt\, \frac{d}{dt} \mbox{Tr} F_t^n =
 n \int_0^1 dt\, \mbox{Tr} \lh (D_t A) F_t^{n-1} \rh }\\
 & & \\
 & = & \dsp{ n d \int_0^1 dt\, \mbox{Tr} \lh  AF^{n-1}_t \rh. }
\ea
\label{3.3.5}
\ee 
In this derivation we have used both (\ref{3.3.4}) and the 
Bianchi identity (\ref{3.3.3}). 

It is now straightforward to compute the forms $\og_5^0$ and $\og_4^1$.
First, taking $n = 3$ in the result (\ref{3.3.5}) gives $Ch_3[A] = 
d \og_5^0$ with 
\be 
I^0_5[A] = \og^0_5[A] = 3 \int_0^1 dt\, \mbox{Tr} \lh (D_t A) F_t^2 \rh 
 = \mbox{Tr} \lh AF^2 + \frac{g}{2}\, A^3 F + \frac{g^2}{10}\,
 A^5 \rh. 
\label{3.3.6}
\ee 
Next, using eqs.(\ref{3.2.14}) the BRST differential of this expression 
gives $\del_{\Og} I_5^0 = d I_4^1$, with 
\be 
I_4^1[A,c] = - \mbox{Tr} \lh c \left[ F^2 + \frac{g}{2}\, \lh A^2 F + 
 AFA + FA^2 \rh + \frac{g^2}{2}\, A^4 \right] \rh.
\label{3.3.7}
\ee 
This expression determines the anomaly up to a  constant of 
normalization $\cN$: 
\be 
\Gam_a[A] = \cN\, \mbox{Tr} \lh T_a \left[ F^2 + \frac{g}{2}\, \lh 
 A^2 F + AFA + FA^2 \rh + \frac{g^2}{2}\, A^4 \right] \rh.
\label{3.3.8}
\ee 
Of course, the component form depends on the gauge group; for example,
for $SU(2) \simeq SO(3)$ it vanishes identically, as is true for 
any orthogonal group $SO(N)$; in contrast the anomaly does not 
vanish identically for $SU(N)$, for any $N \geq 3$. In that case it 
has to be anulled by cancellation between the contributions of chiral 
fermions in different representations of the gauge group $G$.